\documentclass[11pt]{article}
\pdfoutput=1

\usepackage{jheppub}
\usepackage{url,comment}
\usepackage{times}
\usepackage{latexsym}
\usepackage{graphicx, graphics, hyperref, amsmath, amssymb, slashed, color, bbm,amsthm}
\usepackage{array}
 
\newcommand{\nc}{\newcommand}

\nc{\beq}{\begin{equation}}
\nc{\eeq}{\end{equation}}
\nc{\barray}{\begin{eqnarray}}
\nc{\earray}{\end{eqnarray}}
\nc{\barrayn}{\begin{eqnarray*}}
\nc{\earrayn}{\end{eqnarray*}}
\nc{\bcenter}{\begin{center}}
\nc{\ecenter}{\end{center}}
\nc{\mc}{\mathcal}
\nc{\er}[1]{(\ref{eq:#1})} 
\nc{\onehalf}{\frac{1}{2}}
\nc{\partialbar}{\bar{\partial}}
\nc{\psit}{\widetilde{\psi}} 
\nc{\Tr}{\mbox{Tr}}
\nc{\hc}{\mbox{H.c.}}
\nc{\ev}{\;\mathrm{eV}}
\nc{\mev}{\;\mathrm{MeV}}
\nc{\gev}{\;\mathrm{GeV}}
\nc{\tev}{\;\mathrm{TeV}}
\nc{\f}{\frac}

\def\chii0{\chi_i^0}
\def\chij0{\chi_j^0}

\newcommand{\gsim}{\lower.7ex\hbox{$\;\stackrel{\textstyle>}{\sim}\;$}}
\newcommand{\lsim}{\lower.7ex\hbox{$\;\stackrel{\textstyle<}{\sim}\;$}}
\nc{\ttbar}{t\bar t}
\def\ifb{{\ \rm fb}^{-1}}

\def\eref#1{{Eq.~(\ref{eq:#1})}}
\def\fref#1{{Fig.~\ref{fig:#1}}}
\def\sref#1{{Sec.~\ref{sec:#1}}}

\title{Illuminating Dark Photons with High-Energy Colliders}

\author[a]{David Curtin,}
\author[b]{Rouven Essig,}
\author[c]{Stefania Gori,}
\author[d]{and Jessie Shelton}
\affiliation[a]{Maryland Center for Fundamental Physics, University of Maryland, College Park, MD 20742, USA}
\affiliation[b]{C.N.~Yang Institute for Theoretical Physics, Stony Brook University, Stony Brook, NY 11794, USA}
\affiliation[c]{Perimeter Institute for Theoretical Physics, 31 Caroline St. N, Waterloo, Ontario, Canada}
\affiliation[d]{
1110 West Green Street   Urbana, IL 61801, Dept of Physics,
University of Illinois at Urbana-Champaign}

\emailAdd{dcurtin1@umd.edu}
\emailAdd{rouven.essig@stonybrook.edu}
\emailAdd{sgori@perimeterinstitute.ca}
\emailAdd{sheltonj@illinois.edu}

\preprint{YITP-SB-14-49}

\abstract{
High-energy colliders offer a unique sensitivity to dark
  photons, the mediators of a broken dark $U(1)$ gauge theory that
  kinetically mixes with the Standard Model (SM)
  hypercharge. 
  Dark photons can be detected in the exotic decay of the 125 GeV
  Higgs boson, $h\to Z Z_D\to 4\ell$, and in Drell-Yan events, $pp\to
  Z_D\to\ell\ell$.  If the dark $U(1)$ is broken by a  hidden-sector
  Higgs mechanism, then mixing between the dark and SM Higgs bosons
  also allows the exotic decay $h\to Z_D Z_D\to 4\ell$.  We show that
  the 14 TeV LHC and a 100~TeV proton-proton collider provide powerful
  probes of both exotic Higgs decay channels.  
  In the case of kinetic mixing alone, direct Drell-Yan production offers the best sensitivity to $Z_D$, and can probe $\epsilon\gtrsim 9 \times
  10^{-4}$ ($4 \times 10^{-4}$) at the HL-LHC (100~TeV $pp$ collider). The exotic Higgs decay $h\to Z Z_D$ offers slightly weaker sensitivity, but both measurements are necessary to distinguish the kinetically mixed dark photon from other scenarios.
  If   Higgs mixing is also present, then the decay $h\to Z_D Z_D$ can
  allow sensitivity to the $Z_D$ for $\epsilon \gtrsim 10^{-9} - 10^{-6}$ 
  ($ 10^{-10} -  10^{-7}$) for the mass
  range $2 m_\mu < m_{Z_D}<m_h/2$ by searching for displaced dark photon decays.   
   We also compare the $Z_D$ sensitivity at $pp$ colliders to the indirect, but model-independent, 
   sensitivity of global fits to electroweak
  precision observables.   
We perform a global electroweak fit of the dark photon model, substantially updating previous work in the literature.
Electroweak precision measurements at  LEP, Tevatron, and the LHC  exclude $\epsilon$ as low as $3\times 10^{-2}$.   
   Sensitivity can be improved by up to a factor of $\sim 2$ with HL-LHC data, and an additional factor of $\sim 4$ with ILC/GigaZ data.
}

\arxivnumber{1412.0018}

\begin{document}

\maketitle

\section{Introduction}\label{sec:intro}

The Large Hadron Collider (LHC)
is dramatically increasing our
understanding of physics at and beyond the electroweak scale.  
This major advance is not only due to the LHC's unprecedented
center-of-mass energies, but also the large luminosity it is able
to realize. 
This allows for the potential discovery of not just heavy states that carry 
Standard Model (SM) quantum numbers, but also light, weakly coupled states. 
Searches for such hidden-sector degrees of freedom are an
important component of the physics program at the LHC and future
colliders, such as the envisioned 100~TeV proton-proton collider
\cite{Gershtein:2013iqa,fcc-beijing,fcc-cern} (see also \cite{
Zhou:2013raa, Cohen:2013xda, Jung:2013zya, Rizzo:2014xma, Low:2014cba, Cohen:2014hxa, Larkoski:2014bia, Hook:2014rka, Gori:2014oua, Curtin:2014jma, Chang:2014ida}).
Hidden sectors near the weak scale are motivated by naturalness
\cite{Chacko:2005pe, Burdman:2006tz, Craig:2013fga, Craig:2014aea,
  Burdman:2014zta}, thermal dark matter \cite{Silveira:1985rk, Pospelov:2007mp, Feldman:2006wd}, electroweak baryogenesis (see e.g.~\cite{Morrissey:2012db} for a recent review), but also represent a generic expectation for physics beyond
the SM \cite{Strassler:2006im}. 

As a prototypical hidden sector, we consider the compelling possibility of a spontaneously broken ``dark'' $U(1)_D$ gauge symmetry, mediated by a vector boson called the ``dark photon'', $Z_D$.  The dark photon's only renormalizable interaction
with the SM is through kinetic mixing with the hypercharge gauge
boson~\cite{Holdom:1985ag,Galison:1983pa,Dienes:1996zr}.  In addition,
if a dark Higgs mechanism is responsible for the spontaneous breaking
of the $U(1)_D$ gauge symmetry, the dark Higgs boson will in general have a renormalizable coupling to the 125~GeV SM-like Higgs, resulting in a mixing between the two physical scalar states.
The hidden sector's leading interactions with the SM may thus be through either the
{\em hypercharge portal}, via the kinetic mixing coupling, which we denote as $\epsilon$, or through the {\em Higgs portal}, via
the Higgs mixing, which we denote as $\kappa$.
The impressive integrated luminosities achievable by the LHC and
future hadron colliders make them powerful probes of the hidden sector through these two portals, while current and future electron-positron colliders can place interesting limits on kinetic mixing from precision electroweak tests (EWPTs), independently of the detailed spectrum of the hidden sector.

The dark photon mixes through the hypercharge portal with the SM photon and the $Z$ boson. If there are no hidden-sector states below the $Z_D$ mass, this mixing causes the dark photon to decay exclusively to SM particles, with sizable branching ratio to leptons.
We will focus on the dark photon mass range $m_{Z_D} > 2m_e\sim 1$~MeV, where the $Z_D$ can decay to SM fermions.\footnote{Below 1~MeV, the dominant decay mode is the long-lived $Z_D \to 3 \gamma$, which leads to a very different phenomenology that we will not discuss in this
paper~\cite{Jaeckel:2010ni,Hewett:2012ns,Essig:2013lka}.} There are
many experimental probes of dark photons with a mass above 1~MeV that
decay directly to SM particles, including precision QED measurements, rare meson decays, supernova cooling, collider
experiments, and beam dumps~\cite{Bjorken:2009mm,Batell:2009yf,Essig:2009nc,Freytsis:2009bh,Essig:2010xa,Blumlein:2011mv,Andreas:2012mt,Pospelov:2008zw,Reece:2009un,Aubert:2009cp,Hook:2010tw,Bjorken:1988as,Riordan:1987aw,Bross:1989mp,Babusci:2012cr,Archilli:2011zc,Abrahamyan:2011gv,Merkel:2011ze,Dent:2012mx,Davoudiasl:2012ig,Davoudiasl:2012ag,
  Davoudiasl:2013aya,Endo:2012hp,Balewski:2013oza,Adlarson:2013eza,Agakishiev:2013fwl,Blumlein:2013cua,Andreas:2013lya,Battaglieri:2014hga,Merkel:2014avp,Lees:2014xha,Adare:2014ega,Kazanas:2014mca,Echenard:2014lma,Gorbunov:2014wqa,NA482}.
Most of the current effort in the search for dark photons above the
MeV-scale is devoted to $m_{Z_D} \lesssim 10 \gev$, although
see~\cite{Curtin:2013fra,
  Gopalakrishna:2008dv,Davoudiasl:2012ag,Davoudiasl:2013aya,Chang:2013lfa,Falkowski:2014ffa,Cline:2014dwa,Hoenig:2014dsa}
 for recent discussions of exploring heavier $Z_D$.
There is no compelling reason for not exploring the entire mass range
that is experimentally accessible, since $m_{Z_D}$ is a free parameter
of the theory.  Dark photons with sub-GeV masses have received
attention recently as they could explain the $\sim 3.6\sigma$
discrepancy between the observed and SM value of the muon anomalous
magnetic moment~\cite{Pospelov:2008zw,Bennett:2006fi,Davier:2010nc}
and various dark matter related anomalies via new dark matter-$Z_D$
interactions~\cite{ArkaniHamed:2008qn,Pospelov:2008jd,Finkbeiner:2007kk,Fayet:2004bw}.
Several concrete models have also been suggested in which a sub-GeV
mass is generated
naturally~\cite{ArkaniHamed:2008qp,Cheung:2009fk,Baumgart:2009tn,Morrissey:2009ur,Essig:2009nc},
although in many cases masses above 10~GeV are equally natural.
However, part of the reason for the attention to sub-GeV masses has
been practicality --- the high-intensity experiments necessary to
directly probe dark photons, such as the $B$- and $\Phi$-factories and
various  fixed-target and 
beam dump experiments, do not have particle beams with
high-enough energy to effectively probe masses above 10~GeV.

With the advent of the 14~TeV run at the LHC, including the
high-luminosity run (HL-LHC), a possible future 100 TeV proton-proton collider, and various possibilities for future electron-positron colliders, we will have the exciting opportunity to probe dark photons
well above 10~GeV.  In fact, these experiments are the \emph{only}
known probe of dark photons above 10~GeV that explore $\epsilon$
values not disfavored by current EWPT. 
The hypercharge portal allows for direct production of the dark photon in Drell-Yan (DY) events, $pp\to Z_D \to \ell^+ \ell^-$. It also generates the  exotic Higgs decay $h \to Z Z_D$. Higgs mixing allows for a different exotic Higgs decay, $h \to Z_D Z_D$. Importantly, the Higgs portal can give experimental sensitivity to values of $\epsilon$ far below the reach of searches that only rely on the hypercharge portal, allowing us to peer deeply into the hidden sector.

\begin{figure}
\begin{center}
\begin{tabular}{ccc}
\includegraphics[width=0.4 \textwidth]{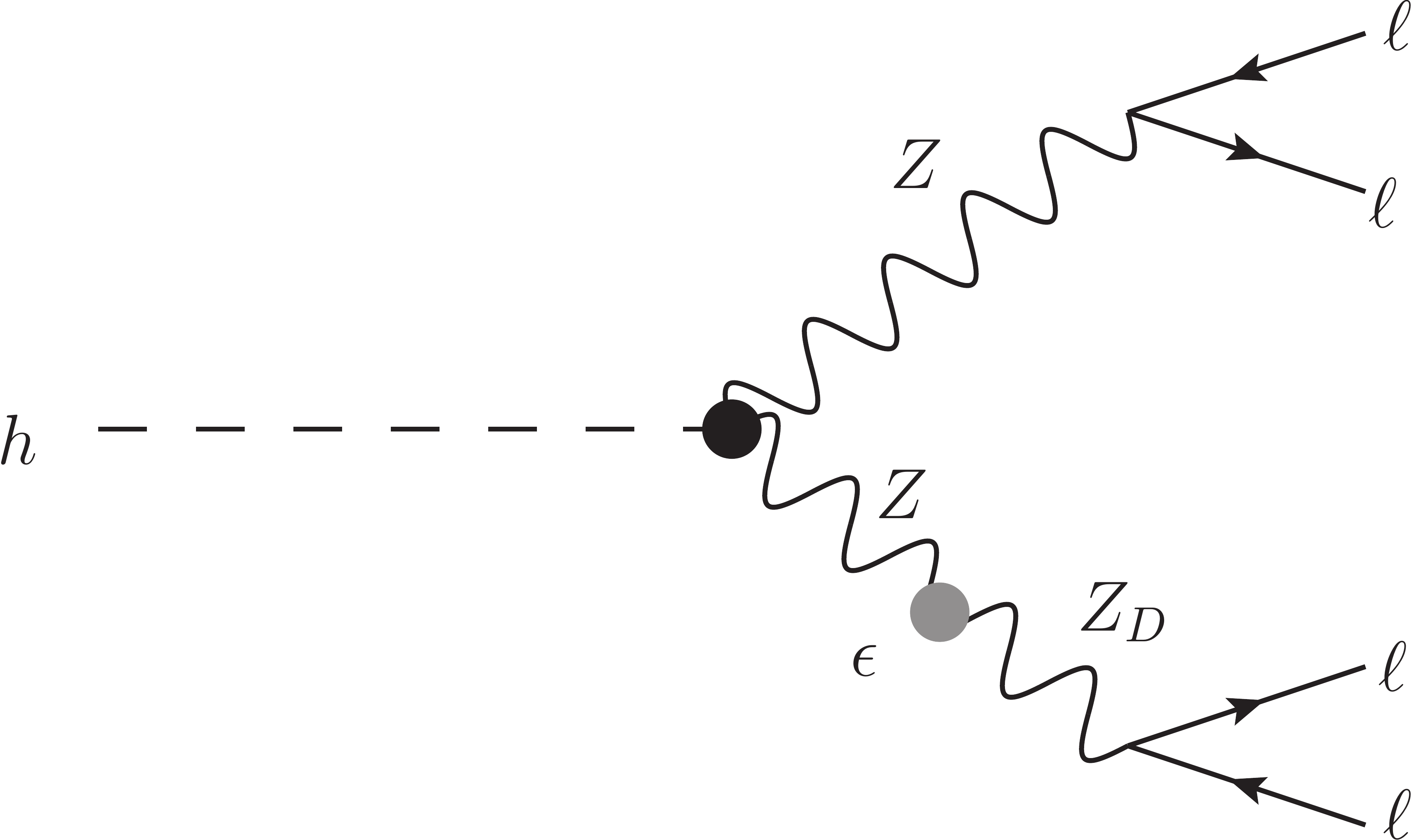}
& \hspace{4mm} &
\includegraphics[width=0.4\textwidth]{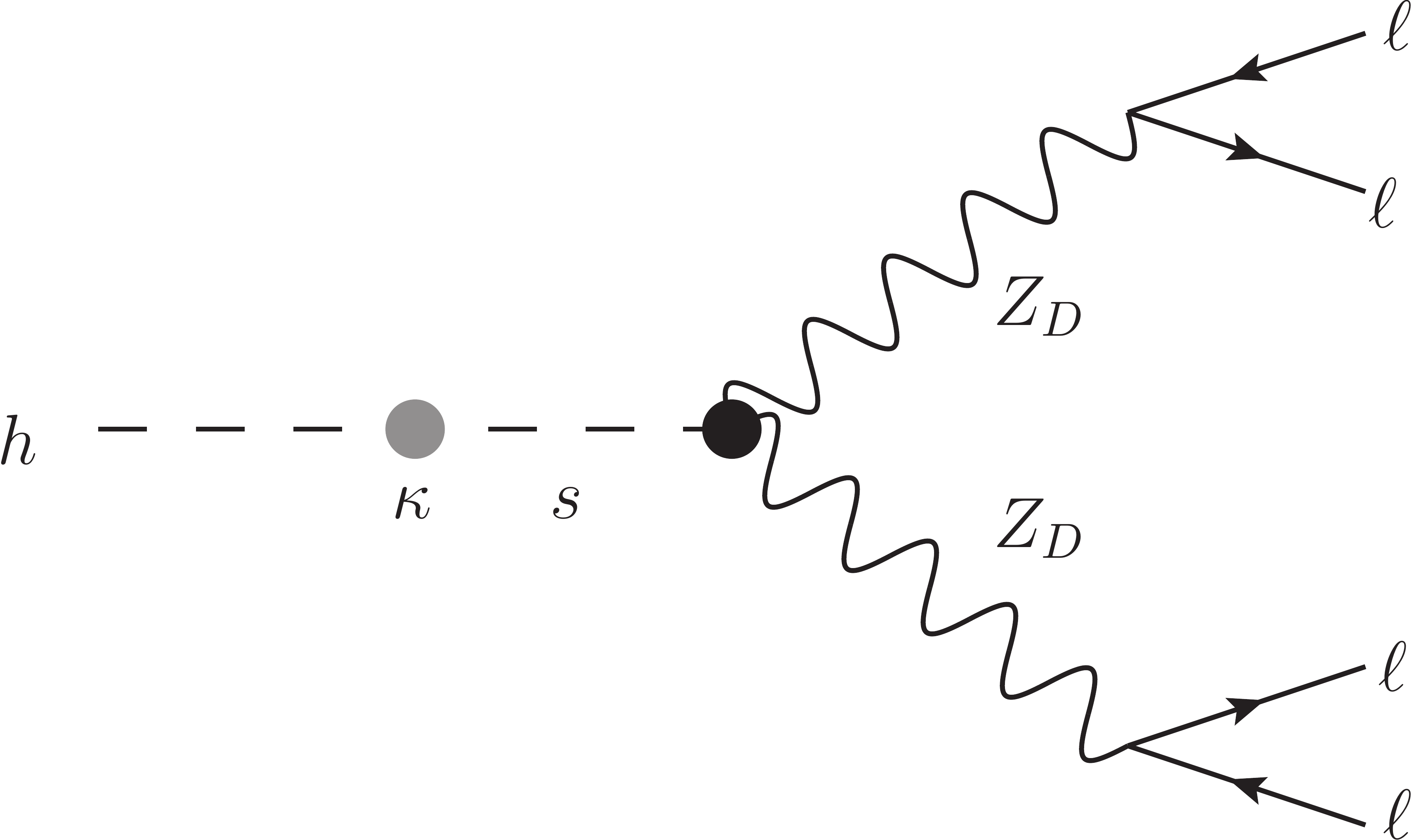}
\end{tabular}
\end{center}
\caption{ Exotic Higgs decays to four leptons induced by intermediate
  dark photons in the higgsed dark $U(1)$ model. \emph{Left:} $h\to Z_D
  Z^{(*)} \to 4\ell$ via the hypercharge portal. \emph{Right:} $h \to Z_D Z_D
  \to 4\ell$ via the Higgs portal. }
\label{fig:feynman}
\end{figure}

Existing data from LHC Run I (7 and 8~TeV run) are already able to set
new limits on dark photons. 
An initial study in~\cite{Curtin:2013fra} used LHC Run I data to set
limits on the exotic Higgs decays $h\to Z Z_D\to 4\ell$ and $h\to Z_D
Z_D \to 4\ell$, shown in \fref{feynman}. While the former decay probes a region in the
$\epsilon-m_{Z_D}$ plane that was already disfavored from EWPTs, the
latter generates the first constraints on Higgs portal couplings for
dark photon masses above a few GeV. Both analyses are
proofs-of-principle that future exotic Higgs decay searches are
sensitive to dark photons.  Meanwhile, experimental searches for the
NMSSM-motivated signal $h\to a a\to 4\mu$, in the region $m_a < 2
m_\tau$, provide limits on Higgs portal couplings for dark photons in
the same mass range \cite{Abazov:2009yi,Chatrchyan:2012cg,Chatrchyan:2011hr,CMS:2013lea} .
Other studies~\cite{Cline:2014dwa,Hoenig:2014dsa} pointed out that
existing LHC data constrains the production of dark photons in DY
events, disfavoring previously open parameter space.  

The upcoming HL-LHC and a future 100~TeV collider will significantly extend the
sensitivity of these direct searches. Furthermore, the LHC and a future ILC/GigaZ collider will improve the measurement of certain important electroweak precision observables (EWPOs). In this paper, we compare the reach of all these
experimental probes.  As part of this comparison, we perform a full
fit to electroweak precision measurements, presenting a new current
bound on dark photons, in addition to forecasting future sensitivity.

The organization of this paper is as follows.  Sec.~\ref{sec:model}
reviews the theory of a kinetically mixed $U(1)_D$.
Secs.~\ref{sec:pew}, \ref{sec:zzd}, and \ref{sec:dy}  analyze existing
constraints and future prospects for dark photons being probed via the hypercharge portal only, using EWPOs,
 the exotic Higgs decay $h\to Z Z_D \to 4\ell$, and  DY events,
respectively.  If the dark Higgs mixes with the ordinary Higgs, then
the decay $h\to Z_D Z_D \to 4\ell$ opens up, which we discuss in
Sec.~\ref{sec:zdzd}.
In Sec.~\ref{sec:improvements} we vary the assumed detector
capabilities at a future 100 TeV proton collider and discuss the
impact this has on our limit projections.
Sec.~\ref{sec:conclusions} contains our conclusions. Supplementary information about calculations in the dark photon model are provided by three Appendices.

\section{A kinetically mixed dark $U(1)$}\label{sec:model}

In this section, we review the theory of kinetic mixing between a
broken dark Abelian gauge symmetry, $U(1)_D$, and the SM hypercharge,
$U(1)_Y$.  The relevant gauge terms in the
Lagrangian are 
\beq\label{eq:KM}
\mathcal{L} \subset -\frac{1}{4} \,\hat B_{\mu\nu}\, \hat B^{\mu\nu} - \frac{1}{4} \,\hat Z_{D\mu\nu}\, \hat Z_D^{\mu\nu}  + \f{1}{2}\,\f{\epsilon}{\cos\theta} \,\hat Z_ {D\mu\nu}\,\hat B^{\mu\nu} + \f{1}{2}\, m_{D,0}^2\, \hat Z_D^\mu \, \hat Z_{D\mu}\, .
\eeq
Here the hatted fields indicate the original fields with non-canonical
kinetic terms, before any field redefinitions. The $U(1)_Y$ and
$U(1)_D$ field strengths are respectively $\hat B_{\mu\nu}
=\partial_\mu \hat B_{\nu} - \partial_\nu \hat B_{\mu}$ and $\hat
Z_{D\mu\nu} =\partial_\mu \hat Z_{D\nu} - \partial_\nu \hat Z_{D\mu}$,
$\theta$ is the Weinberg mixing angle, and $\epsilon$ is the kinetic
mixing parameter.

Since the interaction in Eq.~(\ref{eq:KM}) is renormalizable, the
parameter $\epsilon$ can take on any value. In particular, $\epsilon$
is not required to be small, which is one reason why the hypercharge
portal may provide the dominant interaction between the SM and a hidden 
sector.  Calculable values of $\epsilon$ are obtained in various
scenarios.  For example, if the $U(1)_D$ is embedded in a Grand
Unified Theory (GUT), the mixing is absent above the GUT scale, but
can be generated below it by particles charged under both $U(1)_Y$ and
$U(1)_D$.  If it is generated through a one-(two-)loop interaction,
one naturally obtains $\epsilon \sim 10^{-3}-10^{-1}$ ($\sim
10^{-5}-10^{-3}$)~\cite{ArkaniHamed:2008qp,Holdom:1985ag,Essig:2010ye,Baumgart:2009tn}.
A much larger range of $\epsilon$ has been suggested in certain string
theory
scenarios~\cite{Jaeckel:2010ni,Abel:2008ai,Goodsell:2011wn,Goodsell:2009xc};
see~\cite{Essig:2013lka,Hewett:2012ns,Jaeckel:2010ni} for recent
reviews.

Meanwhile, the general renormalizable potential for the SM and dark
Higgs fields is 
\begin{equation}
\label{eq:HM}
V_0 (H,S) =   -\mu^2|H|^2 +  \lambda |H|^4 -\mu_S^2 |S|^2 + \lambda_S |S|^4 +
   \kappa  |S|^2|H|^2\, .
\end{equation}
Here $H$ is the SM Higgs doublet, while $S$ is the SM-singlet `dark
Higgs' with $U(1)_D$ charge $q_S$.  The Higgs portal coupling, $\kappa$, which links the dark and SM Higgs fields is again a renormalizable
parameter, and may again be sizeable.  After spontaneous symmetry
breaking in the dark and visible sectors, $\kappa$ controls the mixing
between the SM Higgs boson $h_0$ and the uneaten component of the dark
Higgs, $s_0$. 
The importance of an additional Higgs portal coupling to sectors
containing a dark vector boson
has been realized before~\cite{Schabinger:2005ei, Gopalakrishna:2008dv}, particularly in the
context of hidden valley models~\cite{Strassler:2008bv}.
While some collider studies have been performed
\cite{Martin:2011pd, Davoudiasl:2012ig, Chang:2013lfa, Curtin:2013fra}, its
consequences have not been as widely explored as those of the
hypercharge portal.
The physical dark Higgs boson could in principle be produced at
colliders and give an additional experimental handle on the model. However, in this paper we focus on the additional SM Higgs decays to dark photons generated by this interaction, and assume the Higgs decay to dark scalars is kinematically forbidden. 

We have also constructed a fully consistent MadGraph 5 \cite{Alwall:2014hca} implementation of this model using FeynRules 2.0 \cite{Alloul:2013bka}. This MadGraph model consistently implements
all field redefinitions, thereby accurately modeling interference
effects, and has been extensively validated by comparing its output to
various analytical predictions. We utilize this model in the collider
studies of Secs.~\ref{sec:zzd} and \ref{sec:zdzd}, as well as for
the calculation of the three-body decay width $h\to Z_D \ell \ell$
below, and make it publicly available for follow-up
investigations. See Appendix~\ref{sec:mgmodel} for more information.

The minimal model we consider here can be extended to include strongly-coupled hidden sectors, supersymmetry, and mass mixing, among other possibilities; see e.g.~\cite{Strassler:2006im,Strassler:2006qa, Cheung:2009fk, Falkowski:2010cm, Chan:2011aa, Davoudiasl:2012ag,Gabrielli:2014oya} for
related work. The remainder of this section is devoted to a detailed discussion of the properties of the mass eigenstates in the SM and the hidden sector in the minimal higgsed model, including new results for the branching fractions of the  $Z_D$.

\subsection{The gauge sector}\label{subsec:gauge}

We first consider the gauge sector.  The field redefinition
\beq\label{eq:field-redef}
\left(\begin {array}{c} Z_{D,0} \\ B \end{array}\right) = 
       \left(\begin {array}{cc} \sqrt{1-\frac{\epsilon ^2}{\cos^2\theta}} & 0 \\ -\frac{\epsilon}{\cos\theta} & 1 \end{array}\right)
      \left(\begin {array}{c} \hat Z_D \\ \hat B \end{array}\right)\,,
\eeq
diagonalizes the gauge boson kinetic terms in
Eq.~(\ref{eq:KM}) (the
subscript `0' in $Z_{D,0}$ indicates that this is not yet a mass
eigenstate).  We define
\beq
\eta =\frac{\epsilon} {\cos\theta\sqrt{1-\frac{\epsilon ^2}{\cos^2\theta}}}\,,
\eeq
and take the dark vector to have mass $m_{D,0}^2\equiv m_{Z,0}^2\times
\delta^2$ before mixing with SM fields, where $m_{Z,0}$ is the mass of the SM $Z$-boson before mixing.  After electroweak symmetry breaking (EWSB), and
after applying the field redefinition~Eq.~(\ref{eq:field-redef}), we
can write the full mass-squared matrix for the three neutral vectors
as
\beq\label{eq:massmatrix}
\mathcal{M}^2_V = m_{Z,0}^2 \left(\begin {array}{ccc} 0  & 0 &  0 \\  0 & 1 & -\eta \sin\theta \\
                          0 &-\eta\sin\theta & \eta ^ 2\sin ^ 2\theta +\delta ^ 2 \end{array}\right)
\eeq
in the basis $(A^\mu, Z_0^\mu, Z_{D,0}^\mu)$.  Here $A^\mu$ is the
massless SM photon field and $Z_0^\mu$ is the SM $Z$-boson field with
mass $m^2_{Z,0} = (g^2+g_Y^2) v^2/4$, where $v \simeq 246$~GeV is the
SM Higgs vacuum expectation value (vev) and $g$ ($g_Y$) is the
$SU(2)_L$ ($U(1)_Y$) gauge coupling.  Note that $A^\mu$ does not mix
with the other neutral fields and remains massless, since
electromagnetism remains unbroken.  However, the $Z_0^\mu$ and
$Z_{D,0}^\mu$ fields mix, and we can derive the mass eigenstates by
diagonalizing the ($Z_0^\mu, Z_{D,0}^\mu$) submatrix of
Eq.~(\ref{eq:massmatrix}) with
\beq
\left(\begin {array}{c} Z \\ Z_D \end{array}\right) = 
      \left(\begin {array}{cc} \cos\alpha &\sin\alpha \\ -\sin\alpha &\cos\alpha \end{array}\right)
     \left(\begin {array}{c} Z_0 \\ Z_{D,0} \end{array}\right),
\eeq
where the mixing angle is given by\footnote{This convention for the mixing angle is
  chosen so that $\alpha \to 0$ (not $\pi$) when $\epsilon \to 0$,
  regardless of whether $m_{Z_D}$ is larger or smaller than
  $m_{Z}$. We make a similar choice when defining the Higgs mixing
  angle below.} 
\beq\label{eq:mixingangle}
\tan\alpha = \frac{1  - \eta^2 \, \sin^2\theta  - \delta^2 - \mathrm{Sign}(1-\delta^2)\, \sqrt{4\, \eta^2 \, \sin^2\theta + (1  - \eta^2\, \sin^2\theta - \delta^2)^2}}{2\, \eta \, \sin\theta}\,.
\eeq
The eigenvalues of the submatrix, in units of $m_{Z,0}^2$, are
\beq\label{eq:masses}
m^2_{Z, Z_D} = \frac{1}{2}\left( 1+\delta ^ 2+\eta^2 \, \sin^2\theta \pm 
   \mathrm{Sign}(1-\delta^2) \sqrt{(1+\delta^2 +\eta^2 \, \sin^2\theta)^2 - 4\, \delta^2} \right)\,.
\eeq
For $\epsilon\ll 1$ and $\delta \ll 1$, the masses are 
$m^2_{Z_D} \simeq \delta^2 m_{Z,0}^2 \left(1- \epsilon^2 \tan^2\theta \right)$ and $m^2_{Z} \simeq m_{Z,0}^2 \left(1+\epsilon^2\tan^2\theta\right)$.  

Having written the theory in terms of canonically normalized kinetic
terms and mass eigenstates, several important consequences become
apparent.  The interaction between the $Z$-boson and the SM fermions,
$Z\bar f f$, has been modified from the SM expectation,
\begin{eqnarray}\label{eq:Zcoupl}
\mathcal{L}_{Z \bar f f} & = & g_{Z f \bar f} \, Z_{\mu} \bar f \gamma^\mu f\, 
\nonumber \\
g_{Z f \bar f} & \equiv & \frac{g}{\cos\theta}\,\left(\cos\alpha \, (t^3\, \cos^ 2\theta- Y \, \sin^2\theta)
     +\eta \, \sin\alpha \, \sin\theta\, Y \right)\,,  
\end{eqnarray}
where $t^3$ and $Y$ are the weak isospin and hypercharge value,
respectively, of the fermion $f$.  The $Z_D \bar f f$ interaction is
non-zero,
\begin{eqnarray}\label{eq:ZDcoupl}
\mathcal{L}_{Z_D \bar f f} & = & g_{Z_D f \bar f} \, Z_{D,\mu} \bar f \gamma^\mu f
  \nonumber \\
g_{Z_D f \bar f} & \equiv &  \frac{g}{\cos\theta}\, \left(- \sin\alpha \,( t^3 \,\cos^ 2\theta - Y\,\sin^2\theta)
     +\eta \, \cos\alpha \,\sin\theta\, Y \right).
\end{eqnarray}
For $\epsilon\ll 1$, at leading order, the $Z_D$ coupling to fermions
is ``photon-like'' for $\delta \ll 1$: $g_{Z_D f \bar f} \simeq
\epsilon\, e\, Q + \mathcal{O}(\delta^2)$, where $e = \sqrt{4\pi
  \alpha}$ is the electromagnetic coupling and $Q$ the fermion charge,
and ``$Z$-like'' for 
$|\delta| \simeq 1$:
$g_{Z_D f \bar f} \simeq\epsilon
\frac{g}{\cos\theta}(t^3\cos^2\theta-Y\sin^2\theta)$.  Furthermore,
the interaction $Z\bar ff$ receives its first correction at $\mathcal{O}(\epsilon^2)$,
given by 
$g_{Z\bar ff}\simeq g_{Z\bar   ff}^{\rm{SM}}+\epsilon^2\frac{\tan^2\theta}{2}\frac{g}{\cos\theta}(t^3-Q(1+\cos^2\theta) + 2 Y \delta^2)/(1-\delta^2)^2$. 
The admixture of the $Z$-boson in the $Z_D$ mass eigenstate gives rise
to a coupling between the SM Higgs boson to $Z$ and $Z_D$ after EWSB,
\begin{eqnarray}\label{eq:higgsZZD}
\mathcal{L}_{h Z Z_D}& =& \left[\frac{2 i\eta\sin\theta}{v} m^2_{Z_0} \left(\frac{\eta^2\sin^2\theta-1}{ \eta\sin\theta} 2\sin 2\alpha
               - \cos 2\alpha\right) \right]  h Z_\mu Z_D^\mu \nonumber 
\\
 &=&
 \frac {2 i \eta\sin\theta} {v} \frac{m_{Z_D}^2 m_{Z}^2}{m ^ 2_{Z}-m^ 2_{Z_D}}  h Z_\mu Z_D^\mu
          +\mc{O}(\eta^3)\,,
\end{eqnarray}
where, again, $m_{Z,0}$ is the mass of the $Z$ before mixing, and
$m_{Z, Z_D}$ are the {\em physical} $Z, Z_D$ masses.  
At $\mathcal{O}(\epsilon^2)$, this vertex mediates both (i) the decay of
the Higgs to a (potentially off-shell) $Z$ and an on-shell $Z_D$, and
(ii) interference from an off-shell $Z_D$ in the decay $h\to Z^{(*)}
Z^* \to 4 f$.  Sensitivity to $Z_D$ will come almost entirely from its
production on-shell, and thus we ignore the interference contributions
in our collider studies below.   However, post-discovery, the
$Z_D$ interference terms in Higgs decays to four leptons may present a unique opportunity to distinguish the {\em sign} of $\epsilon$, though this would, of course, require much larger integrated luminosities than those needed for discovery.

Note that the overall rate for the SM decay $h\to Z Z^*\to 4 f$ is
also modified at $\mathcal{O}(\epsilon^2)$, owing to the modifications of
the $Z$ mass, $hZZ$ vertex, and $Z\bar f f$ couplings, all of which
receive $\mathcal{O}(\epsilon^2)$ contributions.  However, this effect
is simply an overall numerical suppression of the rate, and does not
change the shape of any lepton distributions. Due to its small size, it is therefore not observable in
the currently forseeable future, given the theoretical uncertainties
on the SM branching fractions Br$(h\to c\bar c)$ and Br$(h\to b\bar
b)$, with additional limitations from experimental precision in the
determination of Br$(h\to ZZ^*)$. For this reason we do not consider these
contributions further.

We first discuss dark photon decays, since this affects the
experimental signatures of all exotic Higgs decay modes under
consideration in this paper.  The lowest order (LO) dark photon decay widths
are 
\begin{equation}\label{eq:width-ZDff}
\Gamma(Z_D \to \bar f f) = \frac{N_c}{24 \pi  m_{Z_D}} \sqrt{1-\frac{4 m_f^2}{m_{Z_D}^2}} \left(m_{Z_D}^2 \left(g_L^2+g_R^2\right)-m_f^2 \left(-6 g_L g_R+g_L^2+g_R^2\right)\right),
\end{equation}
where $g_{L,R} = g_{Z_D f_{L,R} \bar f_{L,R}}$ are given in
Eq.~(\ref{eq:ZDcoupl}) and are proportional to $\epsilon$ for $\epsilon \ll
1$.  
This tree-level parton-level formula is a good approximation for
$m_{Z_D}$ above the $b\bar b$ threshold. For smaller masses, threshold
effects, QCD corrections, and hadronic resonances cannot be
neglected. To obtain consistent predictions for the dark photon total
width and branching fractions across the entire relevant mass range we
must include experimental information and higher order QCD
calculations.

Define the ratio
\begin{equation}
\label{eq:RZD}
R_{Z_D} \equiv  \frac{\Gamma(Z_D \to \ \mathrm{hadrons})}{\Gamma(Z_D \to \mu^+ \mu^-)} \underset{\epsilon \ll 1}{=} R_{Z_D}(m_{Z_D}) \, ,
\end{equation}
which is independent of $\epsilon$ for $\epsilon \ll 1$.
If we knew this function, including higher order corrections, we could
write the total width of $Z_D$ to high accuracy as
\begin{equation}
\label{eq:Zdtotalwidth}
\Gamma_{Z_D} = R_{Z_D} \Gamma(Z_D \to \mu^+ \mu^-) + \sum_{f = e,\mu,\tau, \nu_{e,\mu,\tau}} \Gamma(Z_D \to f \bar f)\,,
\end{equation}
where all the partial widths are computed at LO using
\eref{width-ZDff}. This also gives the leptonic branching fractions
\begin{equation}
\label{eq:Zdbranchingratios}
\mathrm{Br}(Z_D \to \ell \ell) = \frac{\Gamma(Z_D \to \ell
\ell)}{\Gamma_{Z_D}}
\end{equation}
to high accuracy.

\begin{figure}[t]
\begin{center}
\hspace*{-8mm}
\begin{tabular}{cc}
\includegraphics[width=0.5\textwidth]{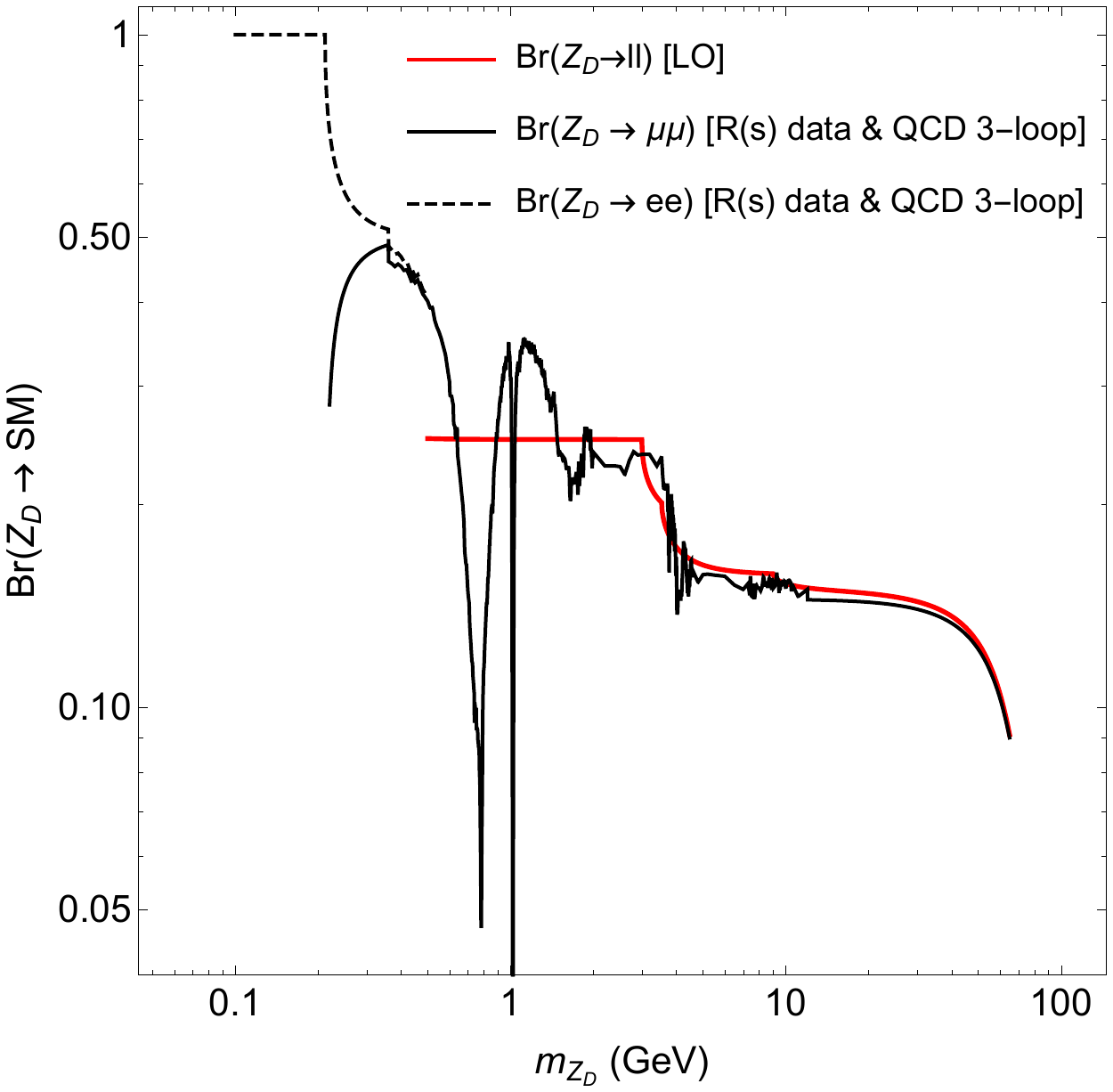}&
\includegraphics[width=0.5\textwidth]{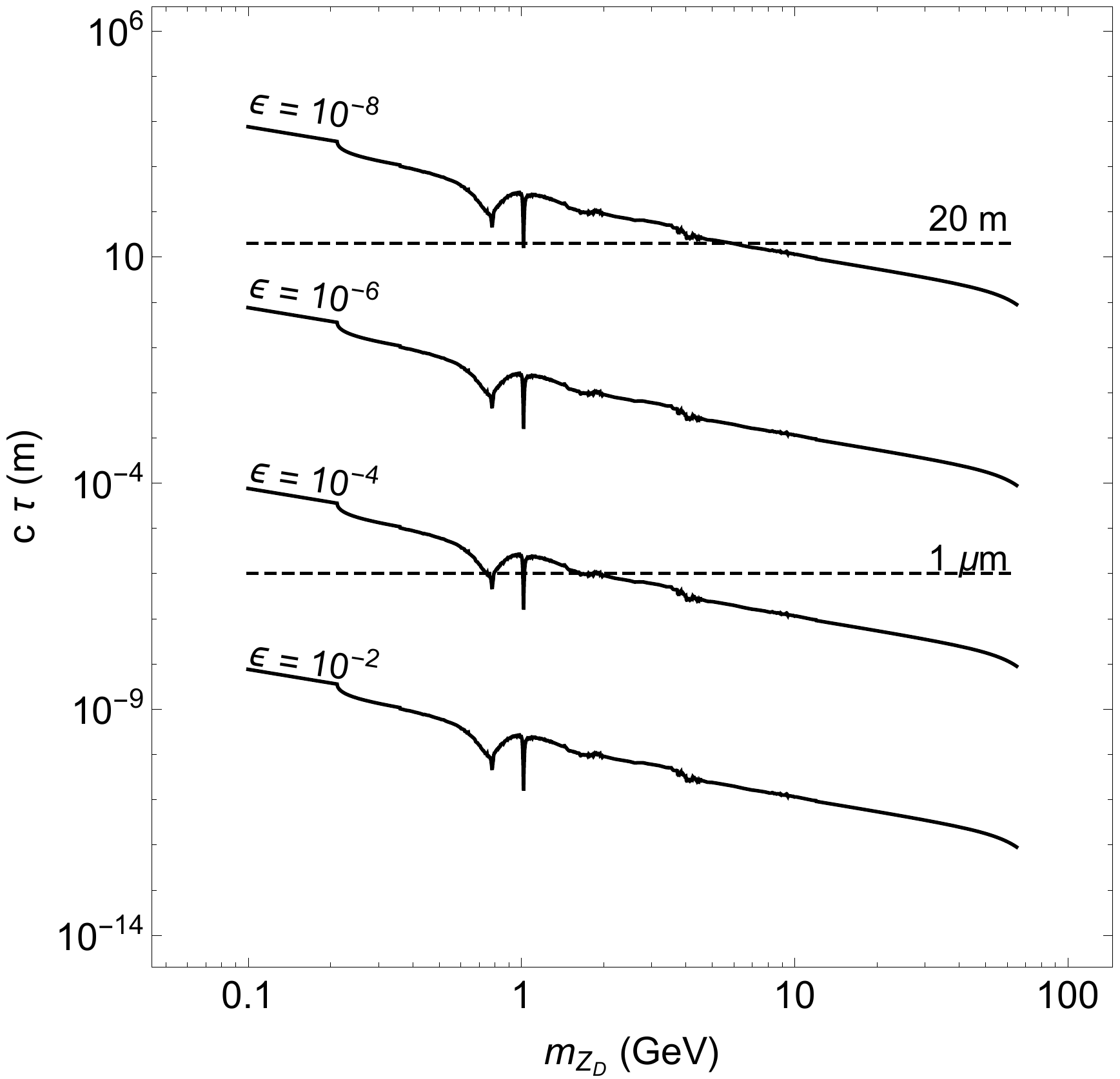}
\end{tabular}
\end{center}
\caption{\small{ \emph{Left:} Leptonic branching fraction of $Z_D$. \emph{Right:}
    Decay length of $Z_D$ for different $\epsilon$. The dashed lines
    indicate boundaries between qualitatively different experimental
    regimes: prompt decay for $c \tau \lesssim 1 \mu$m and likely
    escape from an ATLAS-size detector for $c \tau \gtrsim 20$m.}}
\label{fig:BrZdll}
\end{figure}

In fact, we can obtain $R_{Z_D}(m_{Z_D})$ very accurately. For
$m_{Z_D} < 12 \gev$, the couplings of $Z_D$ to SM fermions are
photon-like up to corrections of order $\delta^2$ ($< 2\%$). Furthermore,
for $\sqrt{s} \ll m_Z^2$, the experimental ratio
\begin{equation}
R(s) \equiv  \frac{\sigma(e^+e^-\to {\rm
    hadrons})}{\sigma(e^+e^-\to {\rm \mu^+\mu^-})}
\end{equation}
is highly dominated by off-shell $\gamma^*\to f \bar f$ in the
$s$-channel. Therefore we can use experimental data~\cite{pdg} to
determine
\begin{equation}
R_{Z_D}(m_{Z_D}) = R(m_{Z_D}^2) \ \ \ \ \  \ \ \ \ \mathrm{for} \ \ \ \  m_{Z_D} < 12 \gev,
\end{equation}
which includes all higher order QCD corrections.\footnote{There is no data below $\sqrt{s} = 0.36 \gev$, so for $2 m_\pi < m_{Z_D} < 0.36 \gev$ we set $R = 0$. This does not affect the results we derive in this paper.}
For higher masses,
the $Z_D$ couplings are different from that of the photon. In this
regime, we use existing 3-loop QCD calculations of $R(s)$
\cite{Chetyrkin:2000zk} to compute $R_{Z_D}$ by replacing the SM
coupling between the (axial) vector current and quarks by the $Z_D$
couplings in \eref{ZDcoupl}.\footnote{See \cite{Chetyrkin:1996ia} for
  a general review on these computations.} In the notation of
\cite{Chetyrkin:2000zk}, we can then determine
\begin{equation}
R_{Z_D}(m_{Z_D}) =  \frac{[R^{(v)} + R^{(a)}]_\mathrm{hadrons}}{[R^{(v)} + R^{(a)}]_{\mu \mu  \ \ \ \ \ \ \ }}
\ \ \ \ \  \ \ \ \ \mathrm{for} \ \ \ \  m_{Z_D} > 12 \gev,
\end{equation}
where the running QCD coupling was computed at $3+$ loop order 
using the \texttt{RunDec} Mathematica package \cite{Chetyrkin:2000yt}. The
resulting leptonic branching fraction and total width of the dark
photon are shown in \fref{BrZdll}. We will use these high-precision
results throughout the paper, but, as the figure shows, the LO expression for
total width and leptonic branching fraction is an excellent
approximation at higher masses: the higher order corrections are 4\%
(1.5\%) at $m_{Z_D} = 12 \gev$ ($60 \gev$). See
Appendix~\ref{sec:brtables} for tables of these branching ratios.

The above interactions Eqs.~(\ref{eq:ZDcoupl}) and (\ref{eq:higgsZZD})
allow the decay $h\to Z_D Z^{(*)} \to 4\ell$, shown in \fref{feynman}
(left).  The partial width for the exotic two-body decay $h\to Z Z_D$
is
\begin{eqnarray}
\Gamma(h\to Z Z_D) &=& \frac{\eta ^2 \, \sin^2\theta \, m_Z^2 \,m_{Z_D}^2}{16 \,\pi \, v^2 \,m_h^3 \,\left(m_Z^2-m_{Z_D}^2\right)^2} \left(-2 m_{Z_D}^2 \left(m_h^2-5 m_Z^2\right)+m_{Z_D}^4+\left(m_h^2-m_Z^2\right)^2\right)  \nonumber 
\\
&&\times \, \sqrt{-2 m_h^2 \left(m_{Z_D}^2+m_Z^2\right)+\left(m_Z^2-m_{Z_D}^2\right)^2+m_h^4}\,.
\label{eq:htoZDZ}
\end{eqnarray}
The partial width for the three-body decay $h\to Z_D Z^*\to Z_D
\ell\ell$ is, to leading order in $\epsilon$,
\begin{equation}
\label{eq:hto4lthreebody}
\Gamma(h\to Z_D Z^{(*)}\to 4\ell) = \frac{\eta^2\sin^2\theta}{64\pi^3}\,\frac{m_Z^4}{m_h^3 v^2}\, (g_{Z,L}^2+g_{Z,R}^2)
  \left(\frac{\delta^2}{1-\delta^2}\right)^2 \; \mathcal{I}(m_Z,m_h,m_{Z_D}),
\end{equation}
where
\begin{eqnarray} 
\mathcal{I}(m_Z,m_h,m_{Z_D}) &\equiv &\int_0^{(m_h-m_{Z_D})^2} dw\, \frac{(m_{Z_D}^4 - 2m_{Z_D}^2 (m_h^2-5w)+(m_h^2-w)^2)}{6m_{Z_D}^2(m_Z^2-w)^2}\nonumber\\
 & &\phantom{smush}\times \sqrt{m_{Z_D}^4+(m_h^2-w)^2-2m_{Z_D}^2(m_h^2+w)}\,,
\end{eqnarray}
and $g_{Z,L},\,g_{Z,R}$ are the (tree-level) couplings of a lepton to
the $Z$ boson, as in \eref{Zcoupl}. For $m_{Z_D} \sim m_h - m_Z$,
finite-width effects of the $Z$ are most easily accounted for by
computing the partial width in MadGraph. \fref{Br4l} (top) shows
$\mathrm{Br}(h\to Z_D Z^{(*)} \to 4\ell)$ for different values of
$\epsilon$.

We note that the kinetic mixing interaction by itself also generates
the decay $h\to Z_D Z_D$.  This decay is highly suppressed, as it
requires that both $Z$'s in $h\to ZZ^{(*)}$ mix with the $Z_D$, see
e.g.~\cite{Curtin:2013fra}, and appears first at
$\mathcal{O}(\epsilon^4)$.  However, if the SM Higgs mixes with the
hidden-sector Higgs, then this decay can proceed through Higgs portal
mixing instead, allowing it to be potentially sizable, as we will now
discuss below.

\begin{figure}[t]
\begin{center}
\includegraphics[width=1\textwidth]{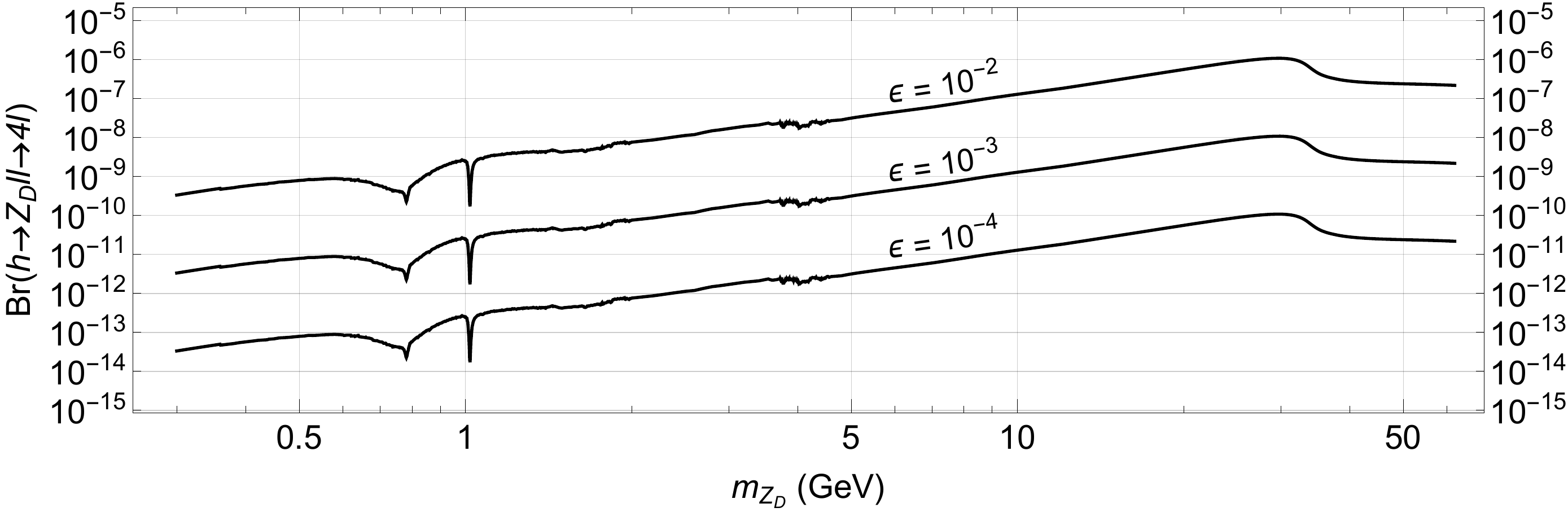} \vspace{3mm} \\
\includegraphics[width=1\textwidth]{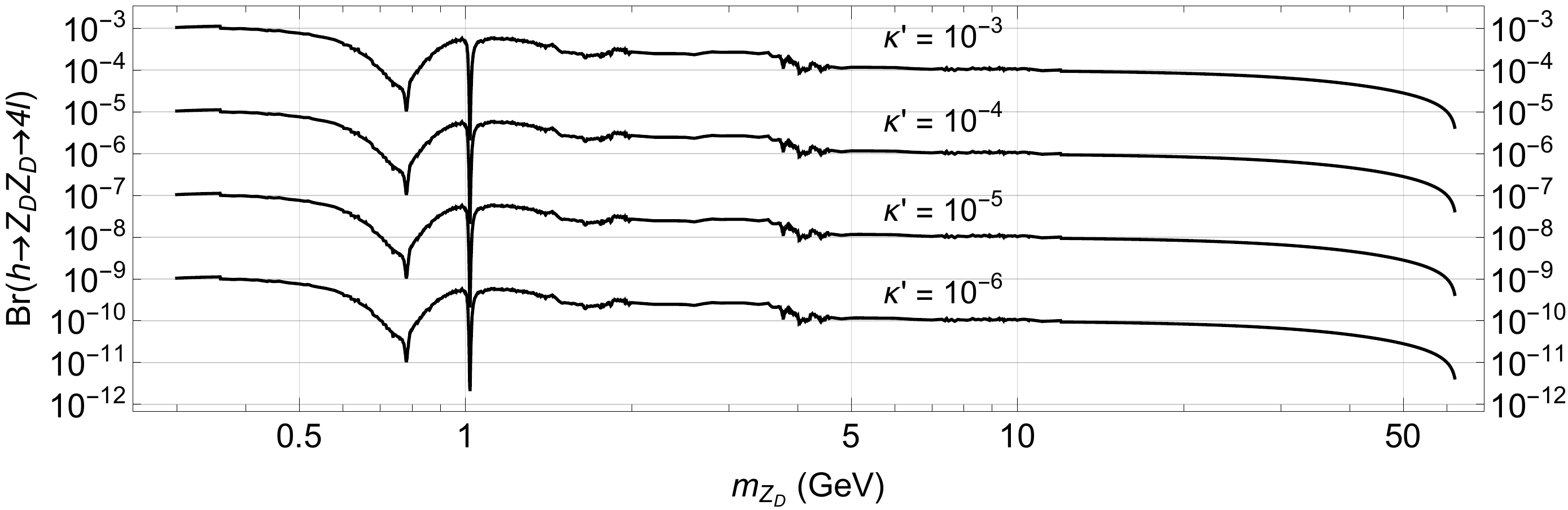} \
\end{center}
\caption{{\small Br$(h\to Z_D Z^* \to 4\ell)$ \emph{(top)} and Br$(h\to Z_D
    Z_D \to 4\ell)$ \emph{(bottom)} for different values of $\epsilon$ and
    $\kappa'$.  }}
\label{fig:Br4l}
\end{figure}

\subsection{The Higgs sector}\label{subsec:higgs}

We now consider the Higgs sector.  Electroweak symmetry is broken by
$\left< H \right> = (0,v/\sqrt{2})$, where $v \approx 246 \gev$. The
singlet acquires a vev $\left< S \right> = v_S/\sqrt{2}$, which
generates the dark photon mass of \eref{KM}:
\begin{equation}
m_{D,0} = g_D q_S v_S.
\end{equation}
Rewriting the scalar mass terms
\eref{HM} in terms of these vevs gives
\begin{equation}
\mu^2 = v^2 \lambda + \frac{1}{2} \kappa v_S^2  \ \ , \ \ \ \ \ 
\mu_S^2 = v_S^2 \lambda_S + \frac{1}{2} \kappa v^2.
\end{equation}
Expanding in small fluctuations $h_0, s_0$ (not yet mass
eigenstates) around the vacuum, the Higgs mass matrix in the $(h_0,
s_0)$ basis is
\begin{equation}
M^2_{h_0 s_0} = \left( \begin{array}{cc}
2 v^2 \lambda & v v_S \kappa \\
v v_S \kappa & 2 v_S^2 \lambda_S
\end{array}\right),
\end{equation}
We  define mass eigenstates $(h, s)$
\begin{equation}
\left( \begin{array}{c} h \\ s \end{array}\right) 
=
\left(
\begin{array}{cc} \cos \theta_h & -\sin \theta_h \\ \sin \theta_h & \cos \theta_h \end{array}\right)
\left( \begin{array}{c} h_0 \\  s_0 \end{array}\right) 
\end{equation}
(note the minus sign). For small mixing angles, $h$ is dominantly
SM-Higgs-like and $s$ is dominantly singlet-Higgs like. The mixing
angle is given by
\begin{eqnarray}
\tan \theta_h &=& \frac{v^2 \lambda - v_S^2 \lambda_S - \mathrm{Sign}(v^2 \lambda - v_S^2 \lambda_S) 
\sqrt{v^4 \lambda^2 + v_S^4 \lambda_S^2 + v^2 v_S^2 (\kappa^2 - 2 \lambda \lambda_S)}}
{v v_S \kappa}\,.
\end{eqnarray}
If we define 
\begin{equation}
s_h \equiv \frac{\kappa}{2} \frac{v v_S}{v_S^2 \lambda_S - v^2 \lambda}\,
\end{equation}
then for small Higgs mixing, 
\begin{equation}
\tan \theta_h \approx \sin \theta_h =  s_h  + \mathcal{O}(\kappa^2)\,.
\end{equation}
The mass eigenvalues are
\begin{equation}
m^2_{h, s} = v^2 \lambda + v_S^2 \lambda_S \pm \mathrm{Sign}(v^2 \lambda - v_S^2 \lambda_S)
\sqrt{v^4 \lambda^2 + v_S^4 \lambda_S^2 + v^2 v_S^2 (\kappa^2 - 2 \lambda \lambda_S)}\,.
\end{equation}
For small Higgs mixing, this reduces to
\begin{eqnarray}
m_h^2 & = & 2 \lambda v^2 + 2 s_h^2 (\lambda v^2 - \lambda_S v_S^2) + \mathcal{O}(\kappa^4)
 \\
m_{s}^2 & = &  2 \lambda_S v_S^2 - 2 s_h^2 (\lambda v^2 - \lambda_S v_S^2) + \mathcal{O}(\kappa^4)
\,.
\end{eqnarray}

Since the $s Z_D Z_D$ coupling is non-zero ($=2g_D q_S m_{Z_D}$), the
mixing between $h$ and $s$ generates a non-zero $h Z_D Z_D$
coupling. To lowest order in $\kappa$, this is
\begin{equation}\label{eq:higgsZDZD}
\mathcal{L}_{h Z_D Z_D}  =  2 \, s_h\, 
 \frac{m_{Z_D}^2}{v_s} \, h\, Z_{D \mu}\, Z_D^\mu\,.
\end{equation}
This allows for the decay $h\to Z_D Z_D$, shown in \fref{feynman}
(right). The partial width to lowest order in $\kappa$ is
\begin{eqnarray}\label{eq:htoZDZD}
\Gamma(h \rightarrow Z_D Z_D) &=& {\kappa'}^2 \ \frac{1}{32 \pi}  \ \frac{v^2}{m_h} \sqrt{1 - \frac{4 m_{Z_D}^2}{m_h^2}}  \  \frac{(m_h^2 + 2 m_{Z_D}^2)^2 - 8 (m_h^2 - m_{Z_D}^2)m_{Z_D}^2}{m_h^4},
\end{eqnarray}
where we have have introduced the dimensionless parameter $\kappa'$,
defined as
\begin{equation}\label{eq:kappaprime}
\kappa' = \kappa \  \frac{m_h^2}{|m_h^2-m_s^2|},
\end{equation}
which, along with $m_{Z_D}$, controls the size of this exotic Higgs decay.
The resulting $\mathrm{Br}(h\to Z_D Z_D \to 4\ell)$ is shown in
\fref{Br4l} (bottom) for different values of $\kappa'$. It does not depend on $\epsilon$, but the \emph{decay length} of the dark photons does. 

An additional interaction exists that allows for $h \to ss$.  We will simply assume that $s$ is heavy enough that this decay is
kinematically forbidden, but see e.g.~\cite{Curtin:2013fra} for a more
comprehensive discussion of the several possibilities. One can also produce the singlet scalar directly via its inherited SM couplings, analogously to the SM Higgs boson. The dominant mode for $m_s > m_h/2$ is gluon fusion, but as we discuss in \sref{zdzd}, the Higgs portal is more sensitively probed by $p p \to h \to Z_D Z_D$ than by $p p \to s \to Z_D Z_D$, even though both processes occur at the same order in $\kappa$. The singlet scalar can also be produced via the process $p p \to Z_D^* \to Z_D s$, which occurs at the same order of $\epsilon$ as the exotic Higgs decay $h\to Z Z_D$. All of these channels should be studied more comprehensively in the future, but are beyond the scope of this paper. 

As demonstrated in \fref{Br4l}, the Br($h\to Z_D Z_D$) can be quite  
sizable.  However, this decay is invisible unless $Z_D$ decays inside the detector, and therefore $\epsilon$ cannot be too small in order for this Higgs portal decay to be observable.
A large
fraction of $Z_D$ will decay inside the detector for $\epsilon \gtrsim
10^{-7}$ (see right panel of Fig. \ref{fig:BrZdll}), but the large luminosity of hadron colliders means that even
$\epsilon \sim 10^{-10}$ could be detected by looking for two displaced
$Z_D \to \ell \ell$ decays.
This presents us with the exciting opportunity to probe very small
values of $\epsilon$ if some Higgs mixing is present, as we discuss in Sec.~\ref{sec:zdzd}.

\section{Constraining the hypercharge portal with electroweak
  precision observables}
\label{sec:pew}

The discovery of a light Higgs boson has been an excellent
confirmation of the self-consistency of the electroweak sector of the
SM.
In fact, global fits of electroweak precision observables
measured at lepton (LEP, SLC) and hadron (Tevatron, LHC) colliders
show that the SM provides a good fit to the data, with a
$p$-value of $\sim 0.2$~\cite{Baak:2014ora} (see
also~\cite{Baak:2012kk,Batell:2012ca,Baak:2013fwa,Wells:2014pga} for earlier fits post-Higgs-discovery). Measurements of
the various EWPOs are in good agreement with the SM prediction, with
the exception of the notorious forward-backward asymmetry of the
bottom quark, $A_{FB}^{b,0}$, as measured at the Z-pole at LEP1, which
differs 
by $\sim 2.5\sigma$ from the SM prediction.

In this context, physics beyond the SM can receive important
constraints from EWPOs. In particular, here we investigate the bound
on the 
hypercharge portal coupling, given by Eq.~(\ref{eq:KM}), that can be obtained
from electroweak precision measurements. (Constraints from EWPOs on the Higgs portal Eq.~\ref{eq:HM} are unimportant, and we thus do not consider them.)
We will perform a fit to the
current measurements of EWPOs and also consider the impact of future
improvements from hadron and lepton colliders.

In contrast to~\cite{Hook:2010tw}, our approach closely mirrors the
procedure performed by the Gfitter group~\cite{Flacher:2008zq}, and
introduces all observables directly related to properties of the electroweak bosons, including observables that are not corrected at tree level in the dark photon model, such as $m_W$.  As we will see, the precision that will be available in future experimental determinations of $m_W$ will make $m_W$ one of the main drivers in future electroweak fits.  To
begin, we implement the SM fit to the EWPO data.  We consider
the following set of independent observables:
\begin{eqnarray}
&m_Z,~ \Gamma_Z,~\sigma_{\rm had}^0, ~R_\ell^0, ~R_c^0, ~R_b^0, 
~ A_{FB}^{\ell, 0},~A_\ell,~A_c,~A_b,~A_{FB}^{c,0},~A_{FB}^{b,0},~\sin^2 \theta_{\rm eff}^\ell(Q_{\rm{FB}}),~& \nonumber \\
&m_W,~\Gamma_W,~m_t,~\Delta\alpha^{(5)}_{\rm had},~m_h,&
\label{eq:data}
\end{eqnarray}
The experimental measurements of these observables are tabulated
in~\cite{Baak:2014ora} (see also Appendix \ref{sec:PEWappendix} for a summary), whose approach to the data we largely follow. 
Note that this fit makes use of the inclusive hadronic charge asymmetry measurements of $\sin^2 \theta_{\rm eff}$, 
which
we call $\sin^2 \theta_{\rm eff}^\ell(Q_{\rm{FB}})$. 
We include this observable to verify our procedure against the GFitter results. However, we will not use $\sin^2 \theta_{\rm eff}^\ell(Q_{\rm{FB}})$ to obtain bounds on the dark photon model, since this measurement is difficult to interpret in theories with vertex corrections to the $Z$ boson coupling (see also~\cite{Falkowski:2014tna}). 

A convenient set of independent input observables is
\begin{equation}
m_h, m_Z, m_t, \alpha_s, \Delta \alpha^{(5)}_{\rm had}, 
\label{fitpar}
\end{equation}
the latter of which replaces the electromagnetic coupling
$\alpha(m_Z^2)$ and is related to the strong coupling constant
$\alpha_s(m_Z)$.  The light quark masses and the Fermi constant,
$G_F$, should in principle also be added to the set of input
observables. Since $G_F$ is very precisely determined from muon decay measurements, we simply fix it
to its measured value.  Likewise, the pole masses for $m_b$ and $m_c$
are very well determined, and the difference between the pole mass and
the less well-determined running mass enters at higher order and makes
a negligible contribution to the fit. Therefore, we simply fix $m_b$,
$m_c$ to their $\overline{\mathrm{MS}}$ masses. We set the light quark
masses $m_u$, $m_d$, and $m_s$ to their world averages. We refer
to~\cite{Awramik:2003rn,Cho:2011rk,Awramik:2006uz,Freitas:2012sy} for
the SM prediction of the $W$ boson mass, $Z$ boson partial
widths\footnote{We do not include the full fermionic two-loop
  corrections $\mathcal O(\alpha^2)$ to the $Z$ partial widths, as
  recently computed in~\cite{Freitas:2014hra}.
  Given the numerically small effects of these corrections on the SM
  fit \cite{Baak:2014ora}, we do not anticipate these corrections to
  substantially impact our results.} (see also Appendix \ref{sec:PEWappendix} for more details), the effective mixing angle
$\sin^2 \theta_{\rm eff}^\ell$, and $R_b^0$, respectively.  We build our
log-likelihood function through the comparison between the SM
prediction and the corresponding measurement, taking into account the
correlation matrices among the $Z$-lineshape and the heavy-flavor
observables,
\beq\label{eq:carrelationMatrix}
\chi^2_{\rm{SM}}=V_{\rm{SM}} \cdot cov^{-1}\cdot V_{\rm{SM}}
\:\:\,\,{\rm{with}}\,\,\:\:cov=\Sigma_{\rm{exp}}\cdot cor\cdot
\Sigma_{\rm{exp}}\,,
\eeq
where $V_{\rm{SM}}$ is the difference vector between the SM prediction
and the experimentally measured value of the observables in
(\ref{eq:data}): $V_{\rm{SM}}={\rm{theory_{SM}}}(m_h,m_Z, m_t,
\alpha_s, \Delta \alpha^{(5)}_{\rm had})-{\rm{exp}}$.  Meanwhile, 
$\Sigma_{\rm{exp}}$ is the vector containing the experimental error on
the corresponding measurements, and the correlation matrix $cor$ can
be found in~\cite{ALEPH:2005ab}. In this definition of $\chi^2$, we
neglect the theoretical uncertainties in the determination of the
various observables. This is a good approximation for all observables
except the top mass, for which the experimental uncertainty ($\delta m_t^{\rm{exp}}=0.76$ GeV) is
comparable to the theory uncertainty (see~\cite{Moch:2014tta} for a
detailed discussion of the latter).
The value of the input observables in
(\ref{fitpar}) is varied around the measured value to minimize the
$\chi^2$. The resulting $p$-value that we obtain for the SM is very
similar to the one obtained by~\cite{Baak:2014ora}: $\chi^2_{SM}/{\rm{d.o.f.}}=16.7/13$ corresponding to a p-value $p_{SM}= 0.21$. Dropping the $\sin^2 \theta_{\rm eff}^\ell(Q_{\rm{FB}})$ measurement gives a very similar result, though the fit is slightly worse, with $\chi^{2\prime}_{SM}/{\rm{d.o.f.}}=16.1/12$, corresponding to a p-value $p_{SM}^\prime= 0.19$.

We now consider the effect of adding a dark
photon. This introduces two independent effects in the electroweak
fit:
\begin{itemize}
\item 
a shift in the $Z$ mass
  observable~\cite{Gopalakrishna:2008dv,Hook:2010tw}, see
  Eq.~(\ref{eq:masses}), from its input value in
  Eq.~(\ref{fitpar}). 
The input value is what we call $m_{Z,0}$ in
    Eq.~(\ref{eq:massmatrix}). This effect also induces a small
  correction to the $Z$ total width and to the hadronic peak cross
  section, $\sigma_{\rm had}^0$ purely through kinematics;
\item a shift of the $Z$ couplings to SM fermions, see
  Eq.~(\ref{eq:Zcoupl}), and, consequently, a new physics effect on the
  heavy-flavor observables, as well as on $\Gamma_Z,\sigma_{\rm
    had}^0,R_\ell^0,A_\ell$, and $A_{FB}^{\ell,0}$.
\end{itemize}
Both effects first appear at $\mathcal{O}(\epsilon^2)$ and are therefore independent of the sign of $\epsilon$. Further details are included in Appendix~\ref{sec:PEWappendix}. Note again that $\sin^2 \theta_{\rm eff}^\ell(Q_{\rm{FB}})$ is not used in this fit. 

\begin{figure}[t]
\begin{center}
\includegraphics[width=0.65\textwidth]{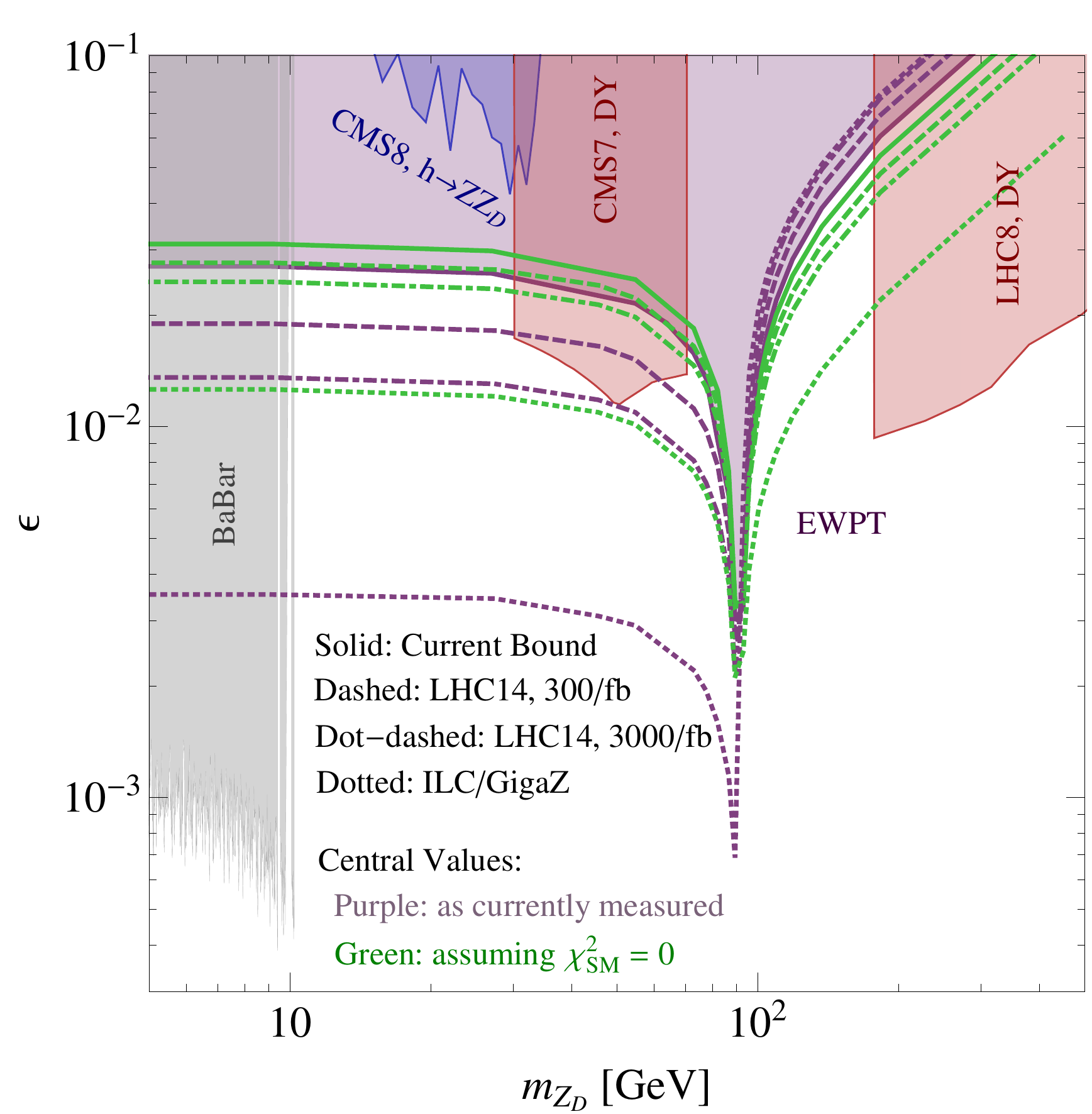}
\end{center}
\caption{{\small Present bound (purple shaded region) on the kinetic
    mixing coefficient $\epsilon$  from the fit to electroweak
    precision observables.  Future projected reach at the 14 TeV LHC
    with 300~fb$^{-1}$ and 3000~fb$^{-1}$ of data, and at the ILC/GigaZ are shown
    by the dashed, dot-dashed, and dotted lines, respectively. Purple
    and green lines, respectively,  represent the bounds obtained by keeping the
    central values of the measurements as they are now, or with
    central values adjusted to the values predicted by the SM best
    fit. For the ILC/GigaZ bound, we also assume the 14~TeV LHC (3000
    fb$^{-1}$ data) precision measurements of $m_h$ and $m_t$. 
    The HL-LHC and ILC/GigaZ projections also include expected improvements in the measurement of $\Delta\alpha^{(5)}_{\rm had}$ from VEPP-2000/Babar data.
    }
\label{fig:EWFitBound}}
\end{figure}
 
Analogously to the fit we perform for the SM, we build the
$\chi^2$ for the theory of a kinetically mixed $U(1)$, denoted
$\chi^2_{Z_D}$, and compare the results to the goodness of the fit
obtained for the SM. 
The  coupling $\epsilon$ can be constrained by imposing
an upper bound on $\chi^2_{Z_D}$,
\beq\label{eq:conditionchi2}
\chi^2_{Z_D}-\chi^{2\prime}_{SM}\lesssim 3.8,
\end{equation}
corresponding to a 95$\%$ CL bound in the case of one degree of
freedom (the $\epsilon$ parameter, once the $Z_D$ mass has been
fixed).  Note that we have chosen to present a bound requiring that
the {\em deviation} of the dark photon model from the SM not exceed
$\approx 2\sigma$, rather than imposing $\chi^2_{Z_D} \lesssim 3.8$ on
the dark photon model alone. This is in order to avoid
overinterpreting the $\sim 1\sigma$ tension between the SM predictions
and the experimental measurements, which is largely driven by
$A_{FB}^{b,0}$. While $A_{FB}^{b,0}$ is shifted in the dark photon model,
other observables, such as leptonic asymmetries, receive comparable
shifts, and those observables show no significant deviation between experimental measurement and SM predictions. Kinetic mixing therefore does not 
preferentially ameliorate the most significant pull in the SM fit.

The solid purple line in Fig.~\ref{fig:EWFitBound} shows the bound
obtained by imposing the requirement of \eref{conditionchi2}. This is
interpreted as the current upper limit on the size of $\epsilon$.  For
comparison, the green solid line in Fig.~\ref{fig:EWFitBound} shows
the $95\%$ CL limit on $\epsilon$ that is obtained from
Eq.~(\ref{eq:conditionchi2}) if we artificially adjust the central
values of the EWPOs to the values that give the best fit to the
SM. 
Above the $Z$ mass the latter fit (in green) gives slightly stronger bounds on $\epsilon$ than 
the fit obtained keeping the central values of the measurements as they are now (in purple).
This is because, for sufficiently small values of $\epsilon$, the effect of the dark photon 
improves the electroweak fit. The opposite is true below the $Z$ mass.
The most important pulls for our theory are the $W$ boson mass, which is the
next-best measured observable after the $Z$ boson mass, and the
asymmetry parameter $A_\ell$.

The LHC will have the potential to significantly increase the
precision with which some of the electroweak observables in the fit
can be measured.  In particular, we expect an improvement in the
determination of $m_W$ by a factor of 2 (3), of $m_t$ by a factor of 2
(4)~\cite{CMS:2013wfa}, and of $m_h$ by a factor of 2.5
(5)~\cite{Dawson:2013bba}, at the 14 TeV LHC, with 300 fb$^{-1}$ (3000
fb$^{-1}$) data. By the end of the HL-LHC's run, we expect to also
have a factor of 2 improvement in the determination of
$\Delta\alpha^{(5)}_{\rm had}$ from BaBar and VEPP-2000
analyses~\cite{Baak:2014ora}. See Tab.~\ref{tab:precision}
for a summary of current and future expected precisions.\footnote{ We thank G.~Wilson for discussions about the possibility of measuring $m_W$ with a precision at the level of $\sim 2 $ MeV at a low energy ILC run. This improves our projected sensitivities only slightly, since the left-right asymmetry $A_\ell$ is the main pull of the fit.}
%

\begin{table}[!t]
\begin{center}
\begin{tabular}{|c||c||c|c||c|}
\hline
  & Present & LHC 14, 300 fb$^{-1}$ &  LHC 14, 3000 fb$^{-1}$ &  ILC (GigaZ)\\
  \hline\hline
$ m_W$ (MeV) & 15 & 8 & 5 & 6 \\\hline
 $m_h$ (MeV) & 240 & 100 & 50 & -- \\\hline
 $m_t$ (MeV) & 760 & 440 & 200 & --\\\hline
 $m_Z$ (MeV) & 2.1 & --  & --  & 1.6 \\\hline
 $\Gamma_Z$ (MeV) & 2.3 & -- & -- & 0.8\\\hline
 $A_b$ & 0.02 & -- & -- & 0.001 \\\hline
 $R_b^0$ ($10^{-5}$) & 69 & -- & -- & 14 \\\hline
 $A_\ell$ ($10^{-4}$) & 18 & -- & -- & $1$ \\\hline
\end{tabular}
\caption{Present and future experimental uncertainty for measurements that will be improved at the LHC and a future ILC/GigaZ. Cases where the experimental precision is not expected
  to significantly improve for a given observable are denoted with a
  dash (--).  In our fits, we also assume an improvement from BaBar and VEPP-2000 data in the
  precision of the measurement of $\Delta\alpha^{(5)}_{\rm had}$ to
  the level of $4.7\times 10^{-5}$, compared to the present $10\times
  10^{-5}$.
    }\label{tab:precision}
\end{center}
\end{table}

The projected bound on the kinetic mixing
parameter $\epsilon$ at the 14 TeV LHC with 300~fb$^{-1}$ (3000~fb$^{-1}$) data is shown as the dashed (dot-dashed) line in
Fig.~\ref{fig:EWFitBound}. This incorporates the above improvements,
including $\Delta \alpha^{(5)}_\mathrm{had}$. We assume progress in theoretical calculations to keep pace with improved experimental measurements, so that the approximation of neglecting theoretical uncertainties in the fit continues to be valid for these future projections.
We show again two possible scenarios for the resulting limits,
corresponding to two limiting assumptions about the future measured
central values.  Purple lines show the results of a fit assuming that
the central values of all measurements remain fixed at their present
values, so only the experimental uncertainties will change.  Green
lines show the results of a fit where the central values of all
measurements are adjusted to their SM-best fit values.  
The mass of the $W$ boson, $m_W$, gives now, by far, the most important pull, followed by  $A_\ell$ and $m_Z$. Fig.~\ref{fig:EWFitBound} shows that we can expect an improvement of the bound on $\epsilon$ by up to $\sim 40\%$ ($\sim 2$) at the at the 14 TeV LHC with 300~fb$^{-1}$ (3000~fb$^{-1}$)
data. We have also verified that the bound is
only weakly dependent on the correlation matrix in
Eq.~(\ref{eq:carrelationMatrix}): assuming a completely uncorrelated
set of measurements would change the bound by at most a few percent.

Beyond the LHC, a possible high-luminosity and low-energy run of
ILC/GigaZ\footnote{ For another discussion on improved new physics reach
  through EWPO at future lepton colliders, see \cite{Fan:2014vta}, which
  focuses on natural supersymmetry scenarios.} would lead to a much more precise measurement of many
precision observables~\cite{Baer:2013cma}. In
particular, measurements of the weak left-right asymmetry $A_\ell$ are
expected to reach a precision of $10^{-4}$, reducing the current uncertainty on this observable by more than an order of magnitude.\footnote{This allows $\sin^2 \theta_{\rm eff}^\ell$ to be determined with a precision of $\sim 10^{-5}$, improving the world average measurement by roughly one order of magnitude as well. However, the improvements in our limit projections derives from the increased precision on $A_\ell$, since $\sin^2 \theta_{\rm eff}$ is not included in the fit.}
Furthermore, ILC/GigaZ will have unprecedented b-tagging capabilities. This will result in improved measurements of the left-right asymmetry, $A_b$, as well as $R_b^0 = \Gamma(Z\to \bar b b)/\Gamma(Z \to \mathrm{hadrons})$, by a factor of 20 and 5, respectively.

In Fig.~\ref{fig:EWFitBound} (dotted curves), we show the bound on $\epsilon$ using the expected
uncertainties on the electroweak observables at ILC/GigaZ as shown in the latter column of
Tab.~\ref{tab:precision}. For this projection, we also assume the 14 TeV LHC (3000~fb$^{-1}$ data) precision measurements of 
$m_h$ and $m_t$, although the bound is very similar if we take their current measured values. 
As seen in the figure, the sensitivity on $\epsilon$ can be increased by up to an additional factor of $\sim 4$ compared to the results from the HL-LHC.
The new $A_\ell$-measurement would then have the potential to provide  the main pull in the fit, followed by $m_Z$, $m_W$,
and $\Delta\alpha^{(5)}_{\rm had}$.

This improvement in the indirect bound on dark photons is notable, and
has the great virtue that it does not depend on how the $Z_D$ decays.  
However, in the minimal model, where the 
$Z_D$ only has SM decays available, the reach from EWPTs is not competitive with hadron colliders due to their 
enormous integrated luminosities, as we will now discuss.

\section{Constraining the hypercharge portal with $h\to Z Z_D$ decays}\label{sec:zzd}

In this section, we estimate the potential sensitivity of the LHC
and a 100 TeV $pp$ collider to the exotic Higgs decay $h\to Z^{ (*)} Z_D\to 4\ell$. This decay mode was previously examined in \cite{Curtin:2013fra}, which recast LHC Run 1 searches for $h\to Z Z^* \to 4 \ell$ \cite{CMS:xwa, ATLAS:2013nma} to set limits on $\epsilon$ for $m_{Z_D} \gtrsim 10 \gev$. While the resulting constraints at the level of $\epsilon \lesssim 0.04$ were weaker than indirect limits from EWPO's, the demonstrated sensitivity motivates a dedicated study of the future reach at hadron colliders.   Crucially, as we also discuss in the next section, measurements of \emph{both} $h\to Z_D Z^{(*)}$ and direct DY production of $Z_D$ are necessary to differentiate a kinetically mixed dark photon from a $Z'$ with very weak gauge coupling.

The sizeable $Z_D$ branching ratio to leptons makes it feasible to examine a broad range of dark photon masses, including the regime where the intermediate $Z$ is off-shell.\footnote{Processes   with an off-shell $Z_D$ are higher order in $\epsilon$, and are   negligible for $\epsilon\lesssim \mathcal O(10^{-2})$.}  When $m_h > m_Z+m_{Z_D} $, the initial Higgs decay is two-body; when $m_h < m_Z + m_{Z_D}$, the initial Higgs decay is  three-body. The branching ratio for this decay is shown in Fig.~\ref{fig:Br4l} (top panel) for several values of $\epsilon$.  At low $m_{Z_D}$, where the $Z_D$ may be produced on shell, the dependence of  Br$(h\to Z Z_D)$ on $m_{Z_D}$ arises because of the mass-dependence of the $Z$-$Z_D$ mixing angle, $\cos\alpha$, see Eq.~(\ref{eq:mixingangle}).  At higher masses the suppression from three-body phase space is evident. 

We impose the following baseline acceptance cuts, which are modeled
after \cite{Chatrchyan:2013mxa} and similar to those in
\cite{ATLAS-CONF-2013-013}:
\begin{itemize}

\item All electrons must satisfy $p_{T,e} > 5$ GeV and $|\eta_e|< 2.5$.

\item All muons must satisfy $p_{T,\mu} > 7$ GeV and $|\eta_\mu|< 2.4$.

\item Any opposite-sign, same-flavor (OSSF) lepton pair must have
  $m_{\ell\ell}>4$ GeV.

\item All events must contain exactly four accepted leptons forming
  two OSSF pairs, with the hardest two leptons satisfying $p_{T,1} >
  20$ GeV and $p_{T,2} >10$ GeV.

\item The total invariant mass of the four leptons must lie in the
  range $120 \gev < m_{4\ell}< 130 $ GeV.

\end {itemize}

Our baseline modeling of lepton efficiencies and resolutions is based
on the 7 and 8 TeV SM $h\to 4\ell$ searches.  In particular, we employ
the $p_T$- and $\eta$-dependent electron efficiencies reported in
\cite{CMS:xwa,CMS:2013hoa}, yielding an average electron efficiency of
$0.87$ in SM $h\to Z Z^*\to 4\ell$ events, while the muon
efficiency is set to a flat $0.96$.  This simulation is validated
against the expected numbers of signal events at 7 and 8 TeV in
\cite{CMS:xwa}.  As the performance achieved by the LHC experiments in
LHC Run I are often quoted as goals for future performance, we will
use these numbers as our benchmark scenario. In
Sec.~\ref{sec:improvements}, we will also comment on the effect of varying assumed lepton $p_T$ thresholds, $\eta$ 
acceptance, and mass resolution on collider reach.

The lepton efficiency above incorporates the probability for a lepton
to be insufficiently separated from a jet to pass isolation
requirements in SM Higgs-like events.  In \cite{CMS:xwa,CMS:2013hoa},
leptons are not allowed to spoil each other's isolation requirements,
and for $h\to 4\ell$ events with SM-like kinematics, the fraction of
events with a nearly collinear pair of leptons is negligibly small. We
impose an additional, explicit requirement that leptons must have a
minimum angular separation: $\mathrm{min} (\Delta R_{e,e}) > 0.02$ for
electrons at both 14 and 100 TeV colliders, while for muons
$\mathrm{min} (\Delta R_{\mu,\mu}) > 0.05$ at 14 TeV \cite{Khachatryan:2010pw} and $\mathrm{min}
(\Delta R_{\mu,\mu}) > 0.02$ at 100 TeV.  This requirement has
negligible impact on the $h\to Z Z_D$ decays considered in this
section, but will become important for the $h\to Z_D Z_D$ decays
considered in Sec.~\ref{sec:zdzd}.  We consider the effect of smearing
lepton energies using a Gaussian distribution with energy-dependent
variance as reported in \cite{CMS:xwa}.  Provided that windows for
cuts on lepton invariant masses are set to reasonable values (see
Eq.~(\ref{eq:dilepwindows}) below), we find that incorporating smearing
changes our limits by only $\mathcal{O}(2\%)$.  For simplicity we
neglect smearing henceforth.

We use our dark photon MadGraph model (see Appendix~\ref{sec:mgmodel}) to simulate gluon fusion Higgs production in MadGraph 5 and shower events in
Pythia 6~\cite{Alwall:2011uj,Sjostrand:2007gs}.  The inclusive Higgs
production cross-section is normalized to the SM prediction of
$\sigma_{ggF}=50.35$ pb at 14 TeV and $\sigma_{ggF}=740.3 $ pb at 100
TeV \cite{HWG}.  The SM value for  Br$(h\to ZZ^*\to 4\ell)$ is taken to
be $1.26\times 10^{-4}$ \cite{Heinemeyer:2013tqa}.

\begin{figure}[t]
\begin{center}
\hspace*{-10mm}
\begin{tabular}{cc}
\includegraphics[width=8cm]{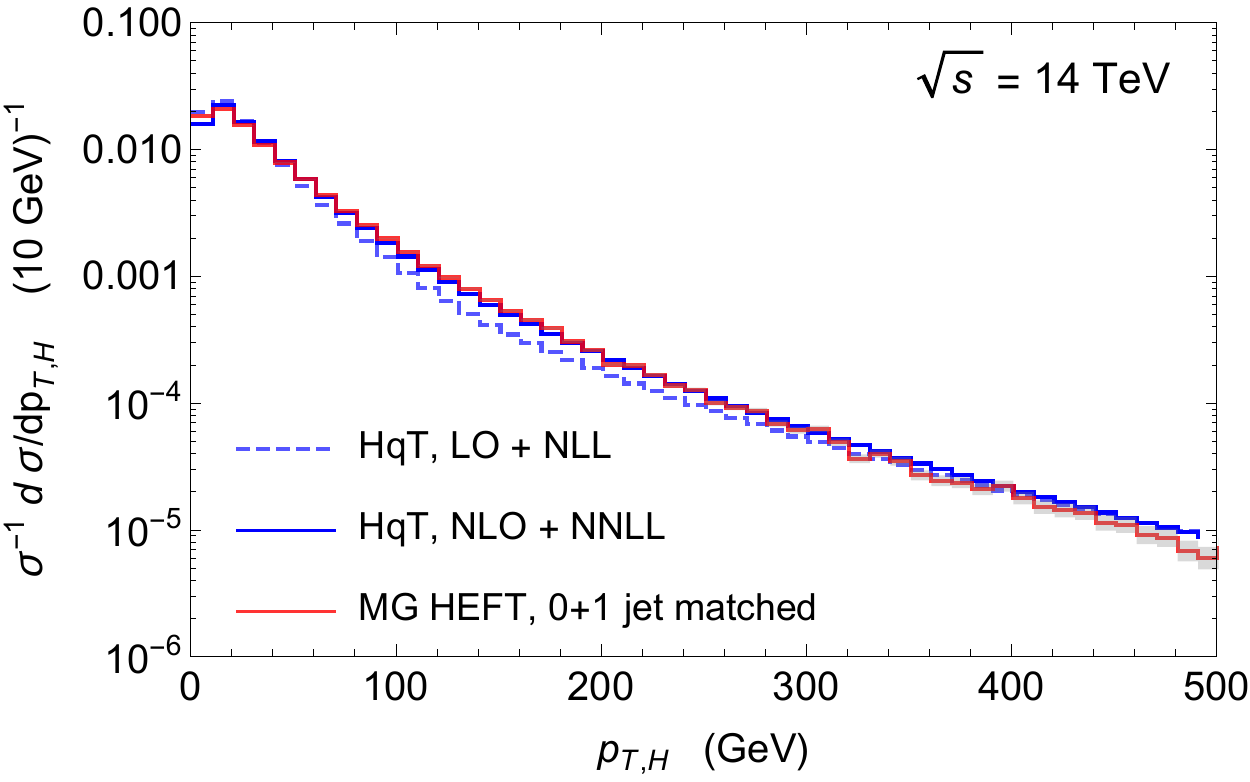}
&
\includegraphics[width=8cm]{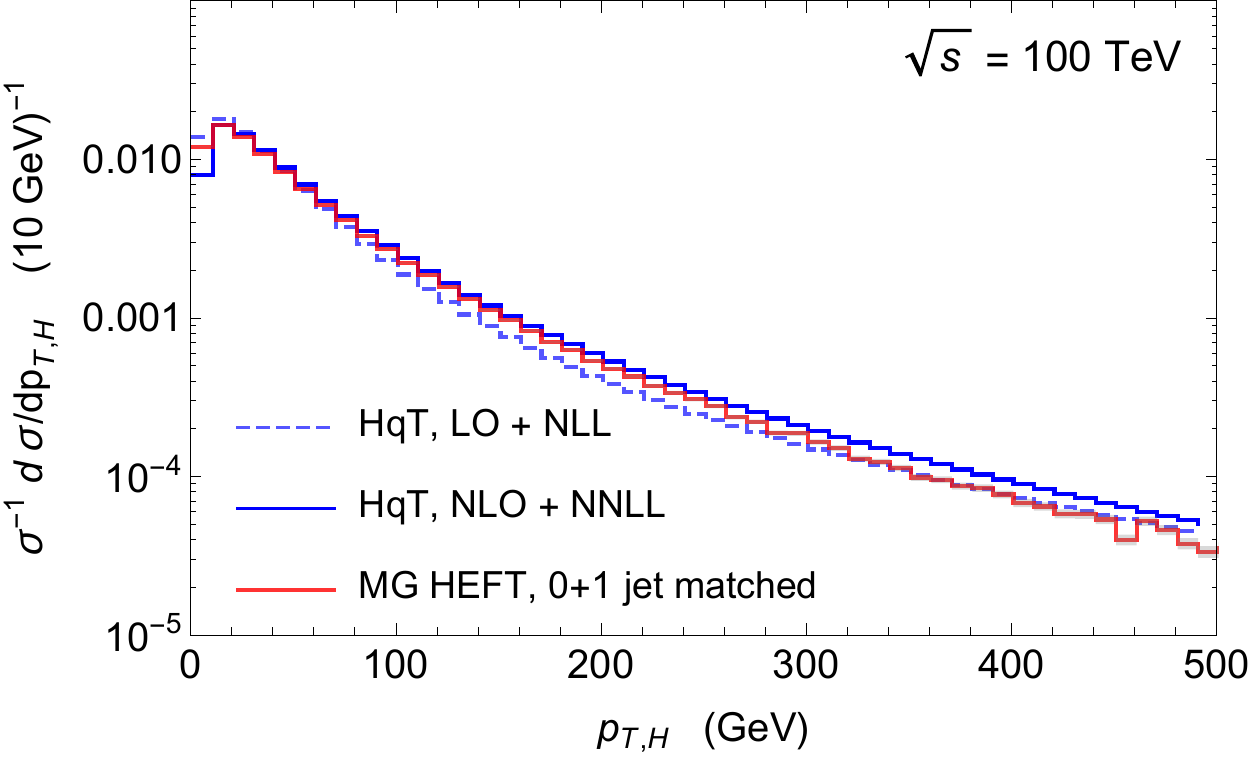}
\end{tabular}
\end{center}
\caption{{\small Predicted Higgs $p_T$ spectra in gluon fusion
    production, as calculated by matched MadGraph 5 + Pythia 6 (red)
    and HqT at LO+NLL (dashed blue) and NLO+NNLL (solid blue), for
    $\sqrt{s} = $ 100 TeV \emph{(left)} and 14 TeV \emph{(right)}. 
      }}
\label{fig:hpt}
\end{figure}

Since the reach for $h\to Z_D Z^{ (*)}\to 4\ell$ depends sensitively
on the Higgs $p_T$ through lepton acceptance, we must have good
control over the Higgs $p_T$ spectrum.  This is especially a concern at
100 TeV, where many more Higgses are produced in the high-$p_T$ tail
where the validity of the effective field theory description of $gg\to
h$, used by MadGraph, breaks down.  We cross-checked the Higgs $p_T$ spectrum from matched
MadGraph/Pythia events with the $p_T$ spectrum predicted at NNLL$+$NLO
by \texttt{HqT 2.0} \cite{Bozzi:2005wk,deFlorian:2011xf}.  The two spectra
are compared in Fig.~\ref{fig:hpt}, and are in good agreement for the
bulk of the distribution.  We reweight events to realize the
NNLL$+$NLO $p_T$ spectrum , but this only gives a fractional change in sensitivity to $\epsilon$ of less than a percent.

For the SM di-$Z^{(*)}/\gamma^*$ background we use MadGraph to
generate tree-level events, which we normalize using a K-factor of
1.2, as computed for events with $m_{4\ell}\in (120, 140)$ GeV using
MCFM 6.8~\cite{Campbell:2011bn} (at both 14 and 100 TeV). 
Following~\cite{ATL-PHYS-PUB-2013-014}, we multiply the diboson background by a
factor of 1.5 to account for reducible backgrounds containing fake
isolated leptons, notably $Z+$jets and $t\bar t$.

We define two different search regions depending on $m_{Z_D}$. First,
we consider the case where $m_{Z_D}< m_h-m_Z$, and consequently the
two-body decay $h\to Z Z_D$ can proceed on-shell.  We designate $M_1$
as the invariant mass of the pair of OSSF leptons that minimizes
$|m_{\ell\ell}-m_Z|$, and $M_2$ as the invariant mass of the remaining
OSSF pair.  Following the SM Higgs analyses, we require
\beq
M_{1,2} > 12\gev,
\eeq
to suppress backgrounds from quarkonia, and concentrate on regions not already probed by 
BaBar in an $e^+e^- \to \gamma Z_D$ search~\cite{Lees:2014xha}.  We then perform a simple bump
hunt in $M_2$, requiring
\beq
\label {eq:dilepwindows}
|M_2-m_{Z_D}|< \left\{ \begin{array}{cc}  0.02 \, M_2 &
    (\mbox{electrons}),\\  2.5 \,(0.026 \mbox{ GeV}+ 0.013\, M_2) &  (\mbox{muons}). \end{array}\right.
\eeq
These mass windows are based on current CMS energy resolutions \cite{Chatrchyan:2011hr}, and
are relatively conservative.  In particular, the muon mass window we
use is based on the mass resolution for {\em forward} muons, $\eta_\mu
> 0.9 $, and is an underestimate of experimental capabilities.

\begin{figure}[t!]
\begin{center}
\includegraphics[width=0.49\textwidth]{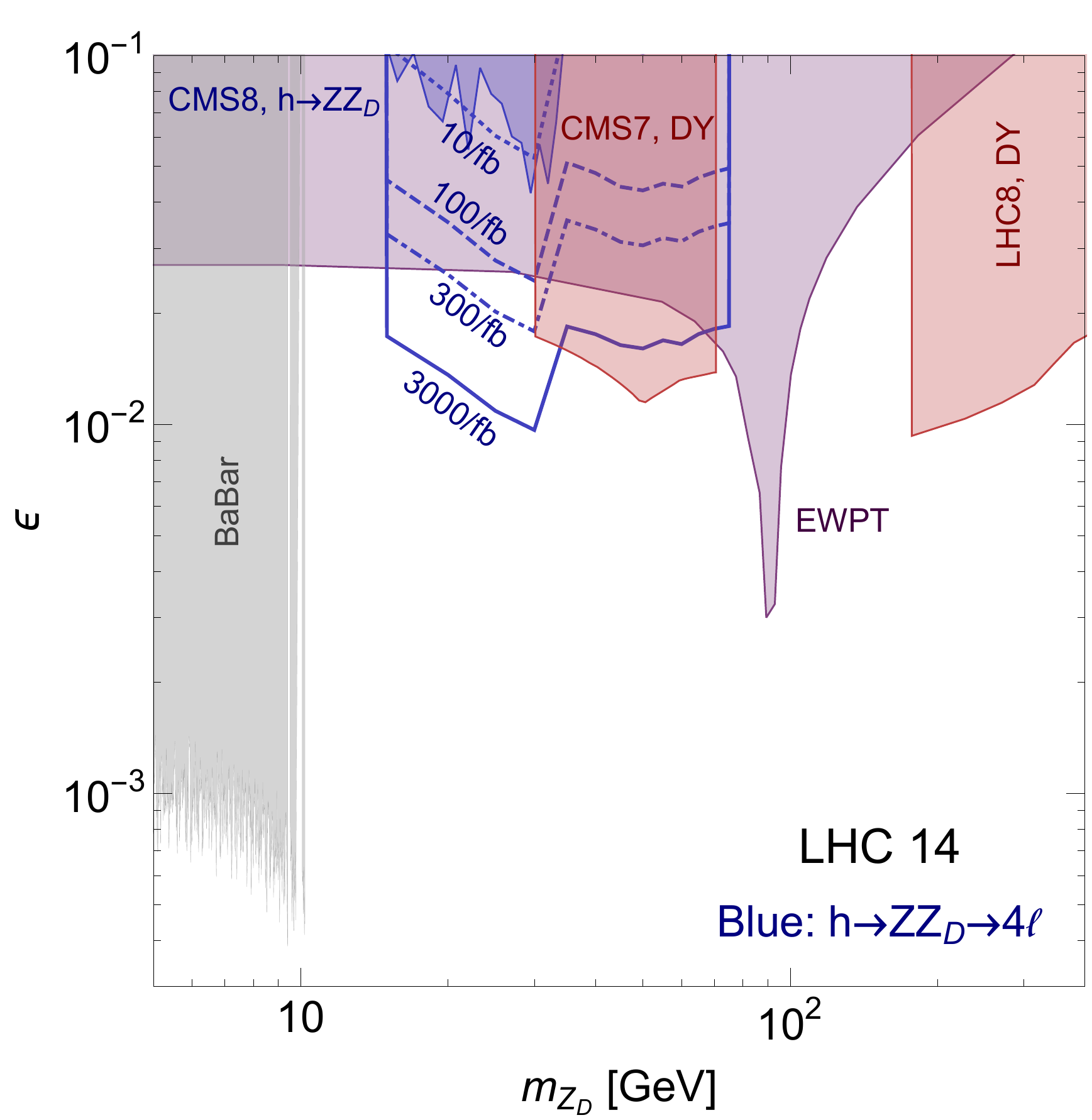}
~~\includegraphics[width=0.49\textwidth]{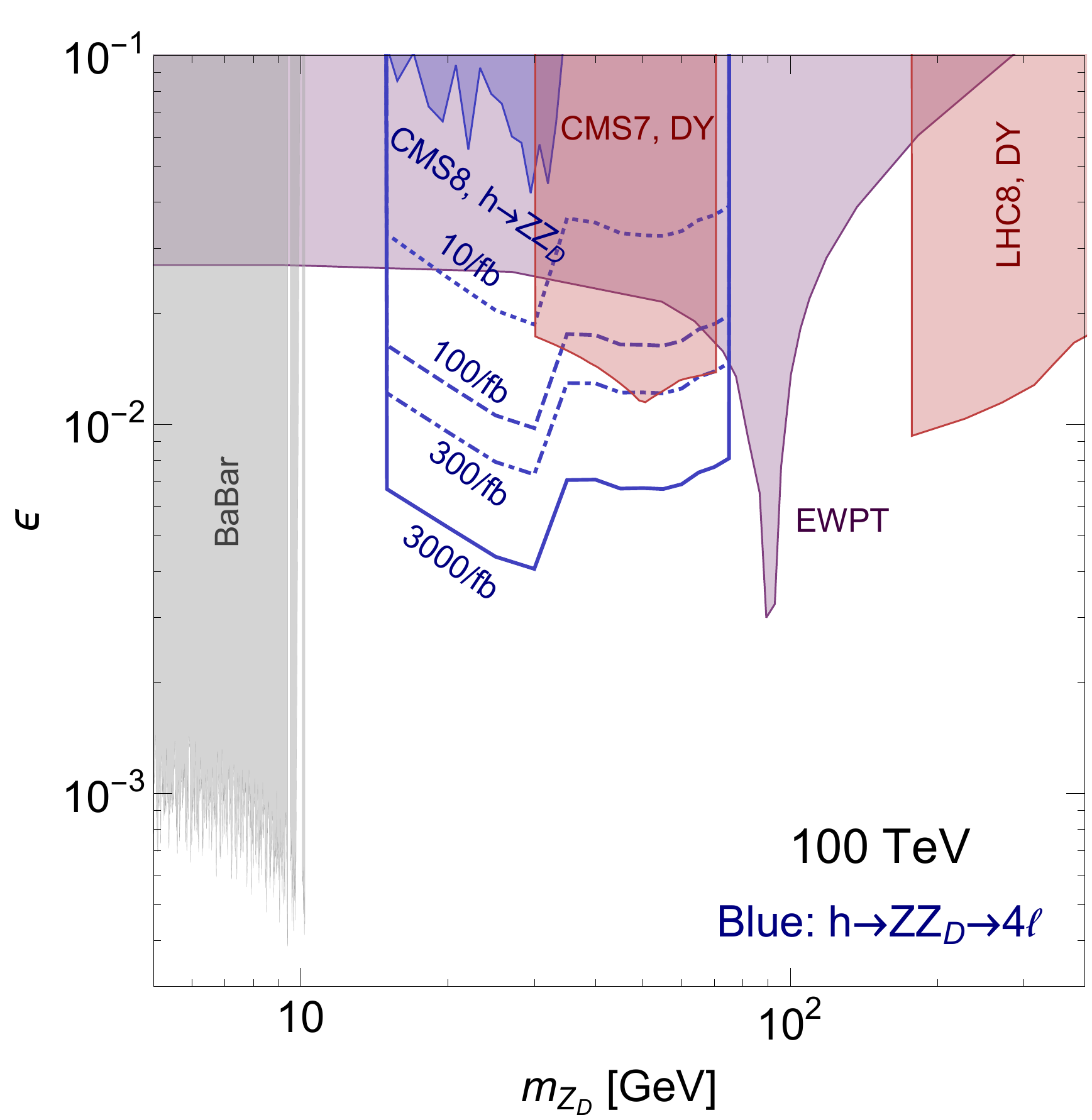}
\end{center}
\caption{{\small The blue lines show expected 95\% CLs limits on
    $\epsilon$ from $h\to Z_DZ^{(*)}\to 4\ell$, at the LHC14 (left)
    and a 100 TeV pp collider (right).
    Limits shown correspond to integrated luminosities of 10 (dotted),
    100 (dashed), 300 (dot-dashed), and 3000 fb$ ^ {-1}$ (solid) in
    both plots.
    A recast~\cite{Curtin:2013fra} of a CMS8 analysis~\cite{CMS:xwa}
    sensitive to $h\to Z Z_D$ is shown in the blue shaded region.
    The purple region shows the current EWPT constraints (this work,
    see Sec.~\ref{sec:pew}), while the gray region is a limit from
    BaBar~\cite{Lees:2014xha}. The red regions are the bounds from
    Drell-Yan production of $Z_D$ \cite{Hoenig:2014dsa,
      Chatrchyan:2013tia, Cline:2014dwa, ATLAS:2013jma} and are
    discussed in Sec.~\ref{sec:dy}.  }
\label{fig:zzdlimits}}
\end{figure}
\begin{figure}[t!]
\begin{center}
\vspace{10mm}
\hspace*{-10mm}
\begin{tabular}{cc}
\includegraphics[width=8cm]{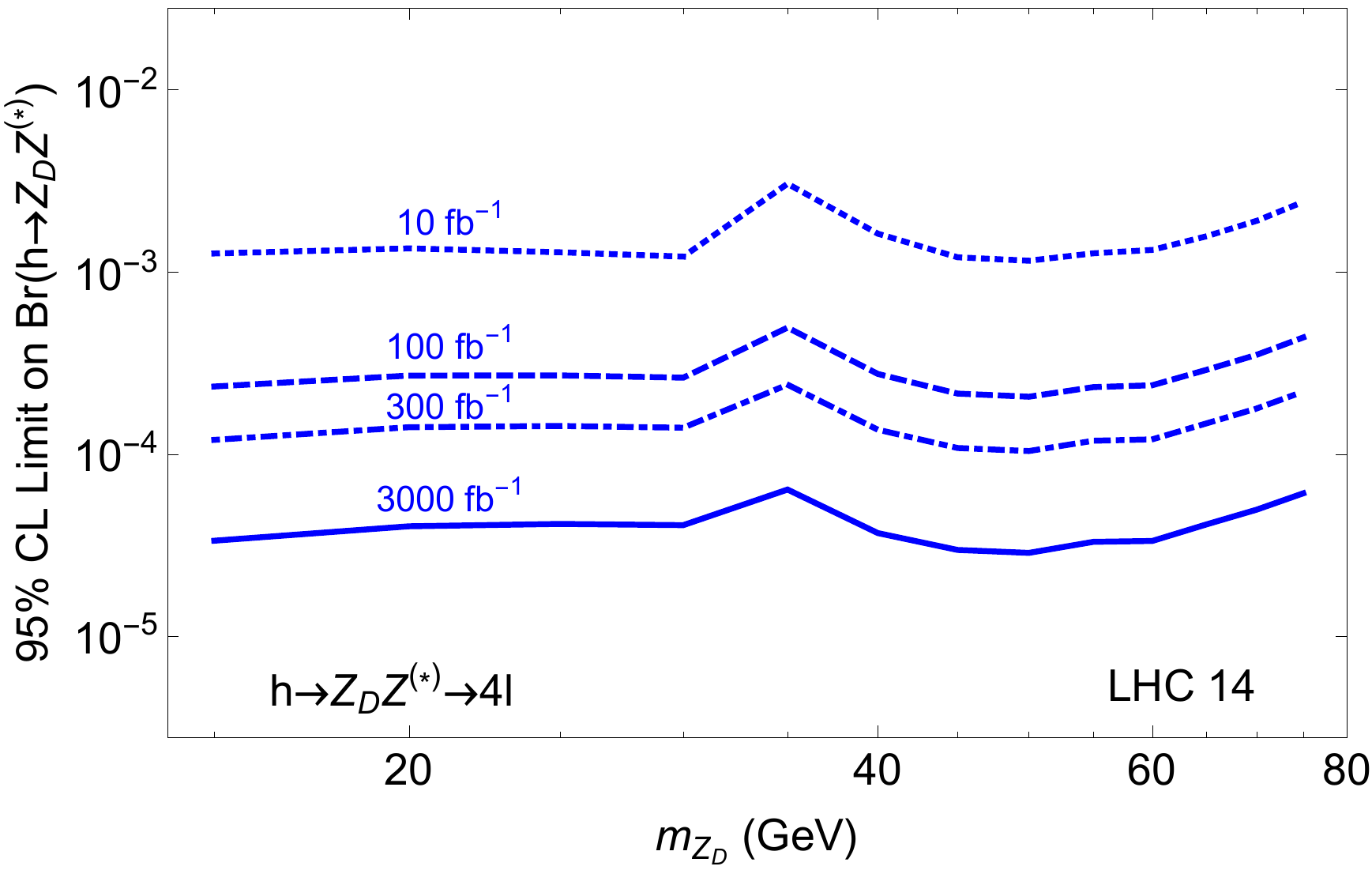}
&
\includegraphics[width=8cm]{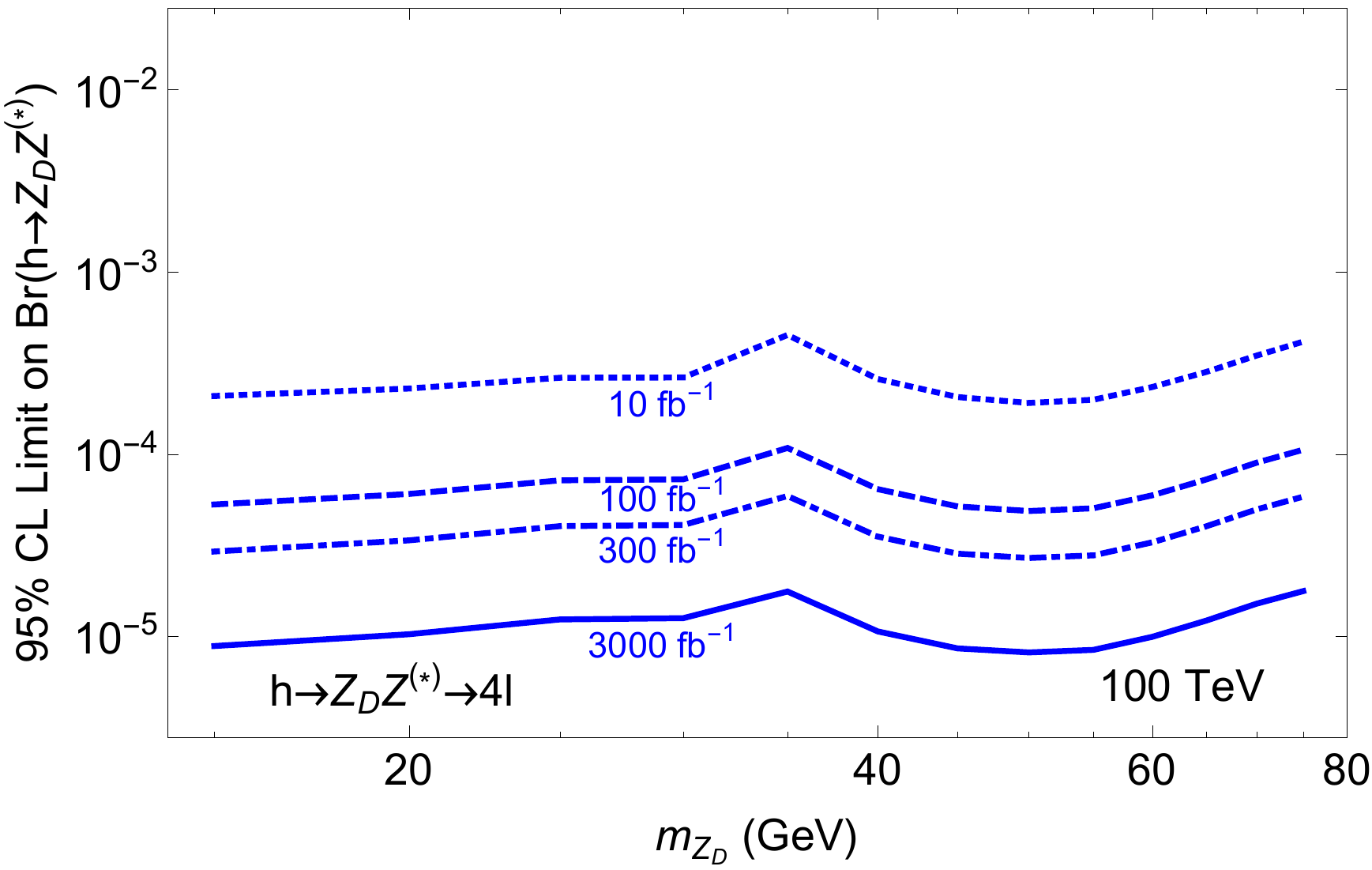}
\end{tabular}
\end{center}
\caption{{\small 
    Projected limits on total exotic Higgs branching ratio $\mathrm{Br}(h\to Z_D Z^*)$ from the $h \to Z_D Z^{(*)} \to 4 \ell$ search at 14 TeV \emph{(left)} and 100 TeV \emph{(right)}. 
      }}
\label{fig:BrZdZlimit}
\vspace{5mm}
\end{figure}

Second, we consider the case where $m_{Z_D} > m_h-m_Z$.  In this case
the three-body process $h\to Z_D \ell\ell$ gives the leading
contribution to the reach. Here we consider all possible divisions of
events into two OSSF lepton pairs, and require that no lepton pair
satisfies either $m_{\ell\ell}<12$ GeV or $|m_{\ell\ell}-m_Z|<15$ GeV.
Events are then selected if at least one OSSF lepton pair lies within
\beq
|m_{\ell\ell}-m_{Z_D}|< M_{cut},
\eeq
where the mass window $M_{cut}$ depends on the flavor and mass of the
lepton pair as in Eq.~(\ref{eq:dilepwindows}).

Fig.~\ref{fig:zzdlimits} shows our expected 95\% CLs exclusions for
both the LHC and a 100 TeV collider.  These limits treat the signal
mass bin as a single Poisson counting experiment, and neglect
systematic uncertainties\footnote{A 10\% upward shift in the
  background leads to a 2.3\% upward shift in the exclusion
  reach.} This limit should be compared to the limit obtained by
recasting the Run I analysis for $h\to ZZ^*\to 4\ell$ (shaded blue
region in the figure)~\cite{Curtin:2013fra}.

It is also instructive to unfold the leptonic branching ratios of
$Z_D$ and the possibly off-shell $Z^{(*)}$, and derive a limit on the
total exotic branching fraction $\mathrm{Br}(h \to Z_D Z^{(*)})$. The
achievable sensitivities, shown in \fref{BrZdZlimit}, are $\sim$ (few
$\times$) $10^{-5}$ at a 100 TeV collider (the HL-LHC).
Since both signal and main background for this search come from Higgs decays, the sensitivity achievable depends mostly on the number of Higgs bosons produced. Therefore, since the number of Higgses produced at the HL-LHC (100 TeV collider) is 2 (3) orders of magnitude higher than at any of the future proposed lepton colliders, we expect this projected limit on $\mathrm{Br}(h\to Z_D Z^{(*)})$ to be at least an order of magnitude more sensitive than anything achievable at a lepton machine.

Our HL-LHC results are less optimistic than those obtained in
Ref.~\cite{Falkowski:2014ffa} using a matrix-element-based likelihood
discriminant, but we have checked that this is due almost entirely
to our use of finite mass resolution in reconstructing the $Z_D$; in
other words, the $Z_D$ mass peak contains almost all of the
information that is useful in the statistics-limited discovery regime.

\section{Constraining the hypercharge portal with Drell-Yan $Z_D$ Production}\label{sec:dy}

The hypercharge portal coupling allows the $Z_D$ to be singly produced in the
$s$-channel via Drell-Yan (DY) production. This gives rise to dilepton
signals $pp \to Z_D \to \ell^+ \ell^-$ that show up in DY dilepton
spectrum measurements, or high-mass $Z'$ searches of the LHC
experimental collaborations.

The sensitivity of DY measurements to a $Z_D$ below the $Z$ mass with
$10 \gev < m_{Z_D} < 80 \gev$ was recently explored by
\cite{Hoenig:2014dsa}. They recast the DY measurement at the 7 TeV
LHC~\cite{Chatrchyan:2013tia} as a $m_{Z_D}$-dependent limit on
$\epsilon$, and give projections for the sensitivity achievable with
optimized analyses at LHC Run 1 (8 TeV, $20 \ifb$) and at the HL-LHC
(14 TeV, $3000\ifb$). The sensitivity of $Z'$ searches for heavier
dark photons ($m_{Z_D}>m_Z$) has been explored most recently in~\cite{Cline:2014dwa} (see also~\cite{Jaeckel:2012yz} for an earlier study), 
where limits were derived on $\epsilon$ in the range
$200 \lesssim m_{Z_D} \lesssim 2800 \gev$ from published ATLAS
$20\ifb$ Run 1 results~\cite{ATLAS:2013jma}. LEP and future lepton
colliders are less sensitive to this channel than the LHC~\cite{Hoenig:2014dsa,Hook:2010tw}.

In this section, we estimate how these expected constraints on
$\epsilon$ from \cite{Hoenig:2014dsa,Cline:2014dwa} change at the
HL-LHC (above the $Z$ mass) and at $\sqrt{s} = 100 \tev$ (above and below $Z$ mass). Rather than repeating the analyses of ~\cite{Hoenig:2014dsa,Cline:2014dwa}, we can estimate the improved
reach by taking into account the change in signal and background cross
sections.

For on-shell $Z_D$ production, the number of expected new physics
events scales with $\epsilon^2$. In the high-statistics limit, a 95\%
CL exclusion in some signal bin is derived by solving
\begin{equation}
\frac{S_1 \epsilon^2}{\sqrt{B_1} }= c,
\end{equation}
for $\epsilon$, where $\epsilon^2 S_1$ ($B_1$) is the number of signal
(background) events for a given search, and $c$ is some constant. The
resulting limit on $\epsilon$ is
\begin{equation}
\epsilon_1^\mathrm{95\%CL} = \left(\frac{c^2 B_1}{S_1^2}\right)^{1/4}.
\end{equation}
Suppose that we want now to rescale this $\epsilon$ limit for a
different integrated luminosity and center of mass energy $\sqrt{s}$. If we know
the ratio by which the signal and background number of expected events
changes,
\begin{equation}
B_2 = r_{B21} B_1 \ \ , \ \ \ \ \ S_2 = r_{S21} S_1 \ \ ,
\end{equation}
we can find the new expected limit on $\epsilon$:
\begin{equation}
\epsilon_2^\mathrm{95\%CL} \  = \ \epsilon_1^\mathrm{95\%CL} \  (r_{B21})^{1/4} \ (r_{S21})^{-1/2}.
\end{equation}
For the HL-LHC and a 100 TeV collider, the $m_{\ell \ell}$ ($m_{Z_D}$)
dependent rescalings $r_{B21}$ $(r_{S21})$ are easily estimated by
computing the differential DY cross section $d \sigma_\mathrm{DY}/d
m_\mathrm{\ell \ell}$ (signal cross section $\sigma_{p p \to Z_D}$) at
different $\sqrt{s}$ in MadGraph at LO parton level. Setting $m_{\ell
  \ell} = m_{Z_D}$, we obtain the rescaled $\epsilon$ limits shown in
\fref{DYlimits}.

\begin{figure}[t]
\begin{center}
\includegraphics[width=0.48\textwidth]{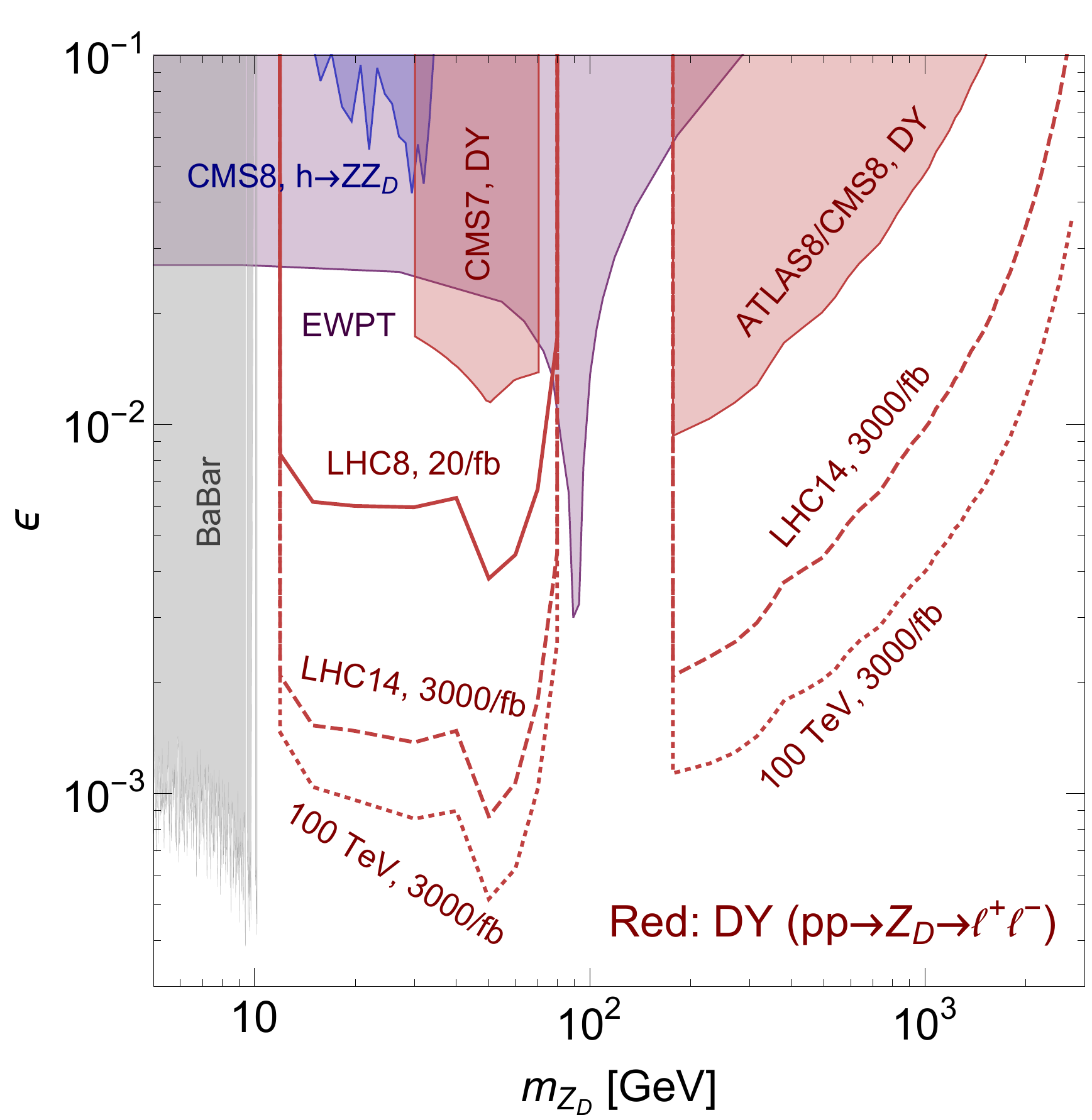} 
\end{center}
\caption{{\small 
Prospects for $Z_D$ searches from DY production (red lines) at
    LHC8 (20~fb$^{-1}$, solid), LHC14 (3000~fb$^{-1}$, dashed), and a
    100~TeV $pp$ collider (3000~fb$^{-1}$, dotted), with limits from
    existing recasts shown in shaded red (from \cite{Hoenig:2014dsa,
      Chatrchyan:2013tia, Cline:2014dwa, ATLAS:2013jma} and our
    rescalings, see text for details).
    A recast~\cite{Curtin:2013fra} of a CMS8 analysis~\cite{CMS:xwa}
    sensitive to $h\to Z Z_D$ is shown in the blue shaded region.
    The purple region shows the current EWPT constraints (this work,
    see \sref{pew}), while the gray region is a limit from
    BaBar~\cite{Lees:2014xha}. }
\label{fig:DYlimits}}
\end{figure}

The figure shows that DY production can be sensitive to
$\epsilon\gtrsim 9 \times
  10^{-4}$ ($4 \times 10^{-4}$) at the HL-LHC (100~TeV $pp$ collider)
  While this is superior to
indirect constraints from EWPTs, it does rely on the $Z_D$ decaying directly to SM particles. 
Direct DY production is
also more powerful than $h \to Z Z_D$ searches,\footnote{Improvements
  of the dilepton mass resolution will not change this conclusion
  \cite{Falkowski:2014ffa}.} but only by a factor of a few
in $\epsilon$.
Crucially, 
a discovery in the DY channel \emph{only} would be unable to distinguish between a kinetically mixed dark photon, and a new $Z'$  which mediates a $U(1)$ gauge interaction with tiny coupling constant. As we have shown in \sref{zzd}, the former scenario would leave comparable traces in the $h\to Z_D Z^{(*)}$ channel, while the latter scenario can only generate $h\to Z_D Z^{(*)}$ (i.e.~$Z' Z^{(*)}$) decays via fermion loops, leading to a much suppressed signal. 
The best-case scenario is, therefore, discovery of
$Z_D$ in both DY and $h\to Z_D Z^{(*)}$ channels, allowing a precise experimental determination of the dark photon's properties.  

We close this section by pointing out that publicly available DY data does not yield any constraints on $\epsilon$ in the range $90 \gev \lesssim m_{Z_D} \lesssim 180 \gev$. The minimal dark photon model provides strong experimental motivation for dedicated dark photon searches  close to the $Z$-peak.

\section{Constraining the Higgs and hypercharge portals with $h\to Z_D Z_D$  decays}\label{sec:zdzd}

In this section, we estimate the potential sensitivity of the LHC
and a 100 TeV $pp$ collider to the exotic Higgs decay $h\to Z_D Z_D\to
4\ell$ (see also \cite{Martin:2011pd}). As for $h\to Z_D Z^{(*)}$, recasts of LHC Run~1 data \cite{CMS:xwa, ATLAS:2013gma, ATLAS:2013nma} were used in \cite{Curtin:2013fra} to set limits on this channel. 
This decay is interesting for
two reasons. Firstly, for a heavy singlet scalar but non-negligible
singlet-Higgs mixing, it is by far the most sensitive test of the $U(1)_D$ hidden 
sector if $\epsilon\lesssim \mathcal O(10^{-3})$. Secondly, while
production of the $Z_D$ pair occurs through the Higgs portal, $Z_D$
decay has to proceed through the hypercharge portal. Therefore,
observation of $h\to Z_D Z_D\to 4\ell$ equires a non-zero value for
$\epsilon$.  As we describe below, the implicit $\epsilon$ sensitivity
is very impressive.

\subsection{Constraining the Higgs portal from prompt $Z_D$ decay}

We consider a broad range of masses for $Z_D$: $2 m_\mu < m_{Z_D} <
m_h/2$.  Both $Z_D$ have to be produced on-shell, as otherwise the Higgs decay will be suppressed by $\epsilon^2$, and thus unobservable.

The collider analysis proceeds in large part identically to the $h\to
Z_D Z^{(*)}$ case outlined in Sec.~\ref{sec:zzd}. The same kinematic
cuts on lepton $p_T$, $\eta$, $\Delta R$, and $m_{4 \ell}$ are applied,
and the detailed signal and background simulation is identical,
including the Higgs $p_T$ reweighing, except that now we generate the
signal process $h\to Z_D Z_D\to 4\ell$.  Recall that our analysis enforces
minimum lepton separation of only $\mathrm{min}(\Delta R_{\mu\mu}) = 0.05$ $(0.02)$ and $\mathrm{min}(\Delta R_{ee}) = 0.02$ $(0.02)$ at the LHC (100 TeV collider), and is therefore sensitive
to lepton jets.  The new features in the $h\to Z_D Z_D$ analysis are
(1) the dilepton invariant mass cuts, and (2) how the final
background estimate is obtained. We divide the analysis into two
regimes: the ``heavy $Z_D$'' case with $10 \gev < m_{Z_D} < m_h/2$,
and the ``light $Z_D$'' case with $2 m_\mu < m_{Z_D} < 10 \gev$.

For the heavy $Z_D$ analysis we divide events into three families
according to the flavor composition of the four lepton final state:
$4e$, $4\mu$, or $2e2\mu$. $Z_D$ candidates are reconstructed by
combining opposite-sign same-flavor dilepton pairs. In the cases of
$4e$ and $4\mu$, the combinatoric uncertainty of this reconstruction is
largely reduced by choosing the pairings that minimize
$|m_{{\ell\ell}_1} - m_{{\ell \ell}_2}|$. The search is conducted
separately for each $m_{Z_D}$, requiring both lepton pairs in the
event to satisfy $|m_{\ell \ell} - m_{Z_D}| < M_{cut}$, with the same
mass windows as \eref{dilepwindows}.

This double-dilepton-mass cut is extremely effective at eliminating
background, to the point where simulating statistically accurate
background samples in the respective signal regions is very
challenging. To circumvent this issue one can make use of the fact
that the background distribution in the $(m_{{\ell \ell}_1},m_{{\ell
    \ell}_2})$ plane is quite smooth, with events that pass the
double-dilepton-mass cut being dominated by coincidental
mispairings. This allows us to estimate the background event
expectation in the small signal regions by interpolation. We construct
the $(m_{{\ell \ell}_1},m_{{\ell \ell}_2})$ distribution for each
background process (and each flavor composition of the four leptons)
with large $5 \times 5 \gev$ bins that each contain sufficient Monte
Carlo events. The signal region is given by a small ($\Delta m_1
\times \Delta m_2)$ rectangular region along the diagonal centered on
$(m_{Z_D}, m_{Z_D})$, where $\Delta m_{1,2}=2 M_{cut}$ are the total
widths of the mass windows for each lepton pair. Rescaling the
contents of the large $5\times5 \gev$ bin along the diagonal by
$\Delta m_1 \Delta m_2/(5 \gev)^2$ therefore gives a suitable estimate
of the background in the small signal bin.

For each flavor channel, separate signal and background expectations
are obtained. 95\% CLs exclusions are obtained for both the LHC
and a 100 TeV collider, treating each flavor bin and the combined bin
as single Poisson counting experiments, and selecting the best limit
obtained from the $4e, 4\mu, 2e2\mu$ or combined channel for each
$m_{Z_D}$.

The light $Z_D$ ($m_{ZD}<10$ GeV) case motivates a lepton jet analysis. Since we do not
enforce lepton isolation in the reconstruction, we perform our
analysis identically to the heavy case, with three
exceptions. Firstly, we only use the $4\mu$ channel, since di-electron
reconstruction at such low masses is more challenging, though it might allow the search to be extended to even lower masses. Secondly, we
find that at these low dilepton masses, the $m_{4\ell}$ cut and the
double-dilepton-mass cut are so restrictive that background can be
neglected completely, making the limits signal-only statistically
limited, though signal efficiency is limited by angular detector
resolution at very low masses. We therefore set our limit at 3.8
signal events. Thirdly, for $m_{Z_D}$ near the $J/\Psi, \Psi(2S)$, and
$\Upsilon$ thresholds, quarkonium background is difficult to estimate
and possibly large. We mark those regions with gray bands in our limit
plots (taken from~\cite{pdg}).

\begin{figure}[t]
\begin{center}
\hspace*{-10mm}
\begin{tabular}{cc}
\includegraphics[width=8cm]{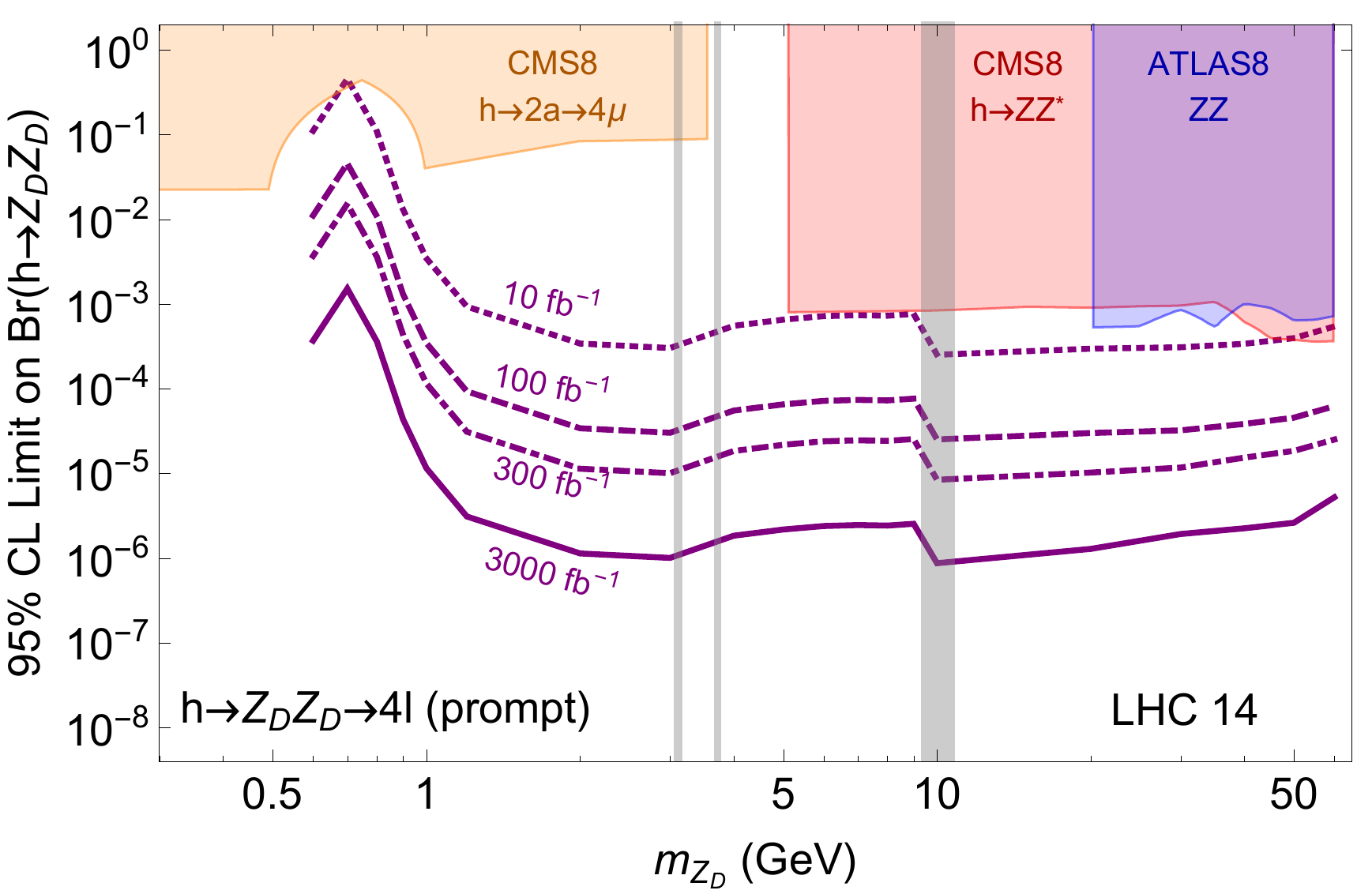}
&
\includegraphics[width=8cm]{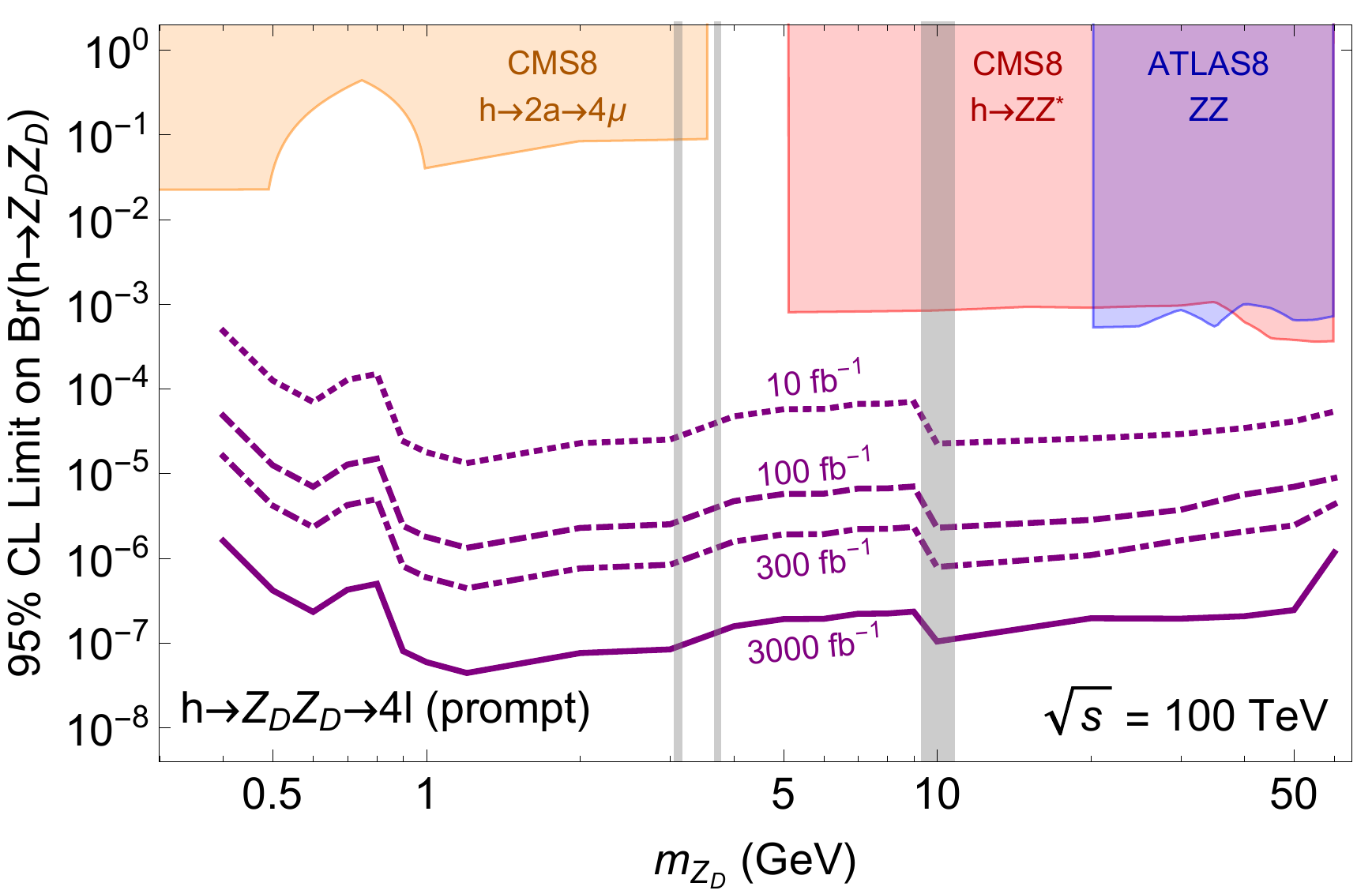}
\\
\includegraphics[width=8cm]
{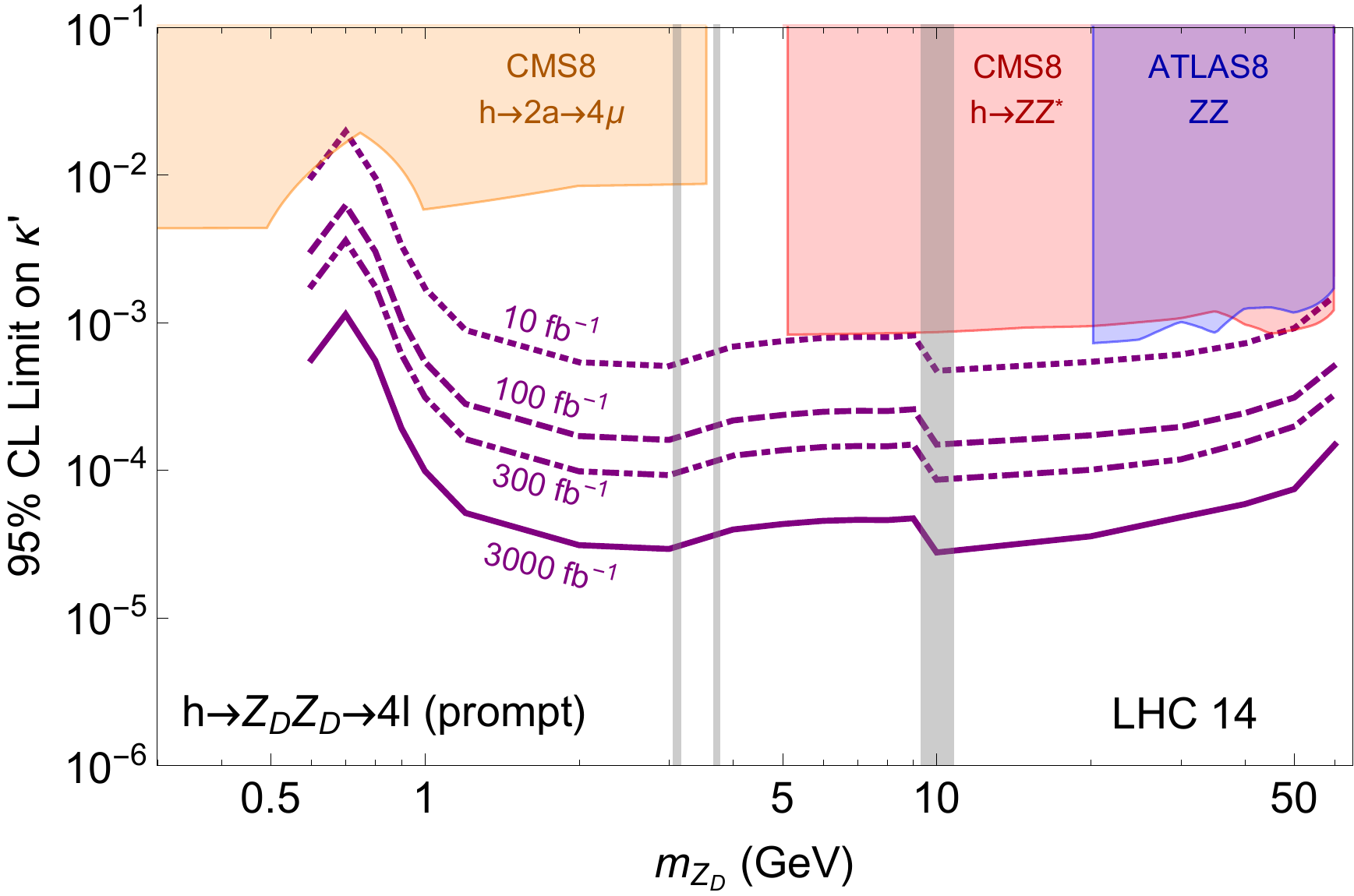}
&
\includegraphics[width=8cm]{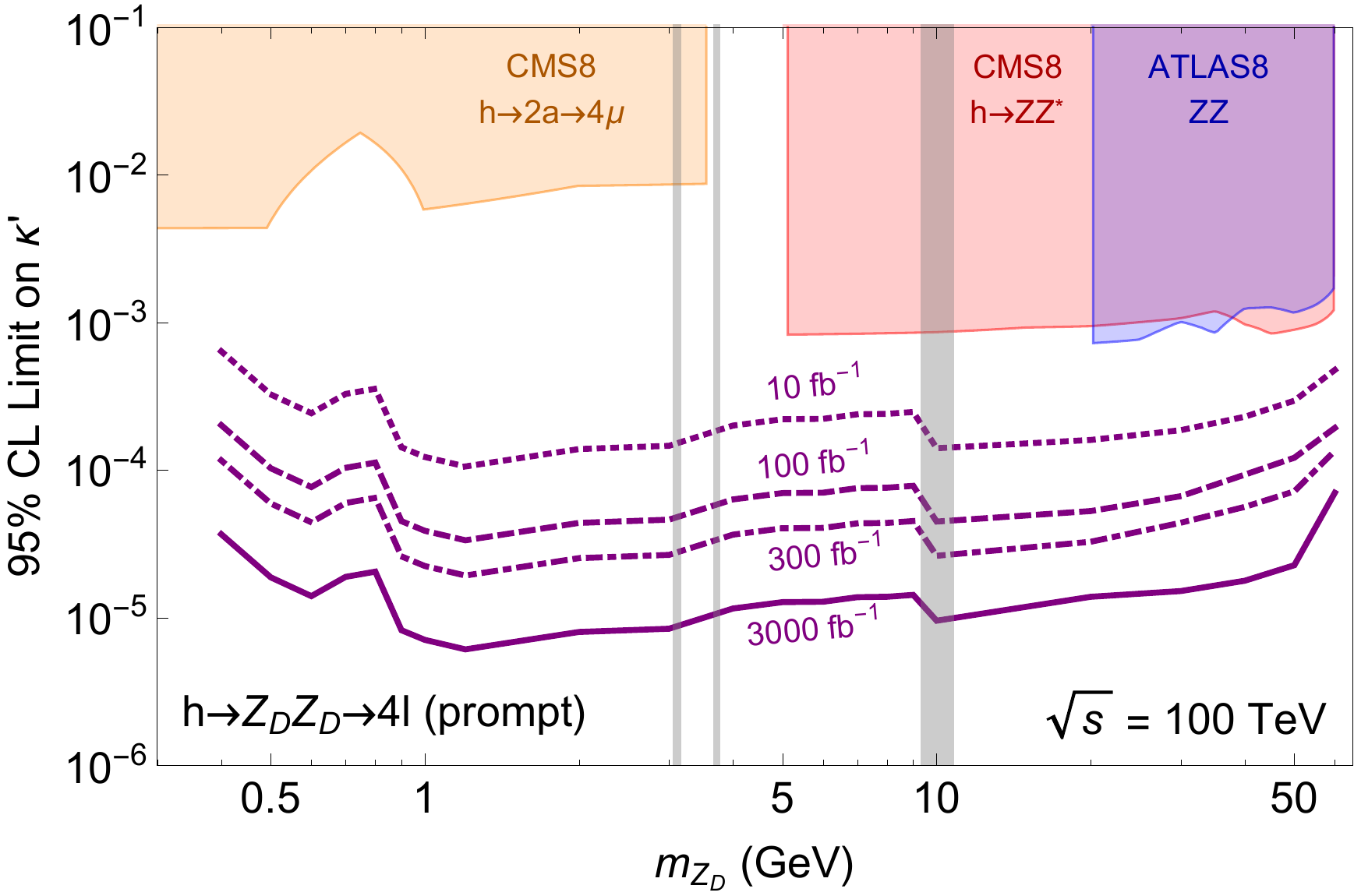}
\end{tabular}
\end{center}
\caption{{\small 
 Expected 95\% CLs limits on the total exotic Higgs decay branching ratio, $\mathrm{Br}(h\to Z_D Z_D)$ (top), and the effective Higgs mixing parameter $\kappa'$ (bottom) at the LHC (left) and a 100 TeV pp collider (right).  Gray bands correspond to regions where quarkonium background may invalidate our analysis. The limits obtained in \cite{Curtin:2013fra} from a recast of LHC Run 1 results are shown in red ($h\to Z Z^*\to 4\ell$ search by CMS \cite{CMS:xwa}) and blue (ATLAS $ZZ$ cross section measurement \cite{ATLAS:2013gma}) shaded regions. The limit from the CMS 8 TeV $h \to 2a \to 4\mu$ search \cite{CMS:2013lea} is shaded in orange, assuming the efficiencies for pseudoscalar and dark photon decay to muons are the same. 
      }}
\label{fig:BrZdZdlimit}
\end{figure}

The projected limits obtained for both the low and high mass $Z_D$ case are
shown in \fref{BrZdZdlimit}. 
Also shown are limits from 8 TeV LHC data, which supersede earlier  limits \cite{Clarke:2013aya, Abazov:2009yi, Chatrchyan:2012cg, Chatrchyan:2011hr}: a CMS search for $h\to2a \to 4\mu$ \cite{CMS:2013lea}, where we assume that efficiencies for pseudoscalar and dark vector decay to di-muon jets are the same, and recasts by \cite{Curtin:2013fra} of the CMS $h\to Z Z^*\to 4\ell$ search \cite{CMS:xwa} and the ATLAS $ZZ$ cross section measurement \cite{ATLAS:2013gma}. It will be possible to improve on these present limits with only 10 $\ifb$ of 14 TeV data. 
The HL-LHC probes Br$(h\to Z_DZ_D) \gtrsim 10^{-6}$, while a 100~TeV collider will this sensitivity by more than an order of magnitude. 
Since this search is signal limited, we expect this projected bound to be 2 to 3 orders of magnitude better than anything achievable by one of the proposed lepton colliders.

The dimensionless parameter $\kappa'$ that determines the exotic
branching fraction depends on both the Higgs portal coupling and the
singlet scalar mass, see \eref{kappaprime}.  Constraints for $\kappa'$ are also shown in
\fref{BrZdZdlimit} (bottom). HL-LHC measurements could probe this
parameter at the few $10^{-5}$ level. A 100 TeV collider could push
the sensitivity by almost another order of magnitude. The results of our study demonstrate the remarkable power of leptonic searches to set bounds on the mixing between the Higgs and an additional scalar. The sensitivity lies many orders of magnitude beyond the  (indirect) sensitivity to non-SM decays from Higgs coupling measurements. 

As mentioned in Sec.~\ref{subsec:higgs}, one could also imagine probing the Higgs portal by making use of direct dark Higgs production, which proceeds through its inherited SM couplings and occurs at the same order in $\kappa$ as $h\to Z_D Z_D$. For small enough $m_{Z_D}$ the dark Higgs would decay to two dark photons, giving a similar signal to the process studied in this section. However, we have checked that the exotic Higgs decay $h\to Z_D Z_D \to 4\ell$ provides the best sensitivity to $\kappa'$ if $h\to ss$ is kinematically forbidden, as we assume throughout this paper. For $m_s > m_h/2$, gluon fusion is the dominant production mode for the dark scalar. Due to the small width of the SM Higgs, 
\begin{equation}
\label{eq:singletproductionxsec}
\sigma (g g \to h)\times BR(h\to Z_D Z_D) \gg \sigma (g g \to s)\times BR(s\to Z_D Z_D)
\end{equation} 
even for $m_h/2 < m_s < m_h$. Since the $h \to Z_D Z_D$ search has very low background it provides the best sensitivity to the Higgs portal coupling $\kappa$.

\subsection{Constraints on kinetic mixing from displaced $Z_D$ decays}
\label{sec:displacedZD}

The $\mathrm{Br}(h\to Z_DZ_D)$ limits of \fref{BrZdZdlimit} assume that
all the $Z_D$ decay promptly. As shown in \fref{BrZdll}, this requires
$\epsilon \gtrsim 10^{-5} - 10^{-3}$ over the $2 m_\mu<m_{Z_D}<  m_h/2$ mass
range. Therefore, discovery of this exotic Higgs decay can also give sensitivity to  much smaller values of $\epsilon$ than any channel
that relies on $\epsilon$ for production, e.g.~the direct DY
production and $h\to Z_D Z^{(*)}$ decays considered in
Secs.~\ref{sec:zzd} and \ref{sec:dy}.   By considering macroscopic decay lengths of the $Z_D$, we can
 extend our sensitivity to even smaller values of $\epsilon$, 
giving an even more impressive sensitivity to the hypercharge
portal.  For example, extending the above analysis to include displaced
$Z_D$ decay to leptons with a decay length of up to $\sim 10$ cm gives
sensitivity to $\epsilon \sim 10^{-8} - 10^{-6}$, assuming a signal reconstruction efficiency similar to prompt $Z_D$ decays . In fact, a recent ATLAS analysis~\cite{Aad:2014yea} has set such limits on displaced dark photons in a supersymmetrized version of the model considered here. As we demonstrate below, similar analyses are highly motivated for the minimal dark photon model.

Let us assume for simplicity
that a displaced dilepton pair search has the same reconstruction
efficiency for all $Z_D$ decays within some length $L$ of the
interaction point as for prompt decays.\footnote{A displaced lepton
  search will have lower background but probably also lower signal
  efficiency, so this is a crude estimate, but it is sufficient to
  illustrate our point.} In that case, the \emph{effective} visible
exotic Higgs branching fraction to four leptons is given by 
\begin{equation}
\label{eq:Breff}
\mathrm{Br}_\mathrm{eff} = 
 \mathrm{Br}(h\to Z_D Z_D) \ \mathrm{Br}(Z_D \to \ell \ell)^2 \ P(L, \sqrt{s}, m_{Z_D},\epsilon),
\end{equation}
where $P(L, \sqrt{s}, m_{Z_D}, \epsilon)$ is the probability that both
$Z_D$ decay before traveling a length $L$:
\begin{equation}
\label{eq:PL}
P(L, \sqrt{s}, m_{Z_D}, \epsilon) =  \int d b_1 d b_2 \ f(\sqrt{s}, m_{Z_D}; b_1, b_2) \ 
\left[1-e^{-L/(b_1 \lambda)}\right] \left[1-e^{-L/(b_2\lambda)}\right].
\end{equation}
Here $\lambda = \lambda(m_{Z_D},
\epsilon) = c/\Gamma_{Z_D}(m_{Z_D}, \epsilon)$ is the proper decay
length of the dark photon, shown in \fref{BrZdll}, $b_{i} = |\vec p_{Z_{Di}}|/m_{Z_D}$ are the
boost factors of each $Z_D$ in the event, and $f(\sqrt{s}, m_{Z_D};
b_1, b_2)$ is the probability distribution for an event to have boost
factors $(b_1, b_2)$ for the two $Z_D$s. 

For the purposes of this estimate we can take our limits on
$\mathrm{Br}(h\to Z_D Z_D \to 4\ell)$ in \fref{BrZdZdlimit} to be
limits on $\mathrm{Br}_\mathrm{eff}$ in \eref{Breff}.  We now estimate
the ability of this search to constrain $\epsilon$ for different
values of $\mathrm{Br}(h\to Z_D Z_D)$ and reasonable choices of $L$,
to demonstrate the reach that might be achieved at HL-LHC or a future
100 TeV collider.

\begin{figure}[t]
\begin{center}
\includegraphics[width=0.48\textwidth]{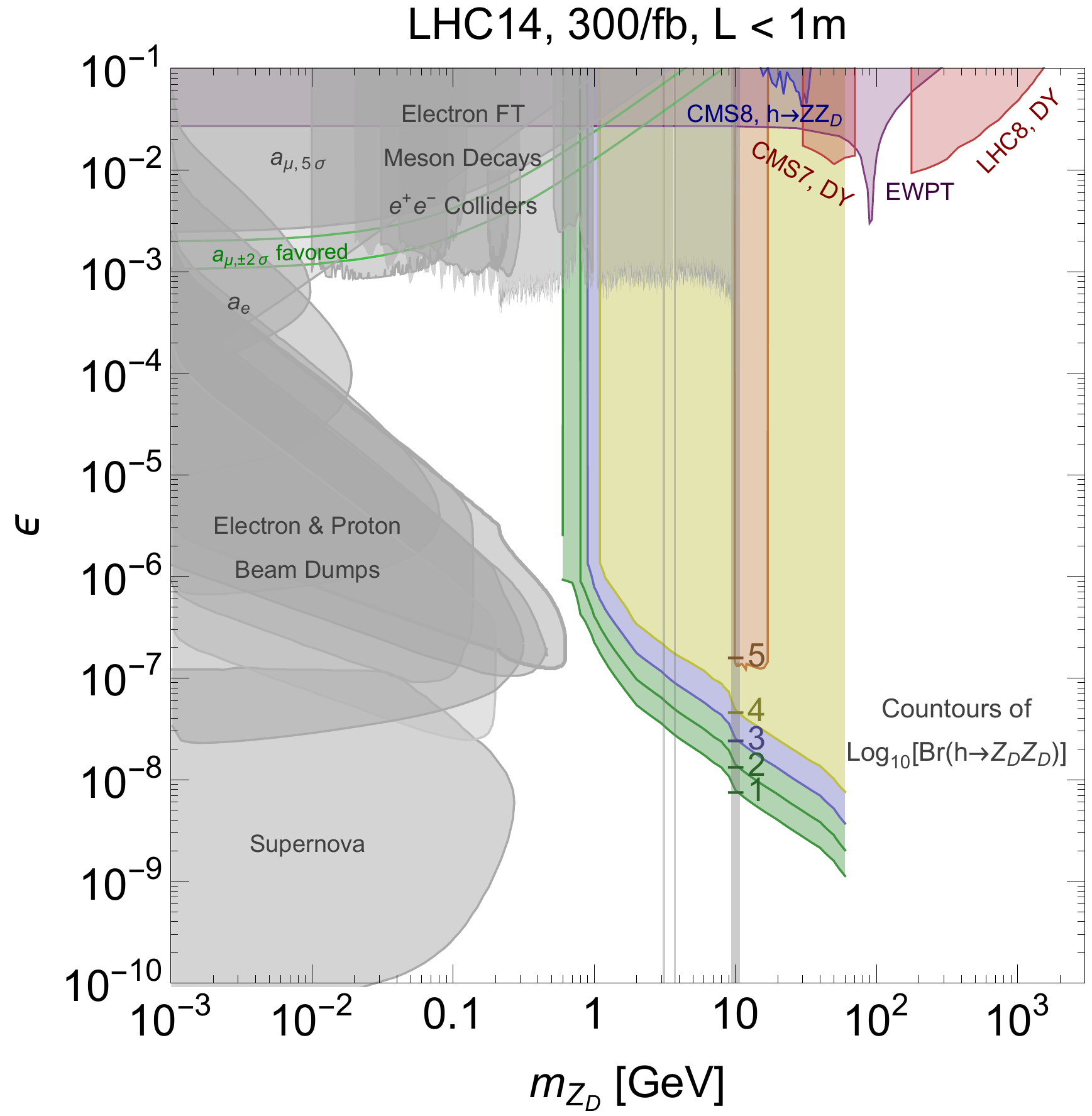} 
~~\includegraphics[width=0.48\textwidth]{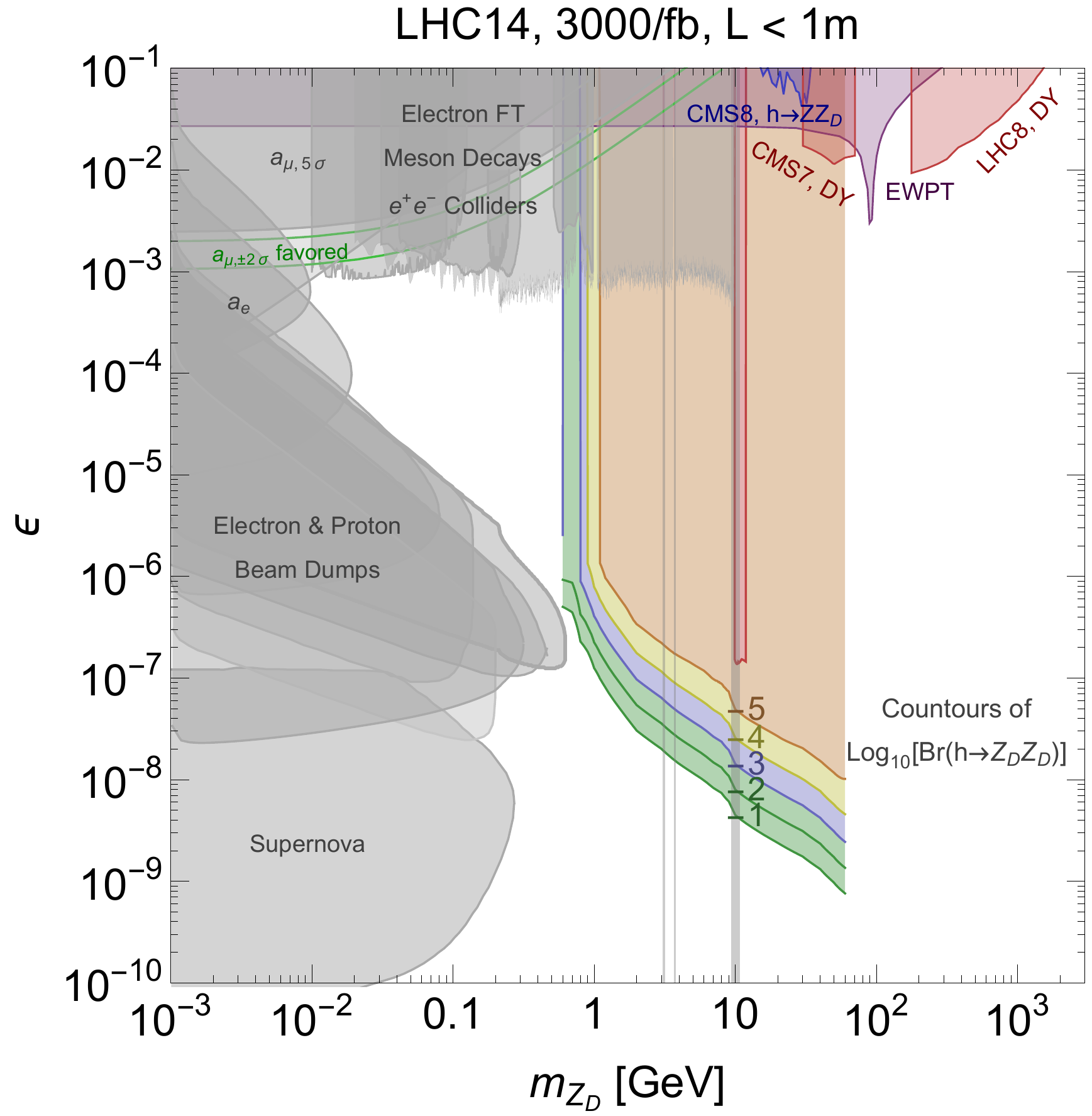} \\
\vskip 4mm
\includegraphics[width=0.48\textwidth]{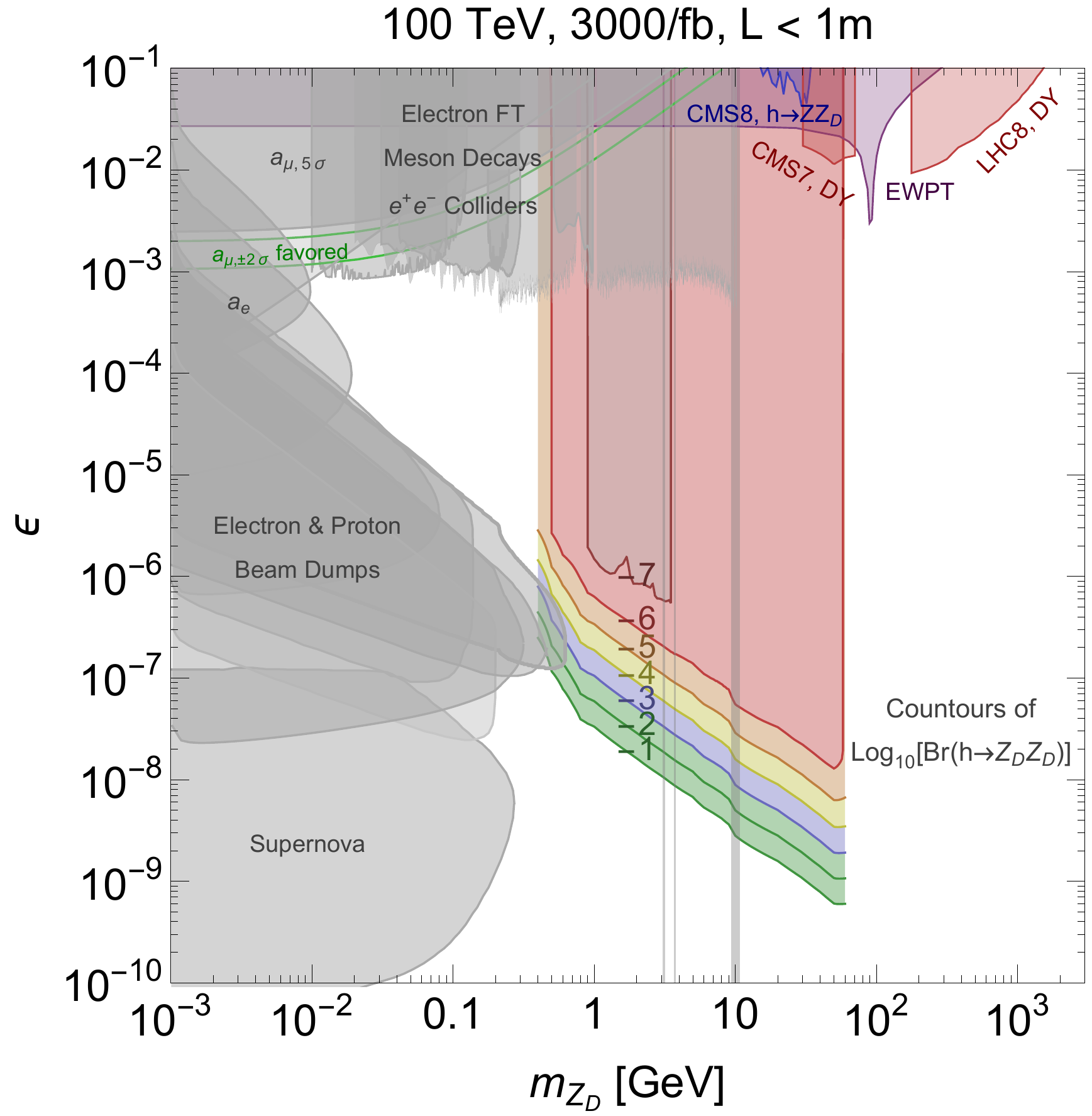} 
~~\includegraphics[width=0.48\textwidth]{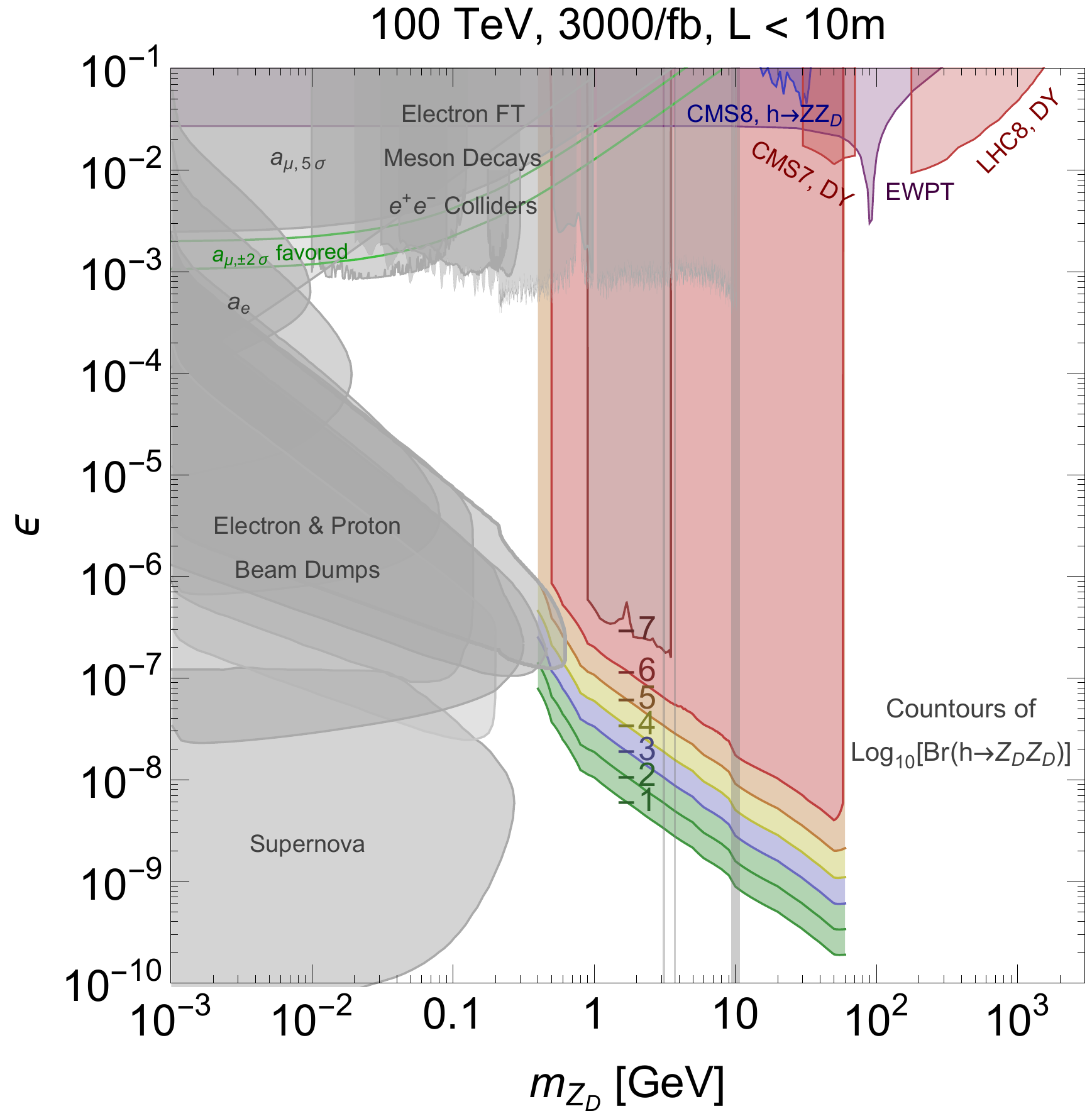} 
\end{center}
\caption{ 
Estimate of expected 95\% CLs limits on $\epsilon$ for
  different $\mathrm{Br}(h \to Z_D Z_D)$ at the LHC (\emph{top left}),
  HL-LHC (\emph{top right}), and a 100 TeV collider (\emph{bottom}),
  assuming a displaced lepton jet search has the same sensitivity to
  decays within the given distance from the interaction point as a
  prompt $Z_D Z_D$ search (see \fref{BrZdZdlimit}).  A detector size
  $L$ of $1$~m is assumed for all plots except for
  the bottom right plot, which assumes 10~m for the 100 TeV collider. Gray shaded regions show current constraints (see Sec.~\ref{sec:intro} for references). 
  }
\label{fig:zdzdlimits2}
\end{figure}

The joint distribution of boosts $f(\sqrt{s}, m_{Z_D}; b_1, b_2)$ can
be obtained from signal Monte Carlo events, but evaluating \eref{PL}
in the $(m_{Z_D}, \epsilon)$ plane is quite computationally expensive,
and neither necessary nor instructive for our estimate. We can
simplify \eref{PL} by first making use of the fact that the two boost
factors $b_1$ and $b_2$ are highly correlated and tend to be similar
in each event. Letting $f(\sqrt{s}, m_{Z_D}; b)$ be the boost factor
distribution of a single $Z_D$ in the signal event sample, $P$ can be
approximated as
\begin{equation}
\label{eq:PL2}
P(L, \sqrt{s}, m_{Z_D}, \epsilon) \approx  \int d b \ f(\sqrt{s}, m_{Z_D}; b) \ 
\left[1-e^{-L/(b \lambda)}\right]^2.
\end{equation}
The most important behavior of $P$ can be captured by the two limits
\begin{equation}
\label{eq:PL3a}
P(L, \sqrt{s}, m_{Z_D}, \epsilon) \approx \left \{
\begin{array}{lll}
1 & \mathrm{for} & L \gg b \lambda\\
\left( \frac{L}{b \lambda}\right)^2 & \mathrm{for}  & L \ll b \lambda
\end{array}
\right.
\end{equation}
where $b$ is a representative boost factor in the event
kinematics for a given $m_{Z_D}$. Expanding \eref{PL2} for small $L$,
\begin{eqnarray}
\label{eq:PL3}
\nonumber P(L, \sqrt{s}, m_{Z_D}, \epsilon) &\approx&  \int d b \ f(\sqrt{s}, m_{Z_D}; b) \ 
\left[ \left( \frac{L}{b \lambda}\right)^2 + \ldots \right] 
\\
\nonumber &=& 
\left(\frac{L}{\lambda}\right)^2 \int d b \ \frac{f(\sqrt{s}, m_{Z_D}; b)}{b^2} + \ldots
\\
&=&
\left(\frac{L}{\tilde b \lambda}\right)^2 + \ldots,
\end{eqnarray}
where we have defined an `effective average boost factor' 
\begin{equation}
\label{eq:btilde}
\tilde b = \tilde b(m_{Z_D}, \sqrt s) \equiv  \left[  \int d b \ \frac{f(\sqrt{s}, m_{Z_D}; b)}{b^2} \right]^{-1/2}.
\end{equation}
We can then write 
\begin{equation}
\label{eq:PL4}
P(L, \sqrt{s}, m_{Z_D}, \epsilon) \approx  \left[1-e^{-L/(\tilde b \lambda)}\right]^2, 
\end{equation}
 which gives the correct limit, \eref{PL3a}, for small $L$. 
The $m_{Z_D}$ dependence of $\tilde b$ is
nearly identical for
$\sqrt s = 100$ and $14 \tev$, since the $Z_D$ kinematics are dominated by
the decay of a Higgs particle produced mostly near threshold. In fact,
assuming both $Z_D$ come from the decay of a stationary Higgs gives
\begin{equation}
b = \sqrt{\frac{m_h^2}{4 m_{Z_D}^2} - 1},
\end{equation}
which is a very good approximation for $\tilde b$ everywhere except
the near threshold region where $m_{Z_D}\approx m_h/2$.

Using \eref{PL4}, it is straightforward to convert limits on
$\mathrm{Br}_\mathrm{eff}$ (\eref{Breff}) from \fref{BrZdZdlimit} to
$\epsilon$ limits as a function of $m_{Z_D}$ for different
$\mathrm{Br}(h \to Z_D Z_D)$. This is shown in \fref{zdzdlimits2} for
the LHC and HL-LHC, assuming displaced vertices out to 1~m from the
interaction point can be reconstructed, as well as for a 100 TeV
collider with $3000\ifb$, assuming displaced vertex reconstruction out
to either 1 or 10~m.

A 10\% invisible Higgs branching ratio to two long-lived $Z_D$ is not presently excluded \cite{ATLAS:2013pma, Chatrchyan:2014tja}, and even future lepton colliders like ILC and TLEP would only constrain such an invisible decay at the $0.5\%$ level \cite{Dawson:2013bba, Peskin:2013xra}. For such relatively large $\mathrm{Br}(h\to Z_D Z_D)$, the HL-LHC (100 TeV collider) offers sensitivity to $\epsilon \gtrsim 10^{-9} - 10^{-6}$  ($ 10^{-10} -  10^{-7}$). This is is many orders of magnitude beyond anything achievable with searches
that rely on the hypercharge portal for $Z_D$ production. Even a very
small Higgs portal can allow us to glimpse deeply into the dark
sector.

\section{Impact of future detector design}\label{sec:improvements}

Detector capabilities are important for assessing the detailed reach
of $pp$ colliders for both decays $h\to Z Z_D\to 4\ell$ and $h \to Z_D
Z_D \to 4\ell$.  Our forecasts are based on LHC8 lepton performance,
which may differ in several aspects from the ultimate detector
performance at a 100 TeV collider.  To illustrate the importance of
detector design on the reach, we examine in this section how our
expected limits depend on key assumptions about lepton identification
and reconstruction.

A future detector could perform \emph{worse} than an LHC8 detector with regards to lepton reconstruction $p_T$ thresholds.  In the
analyses of Secs.~\ref{sec:zzd} and~\ref{sec:zdzd}, we implicitly
assumed $100\%$ trigger efficiency for $p^T_{L1} > 20 \gev$ and
$p^T_{L2} > 10 \gev$. As we show in \fref{etamax} (left), these thresholds can be raised by 10 or even 20 GeV (for $\ell_2$) with relatively little loss of signal acceptance. More serious is the dependence on the minimum lepton reconstruction threshold, which was assumed to be $p_T > 7, 5 \gev$ for electrons and muons respectively. Raising this threshold to $10
\gev$ degrades signal efficiency by about $50\%$ in both $h\to Z
Z_D\to 4\ell$ and $h \to Z_D Z_D \to 4\ell$ analyses, with a $\sim
20\%$ loss of $\epsilon$ and $\kappa'$ sensitivity. This is shown in
\fref{varyinput100pT} for a 100 TeV collider.

\begin{figure}[t]
\begin{center}
\hspace*{-10mm}
\begin{tabular}{ccc}
\includegraphics[height=3.5cm]{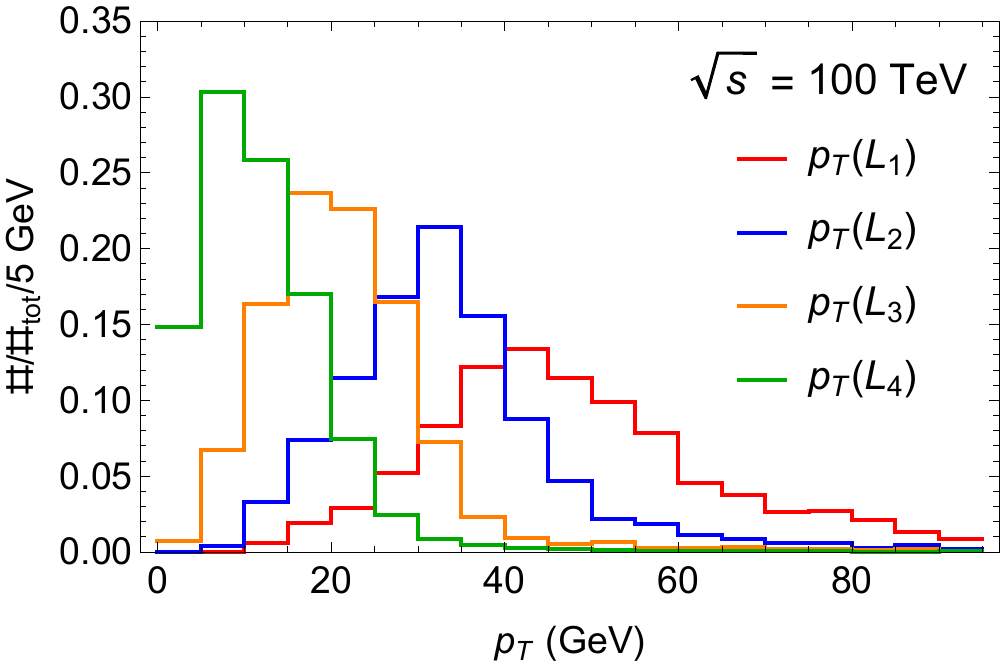}
&
\includegraphics[height=3.5cm]{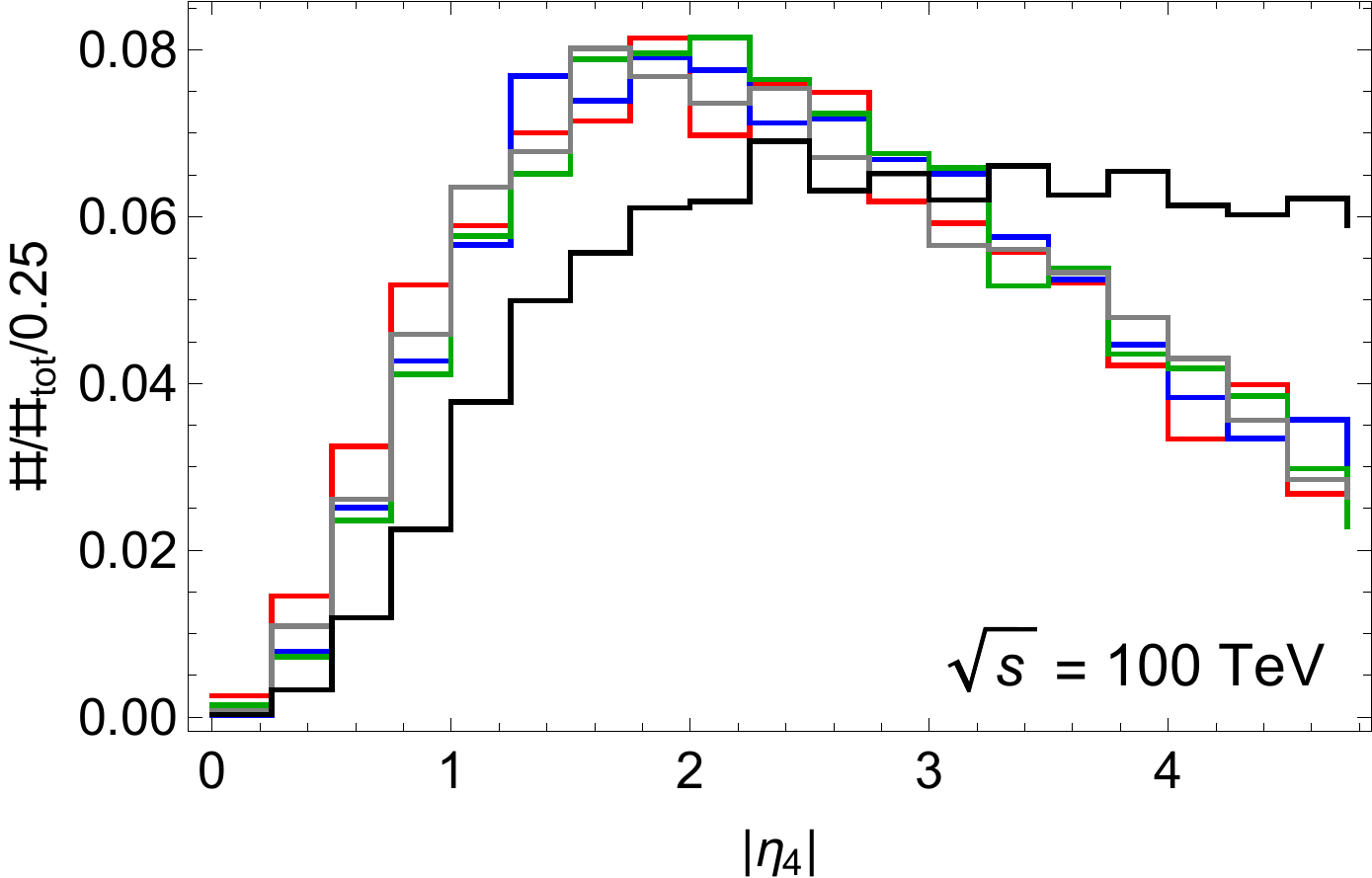}
&
\includegraphics[height=3.5cm]{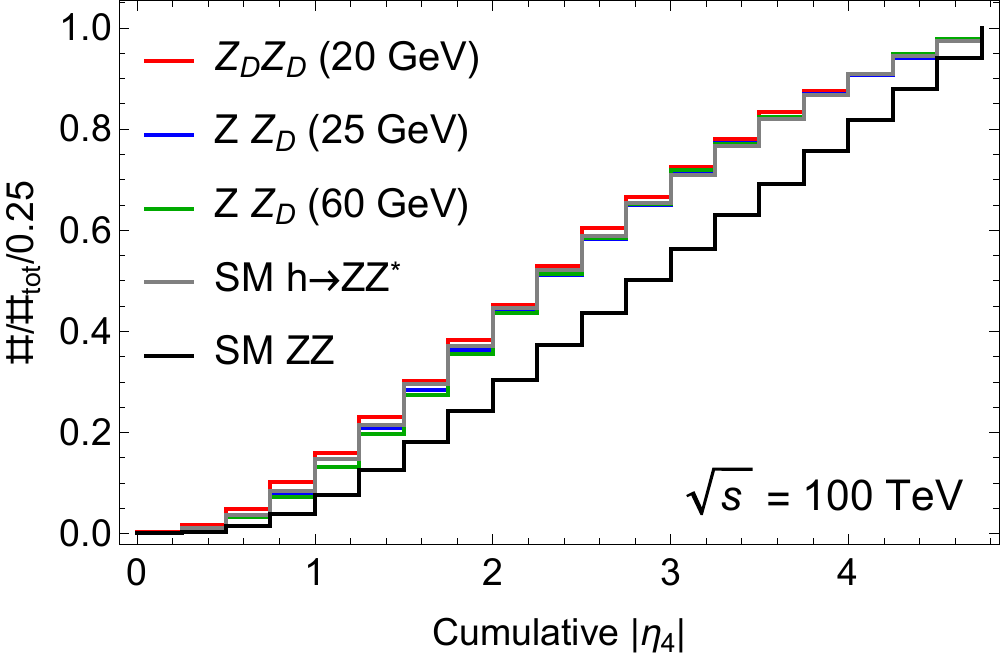}
\end{tabular}
\end{center}
\caption{{\small  \emph{Left:} Distribution of $p^T_{L1,2,3,4}$ for $h\to Z_D Z_D \to 4 \ell$ with $m_{Z_D} = 20 \gev$. 
\emph{Middle} and \emph{right}: $|\eta_4|$ and cumulative $|\eta_4|$ distribution for $h\to Z_D Z_D \to 4\ell$ $(m_{Z_D} = 20 \gev)$, $h \to Z Z_D \to 4 \ell$ $(m_{Z_D} = 25, 60 \gev)$ and SM backgrounds $h \to Z Z^*$, $ZZ$. The only applied cut is $120 < m_{4 \ell} < 130 \gev$.  
}
\label{fig:etamax}}
\end{figure}

On the other hand, a future detector could perform \emph{better} than an LHC8 detector with regards to rapidity acceptance and mass resolution. 
Signal acceptance can be notably improved if lepton
coverage is extended to higher values of $|\eta|$.
Fig.~\ref{fig:etamax} (middle and right) shows the distribution of the
maximum $|\eta|$ among the four leptons in both SM and BSM $h\to
4\ell$ events, along with the distribution in the main background $Z^
{(*)}Z^ {(*)} $, before any $p_T$ requirements are imposed.  Note
that, for both SM and BSM Higgs bosons, requiring all four leptons to
satisfy $ |\eta | <2.5$ eliminates approximately half of the signal
events. We therefore investigate the possibility of raising the maximum rapidity to 4. Conversely, background rejection could be improved by improving the dilepton mass resolution. We consider the change in reach if it is possible to employ an optimistic mass window of 
\beq
\label{eq:newwindow}
|M_{\ell\ell} - m_{Z_D} | < 0.015 M_{\ell\ell},
\eeq
for both electrons and muons.  The impact of these two possible improvemenets on $\epsilon$ and $\kappa'$ sensitivity for a 100 TeV collider is shown in Fig.~\ref{fig:varyinput100}.

Extending lepton $\eta$ coverage by itself does not necessarily
improve the sensitivity to $h\to Z^{ (*)} Z_D$, as acceptance for the
main background, SM $h\to ZZ^*\to 4\ell$, increases as well.  This is
especially notable for $m_{Z_D}> m_h-m_Z$, where SM background is
non-negligible.  In this region, extending lepton $\eta$ coverage is
actually detrimental without other improvements, as can be seen from
the dotted line in Fig.~\ref {fig:varyinput100} (left).  The importance of
reducing SM background can be seen from the dashed line, which shows
the improvement in sensitivity given the improvement in lepton mass
resolution according to Eq.~(\ref{eq:newwindow}).  Notably, combining
the improvement in mass resolution with extended $\eta$ coverage
yields a $\gtrsim 20\%$ improvement in the reach in $\epsilon$ (see dot-dashed line in the figure).
We also consider the possibility of increasing the rapidity coverage at the HL-LHC \cite{CMS-PAS-FTR-13-003}.  For $h\to Z Z_D$, this improves signal
acceptance by $\approx 25\%$; however without an improvement in mass
resolution over our existing projections, we find it does not help
to improve limits on $h\to Z_D \ell\ell\to 4\ell$.

\begin{figure}[t!]
\begin{center}
\hspace*{-10mm}
\begin{tabular}{cc}
\includegraphics[height=4.3cm]{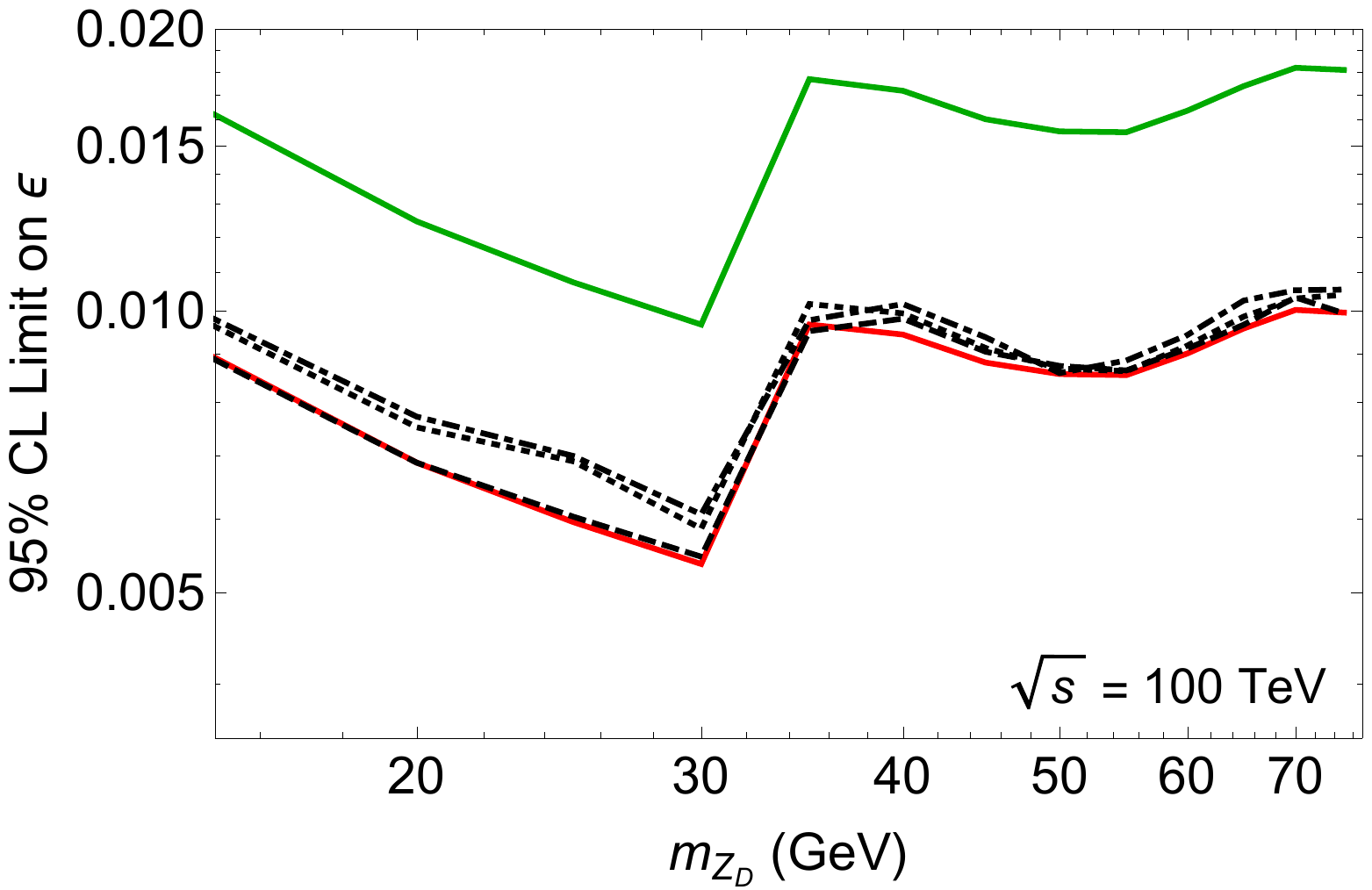}
&
\includegraphics[height=4.3cm]{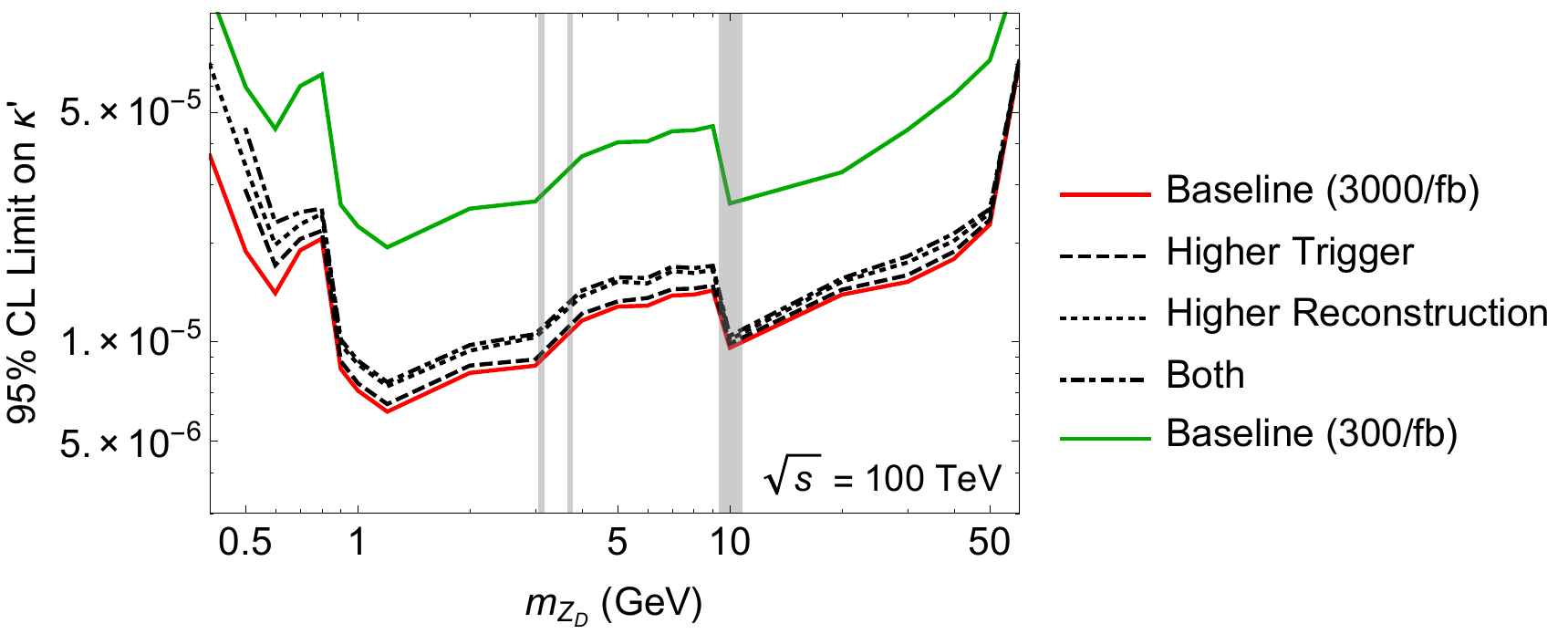}
\end{tabular}
\end{center}
\caption{{\small Estimated 95\% CLs sensitivity to $\epsilon$ from
    $h\to ZZ_D\to 4\ell$ (left) and $\kappa'$ from $h\to Z_D Z_D \to
    4\ell$ (right) at a 100 TeV collider with 3000 $\ifb$.  Baseline
    selection criteria in red, raising trigger thresholds
    $p_{L_{1,2}}^T$ from $(20,10) \gev$ to $(30, 20) \gev$ gives the
    dashed line, raising lepton reconstruction thresholds
    $p_T^{e,\mu}$ from $(7,5) \gev$ to $10 \gev$ gives the dotted
    line, dot-dashed shows both.
    The projected $300\ifb$ baseline limit is shown for scale in green.
     }
\label{fig:varyinput100pT}}
\end{figure}

\begin{figure}
\begin{center}
\hspace*{-10mm}
\begin{tabular}{cc}
\includegraphics[height=4.3cm]{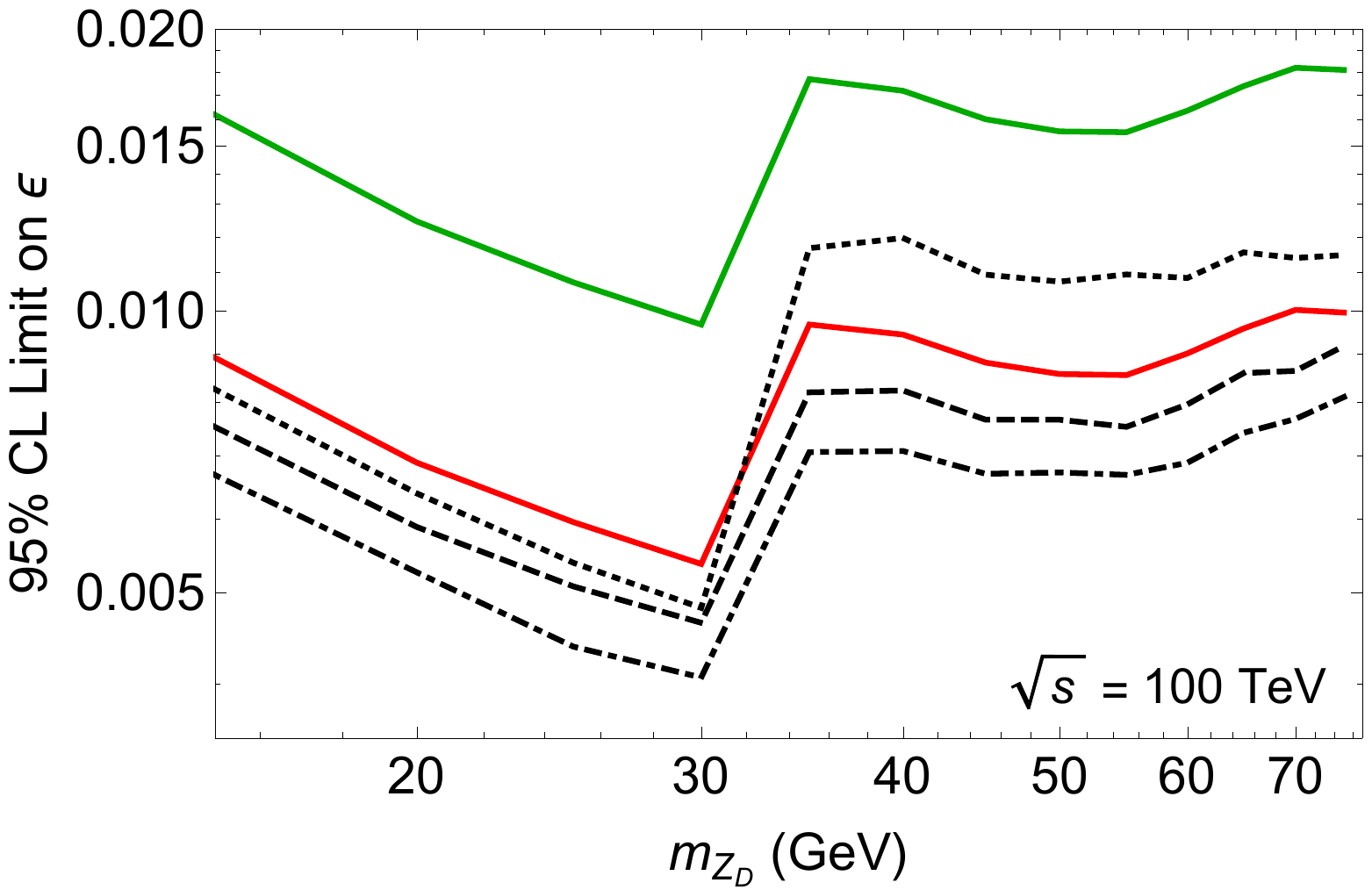}
&
\includegraphics[height=4.3cm]{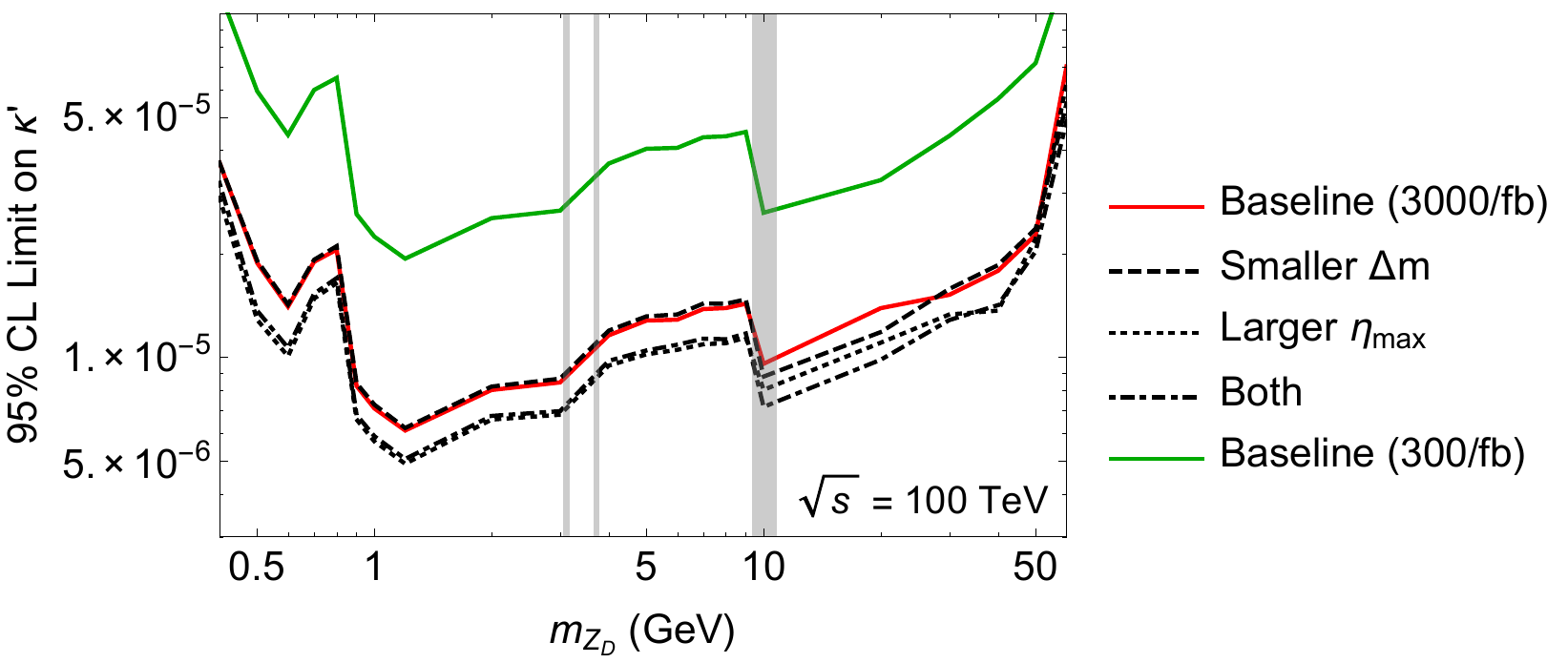}
\end{tabular}
\end{center}
\caption{{\small Estimated 95\% CLs sensitivity to $\epsilon$ from
    $h\to ZZ_D\to 4\ell$ (left) and $\kappa'$ from $h\to Z_D Z_D \to
    4\ell$ (right) at a 100 TeV collider with 3000 $\ifb$.  Baseline
    selection criteria in red, improved mass resolution (according to
    Eq.~(\ref{eq:newwindow})) in the dashed line, increased lepton
    acceptance ($|\eta|<4$) in the dotted line, dot-dashed shows
    both. The projected $300\ifb$ baseline limit is shown for scale in
    green.  }
\label{fig:varyinput100}}
\end{figure}

The situation is simpler for $h\to Z_D Z_D$. Due to the
double-dilepton-mass cut, backgrounds are so low that the increased
signal acceptance for larger $\eta$ coverage more than offsets the
elevated background levels. The improved mass window by itself
slightly improves higher-mass limits on $\kappa'$ (where there is some
background) but has no effect on the background-free low-mass
limits. Best results are achieved by utilizing both improvements,
which increases $\kappa'$ sensitivity by about $25\%$.

\begin{figure}[!t]
\begin{center}
\includegraphics[width=0.95\textwidth]{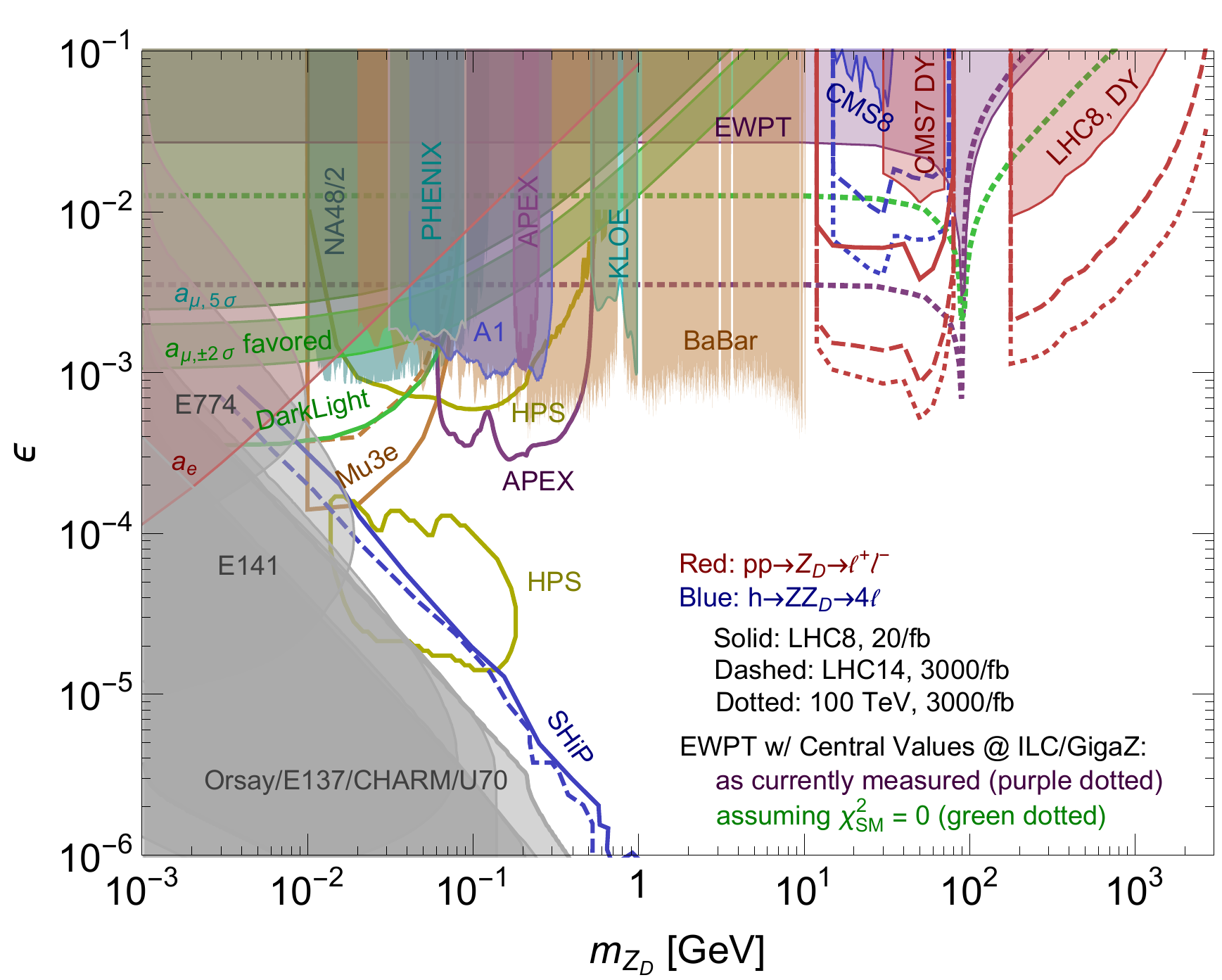} 
\end{center}
\caption{Summary of dark photon constraints and prospects (see Sec.~\ref{sec:intro} for references).  
High-energy colliders (LHC14, 100~TeV, ILC/GigaZ) are uniquely sensitive to dark photons with $m_{Z_D}\gtrsim 10$~GeV, 
while precision QED observables and searches at $B$- and $\Phi$-factories, beam dump experiments, and 
fixed target-experiments probe lower masses.  
Dark photons can be detected at high-energy colliders in a significant part of open parameter space 
in the exotic decay of the 125 GeV Higgs boson, $h\to Z Z_D\to 4\ell$, (blue curves) 
in Drell-Yan events, $pp\to Z_D\to\ell\ell$, (red curves) and through 
improved measurements of electroweak precision observables (green/purple dashed curves).  
Note that all constraints and prospects assume that the dark photon decays directly to SM particles, except 
for the precision measurements of the electron/muon anomalous magnetic moment and the electroweak observables.  
If, in addition to kinetic mixing, the 125 GeV Higgs mixes with the dark Higgs that breaks the dark $U(1)$, then the decay 
$h\to Z_D Z_D$ would set constraints on $\epsilon$ that are orders of magnitude more powerful than other searches down to 
dark photon masses of $\sim 100$~MeV, see Fig.~\ref{fig:zdzdlimits2}. 
}
\label{fig:summaryplot}
\end{figure}
%

\section{Discussion and Conclusions}\label{sec:conclusions}
%
Dark sectors with a broken $U(1)_D$ gauge group that kinetically mixes with the SM hypercharge 
are well motivated and appear in a variety of new physics scenarios.
In this paper, we showed that high-energy proton-proton and electron-positron colliders, like the LHC14, a  
100~TeV collider, and an ILC/GigaZ, have excellent sensitivity to dark photons.   
In fact, they may provide the {\it only} probe for dark photons  with masses above $10$~GeV, as high-intensity beam-dump experiments 
or $B$-factories do not have enough energy to probe this mass region.   
Moreover, the 125~GeV Higgs boson plays a pivotal role in these searches, providing additional motivation to search for 
its possible exotic (non-standard) decays.  

If the only connection between the dark and SM sectors is kinetic mixing (i.e. the the hypercharge portal), then the 
dark photon can be produced in Drell-Yan events and in the exotic Higgs decay $h\to Z_D Z^{(*)}$.  
In addition, it would change the SM expectation for electroweak precision observables.  
A renormalizable mixing between the 125~GeV Higgs and the hidden-sector Higgs (i.e.~a Higgs portal) is expected to be present if the dark photon mass is generated by a Higgs mechanism. This would  allow for the exotic Higgs decay $h\to Z_D Z_D$.  
We have investigated these various possibilities, summarizing them in \fref{summaryplot} for the pure hypercharge 
portal case and in \fref{zdzdlimits2} if there is an additional Higgs mixing.  

Our conclusions are the following: 
\begin{itemize}
\item Drell-Yan production is the most promising discovery channel of dark photons if the dark Higgs does not mix with SM-like Higgs boson (see Fig.~\ref{fig:DYlimits} and \cite{Cline:2014dwa, Hoenig:2014dsa}).  Recasts of existing LHC Run 1 data already set some of the best limits for some ranges of dark photon masses above 10~GeV, and especially for masses above about 180~GeV.   Data from the upcoming HL-LHC run and a potential future 100~TeV collider can probe $\epsilon\gtrsim 9 \times
  10^{-4}$ and $4 \times 10^{-4}$, approaching the same sensitivity to dark photon masses {\it above} 
10~GeV that BaBar data achieved {\it below} 10~GeV. Additional experimental analyses of the DY dilepton mass spectrum near the $Z$-peak are motivated to help  fill in the gap between the high- and low-mass DY bounds.
\item Exotic Higgs decays $h \to Z_D Z^{(*)} \to 4\ell$ provide an additional powerful probe of dark photons 
with masses below the $Z$-boson (see Fig.~\ref{fig:BrZdZlimit}), and serve as complementary discovery channels to 
DY production.
Moreover, a discovery in the Drell-Yan channel alone would  not be sufficient to pinpoint the properties of a new vector boson; the sizeable branching ratio predicted for 
$h \to Z_D Z^{(*)}$ in the case of a kinetically-mixed $Z_D$ makes this exotic Higgs decay a key diagnostic in establishing the properties of any newly discovered vector boson.
\item Electroweak precision constraints have the distinctive advantage of being independent of the dark photon decay mode (see 
Fig.~\ref{fig:EWFitBound}).  
Existing constraints require $\epsilon\lesssim 3\times 10^{-2}$ for masses below $\sim 80$~GeV.  
The upcoming HL-LHC can probe $\epsilon$ down to almost $10^{-2}$, while an ILC/GigaZ can probe down to almost 
$3\times 10^{-3}$ in the same mass range.  Above the $Z$-pole, the constraint and prospects weaken, but are stronger than 
any other existing constraint up to about 180~GeV. {\it If} the dark photon decays directly to SM particles, the above-mentioned searches in DY events and exotic Higgs decays will 
be significantly more powerful in the entire mass range above 10~GeV than measurements of electroweak observables.  
\item For $\epsilon \lesssim 10^{-3}$, direct production of the dark photon through the hypercharge portal is very unlikely at current or future planned colliders. 
However, if the dark Higgs mixes with the 125~GeV Higgs, the spectacular exotic decay  $h\to Z_D Z_D \to 4\ell$ gives an additional probe into the hidden sector through the Higgs portal. 
The effective Higgs mixing parameter, see Eq.~(\ref{eq:kappaprime}), can be constrained by the HL-LHC (100 TeV collider) to be  $\kappa' \lesssim \rm{few}\times 10^{-5}$ ($\rm{few}\times 10^{-6}$), see \fref{BrZdZdlimit}.
Since the detection of the $h\to Z_D Z_D \to 4\ell$ decay relies on the dark photon decaying directly to leptons, any such discovery would also yield sensitivity to the hypercharge portal at the $\epsilon \lesssim 10^{-7} - 10^{-6}$ level, which is the smallest kinetic mixing for which almost all dark photons decay inside of the detector. 
\item Future lepton colliders will constrain the invisible Higgs decay branching fraction at the $0.5\%$ level \cite{Dawson:2013bba, Peskin:2013xra}. If $\mathrm{Br}(h\to Z_D Z_D)$ is of this size, then then $\epsilon$ values as low  $10^{-9} - 10^{-6}$  ($ 10^{-10} -  10^{-7}$) can be probed at the HL-LHC (100 TeV collider) by looking for highly displaced dark photon decays, see \fref{zdzdlimits2}. 
\end{itemize}

Our forecasts for the sensitivity of exotic Higgs decays are based on LHC8 lepton performance, which may differ in several aspects from the ultimate detector performance at a 100 TeV collider. We have investigated the sensitivity of our conclusions to varying those assumptions, and find that the plausible range of lepton $p_T$ thresholds, mass resolutions, and rapidity acceptances can affect the limits on $\epsilon$ and $\kappa'$ at the $\sim 20\%$ level. 
Our results can also be applied to estimate sensitivity to the rare SM $h\to 4\ell$ decays  via exclusive quarkonia decays \cite{Isidori:2013cla, Kagan:2014ila, Gonzalez-Alonso:2014rla}.  

Finally, we have made a fully consistent MadGraph implementation of the minimal dark photon model publicly available for future investigations, see Appendix \ref{sec:mgmodel}.

This work showcases one example of the impressive sensitivity to light hidden sectors provided by future colliders. 
Discovery requires both large center of mass energies and enormous production rates for relatively light particles, and in particular the SM Higgs boson.   Future hadron colliders will offer unique discovery avenues onto both frontiers, {\em provided}  sensitivity to relatively low-$p_T$ objects is maintained.

\bigskip

{\em Acknowledgements:} We thank Wolfgang Altmannshofer, Brian Calvert, Aviana Essig, Eder Izaguirre,
Michael Peskin, Kevin Pedro, and Graham W. Wilson for helpful discussions.  
DC is supported in part by the
NSF under Grants PHY-PHY-0969739 and PHY-1315155, and by the Maryland Center for Fundamental Physics. 
RE is supported by the DoE Early Career research program DESC0008061 and 
through a Sloan Foundation Research Fellowship.  
JS is grateful to the Mainz Institute for Theoretical
Physics for its hospitality and partial support during the completion
of this work. SG would like to thank the Center for Future High Energy Physics
(CFHEP) in Beijing for hospitality and partial support.
Research at Perimeter Institute is supported by the Government of Canada through Industry
Canada and by the Province of Ontario through the Ministry of Economic Development
$\&$ Innovation.  Part of this work was completed at the Aspen Center for Physics, which operates under the NSF Grant 1066293

\appendix

\section{Tables of Branching Ratios and $Z_D$ full width}\label{sec:brtables}

For $m_{Z_D} > 12 \gev$, Tab.~\ref{tab:brtables} shows the dark photon
total width, leptonic branching fraction, and the exotic Higgs decay
branching fractions. This includes 3-loop QCD corrections using the
results of~\cite{Chetyrkin:2000zk}.  In each case, $\ell$ is taken to
mean \emph{both} $e$ and $\mu$. ($Z_D$ partial widths to each are
identical in this mass range.) See Sec.~\ref{subsec:gauge} for more
details.

For $m_{Z_D} < 12 \gev$, the same information can be computed using
$R(s)$ data from~\cite{pdg} and LO leptonic $Z_D$ widths from
\eref{width-ZDff} to compute the total $Z_D$ width via
\eref{Zdtotalwidth}.  The resulting leptonic branching ratios are then
given by \eref{Zdbranchingratios}, while the exotic Higgs decay
branching fractions for $h\to Z_D Z_{(D)} \to 4\ell$ can then be
computed analytically with Eqns. (\ref{eq:htoZDZ}) and (\ref{eq:htoZDZD}).

\begin{table}[th!]
\vspace*{-10mm}
\begin{center}
\begin{tabular}{|ccccc|}
\hline
$\displaystyle \frac{m_{Z_D}}{\gev}$ & 
$\displaystyle \frac{\Gamma_{Z_D}}{\epsilon^2 (\gev)}$ & 
$\displaystyle \mathrm{Br}(Z_D \to \ell \ell)$ & 
$\displaystyle \frac{\mathrm{Br}(h\to Z_D Z^{(*)} \to 4\ell)}{\epsilon^2} $ & 
$\displaystyle \frac{ \mathrm{Br}(h\to Z_D Z_D \to 4\ell)}{{\kappa'}^2}$ \\ 
\hline 
\hline
 12 & 0.217 & 0.289 & 0.00180 & 93.6 \\
 14 & 0.253 & 0.288 & 0.00252 & 91.4 \\
 16 & 0.290 & 0.288 & 0.00338 & 89.0 \\
 18 & 0.327 & 0.287 & 0.00439 & 86.4 \\
 20 & 0.365 & 0.286 & 0.00555 & 83.4 \\
 22 & 0.403 & 0.285 & 0.00681 & 80.2 \\
 24 & 0.442 & 0.284 & 0.00814 & 76.9 \\
 26 & 0.482 & 0.283 & 0.00940 & 73.3 \\
 28 & 0.522 & 0.281 & 0.0104 & 69.6 \\
 30 & 0.564 & 0.280 & 0.0108 & 65.8 \\
 32 & 0.607 & 0.278 & 0.00961 & 61.9 \\
 34 & 0.651 & 0.275 & 0.00599 & 58.0 \\
 36 & 0.697 & 0.273 & 0.00380 & 54.1 \\
 38 & 0.746 & 0.270 & 0.00312 & 50.2 \\
 40 & 0.797 & 0.267 & 0.00280 & 46.4 \\
 42 & 0.851 & 0.263 & 0.00263 & 42.6 \\
 44 & 0.909 & 0.259 & 0.00253 & 38.9 \\
 46 & 0.972 & 0.254 & 0.00247 & 35.3 \\
 48 & 1.04 & 0.249 & 0.00242 & 31.7 \\
 50 & 1.12 & 0.244 & 0.00238 & 28.2 \\
 52 & 1.20 & 0.238 & 0.00235 & 24.8 \\
 54 & 1.29 & 0.231 & 0.00232 & 21.4 \\
 56 & 1.40 & 0.223 & 0.00229 & 17.9 \\
 58 & 1.53 & 0.215 & 0.00225 & 14.2 \\
 60 & 1.68 & 0.206 & 0.00221 & 10.1 \\
 62 & 1.86 & 0.196 & 0.00217 & 4.28 \\
 64 & 2.09 & 0.186 & 0.00212 & - \\
 66 & 2.37 & 0.174 & 0.00208 & - \\
 68 & 2.73 & 0.163 & 0.00203 & - \\
 70 & 3.21 & 0.150 & 0.00198 & - \\
 72 & 3.87 & 0.138 & 0.00193 & - \\
 74 & 4.79 & 0.125 & 0.00189 & - \\
 76 & 6.17 & 0.113 & 0.00184 & - \\
 78 & 8.34 & 0.102 & 0.00179 & - \\
 80 & 12.0 & 0.0914 & 0.00175 & - \\
 82 & 18.9 & 0.0827 & 0.00170 & - \\
 84 & 34.1 & 0.0754 & 0.00165 & - \\
 86 & 78.2 & 0.0695 & 0.00161 & - \\
 88 & 308. & 0.0647 & 0.00156 & - \\
 \hline
\end{tabular}
\end{center}
\vspace*{-5mm}
\caption{
  Branching ratios and total widths in the dark photon model as a function of dark photon mass $m_{Z_D}$, for $\epsilon, \kappa' \ll 1$. Three-loop QCD corrections are included using the results of \cite{Chetyrkin:2000zk}. See Sec.~\ref{subsec:gauge} for  details. 
}
\label{tab:brtables}
\vspace{-10mm}
\end{table}

The methods we employ in Sec.~\ref{subsec:gauge} can also be
applied to compute the various widths and branching ratios for
$m_{Z_D} > m_Z$. However, the LO approximation for $Z_D$ partial
widths, \eref{width-ZDff}, is an excellent approximation for such high
masses. The three-body width $\Gamma(h\to Z_D Z^{(*)}\to 4\ell)$ can
be computed with \eref{hto4lthreebody} or in MadGraph, see Appendix
\ref{sec:mgmodel}.

We also make tables of all branching ratios and partial widths used in
this paper, for all $m_{Z_D} < m_Z$, available for download at the
Exotic Higgs Decay Working Group website and the website for the
Madgraph model. See Appendix \ref{sec:mgmodel} for the urls.

\section{$Z_D$ contributions to precision electroweak observables}\label{sec:PEWappendix}
In this Appendix, we give more details on computing the effects of $Z_D$ on several of the electroweak precision observables 
used for our fit (see Eq. (\ref{eq:data}), with the exception of $\sin^2 \theta_{\rm eff}^\ell(Q_{\rm{FB}})$).

The first set of observables, the mass of the $Z$ and of the $W$
bosons and the total width of the $W$ boson, $\Gamma_W$, are only affected by the
shift in the $Z$ mass. In particular, the physical mass of the $Z$
boson does not correspond anymore to the input value in (\ref{fitpar})
but it is given by the expression in Eq. (\ref{eq:masses}) where
$m_{Z,0}$ is our input value, over which we are marginalizing. The
computation of the $W$ boson mass follows closely the computation in
the framework of the SM. More specifically, for the $W$ mass, we have
to solve iteratively the equation
\begin{equation} \label{eq:mW}
  m_W^2\left(1-\frac{m_W^2}{m_Z^2}\right)=\frac{\pi\alpha}{\sqrt 2
    G_F}(1+\Delta r),
\end{equation} 
where $G_F$ is the Fermi constant
and $\alpha$ is the fine structure constant. $\Delta r$, which also depends on the $W$ mass, summarizes all
radiative corrections, computed fully at the two-loop level in the
Standard Model~\cite{Awramik:2003rn}.  The leading NP effect in our
theory comes from the shift of $Z$ mass, compared to the input
value $m_{Z,0}$ that enters  (\ref{eq:mW}).  For the $W$ boson
width, $\Gamma_W$, we employ the one loop parametrization
in~\cite{Cho:2011rk}. From there we can see that, again, the main NP
effect comes from the shift in the $Z$ mass through $m_W$.

 Next, we
discuss those observables measured at the $Z$ peak that are affected both by the shift in the $Z$ mass and by the shift of the $Z$
couplings.
The partial widths of the $Z$ into fermions can be expressed in terms
of the effective vector- and axial-vector couplings, $v_f$ and $a_f$,
of the Z boson to leptons at
the Z-pole $i \bar f \gamma^\mu (v_f+a_f\gamma_5)f\,Z_\mu$,
as~\cite{Cho:2011rk}
\beq\label{eq:GammaZf}
\Gamma_f=\frac{G_F m_Z^3}{6\sqrt 2 \pi}\left[\left((v_f^2+\delta_{im\,\kappa}^f\right)C_{fV}+a_f^2C_{fA}\right]\left(1+\frac{3}{4}Q^2\frac{\hat \alpha(m_Z)}{\pi}\right)+\Delta_{EW/QCD}^f\,,
\eeq
where $Q$ is the electric charge of the fermion $f$, while $C_{fV}$ and
$C_{fA}$ describe corrections to the color factor in the vector
and axial-vector currents, and $\Delta_{EW/QCD}^f$ are mixed QED and QCD
corrections taken from~\cite{Bardin:1999yd}. Finally,
$\delta_{im\,\kappa}^f$ is the correction from the imaginary part of the loop-induced mixing of the photon and the Z boson. In our theory,  several partial widths of the $Z$ boson will be affected due to the
shift in the $Z$ couplings $v_f,a_f$, as well as by kinematics, since $\Gamma_f\propto m_Z$ and the physical $m_Z$ is not given by the input value $m_{Z,0}$, see \eref{massmatrix}.

From these partial widths, it is easy to compute the remaining
electroweak observables: the Z-peak hadronic cross-section,
$\sigma_{\rm had}^0$ and the partial width ratios, $R_\ell^0,R_c^0,R_b^0$:
\begin{eqnarray}
\sigma_{\rm had}^0&=&\frac{12\pi}{m_Z^2}\frac{\Gamma_e\Gamma_{\rm had}}{\Gamma_Z^2},\\
R_\ell^0&=&\frac{\Gamma_{\rm had}}{\Gamma_\ell},\,
R_q^0=\frac{\Gamma_{q}}{\Gamma_{\rm had}},
\end{eqnarray}
where $q=c,b$,  $\Gamma_{\rm
  had}=\Gamma_u+\Gamma_d+\Gamma_c+\Gamma_s+\Gamma_b$, and $\Gamma_Z=\Gamma_{\rm
  had}+\Gamma_\ell+\Gamma_\nu$, with $\Gamma_\nu$ being the partial width of the $Z$ into neutrinos.

\begin{table}\begin{center}
\begin{tabular}{|c|c||c|c|}
\hline
 Observable & Measurement &Observable & Measurement  \\
  \hline\hline
$ m_Z$  & $(91.1875\pm 0.0021)$ GeV  & $A_\ell$ & $0.1499\pm 0.0018$  \\\hline
 $\Gamma_Z$  & $(2.4952\pm 0.0023)$ GeV & $A_b$ & $0.923\pm 0.020$ \\\hline
 $\sigma_{\rm had}^0$ & $(41.540\pm 0.037)$ nb  & $A_c$ & $0.670\pm 0.027$  \\\hline
 $R_\ell^0$ & $20.767\pm  0.025$ &  $A_{FB}^{\ell,0}$ & $0.0171\pm 0.0010$  \\\hline
 $R_b^0$  & $0.21629 \pm 0.00066 $ & $A_{FB}^{b,0}$ & $0.0992\pm 0.0016$ \\\hline
 $R_c^0$ & $0.1721 \pm 0.0030$ & $A_{FB}^{c,0}$ & $0.0707 \pm 0.0035$  \\\hline
  $m_W$ & $(80.385 \pm 0.015)$ GeV & $\Gamma_W$ & $(2.085 \pm 0.042)$ GeV \\\hline
  $m_t$ & $(173.34 \pm 0.76)$ GeV& $m_h$ & $(125.14\pm 0.24)$ GeV\\\hline
   $\Delta\alpha^{(5)}_{\rm had}$ & $0.2757 \pm 0.0001$ & &
 \\\hline
\end{tabular}
\caption{\label{exp-values} Experimental values, as measured at LEP, SLC, Tevatron and LHC, of the several EWPOs. Note that for $A_\ell$ we are using an average of LEP and SLC measurements, for the top mass we are using the newest world average, combination of Tevatron and LHC results~\cite{ATLAS:2014wva} and not the newest CMS result in~\cite{CMS:2014ima}.}\label{tab:measurements}
\end{center}
\end{table}

Furthermore, the left-right asymmetry parameters $A_\ell,A_c,A_b$ can be
expressed at the tree level by
\begin{equation} 
A_f^{\rm{tree}}=\frac{2v_f/a_f}{1+(v_f/a_f)^2}\,.
\end{equation}
 To take into account higher order SM corrections, we express the lepton asymmetry parameters as functions of the effective weak mixing angles $\sin^2 \theta_{\rm eff}^f$, 
\beq
A_f\equiv\frac{2(1-4 |Q|\sin^2 \theta_{\rm eff}^f)}{1+(1-4 |Q|\sin^2 \theta_{\rm eff}^f)^2}\,.\eeq
In the SM, $\sin^2 \theta_{\rm eff}^f$ is the solution to the equation
\begin{equation} \label{eq:sin2theta}
\sin^2 \theta_{\rm eff}^f=\left(1-\frac{m_W^2}{m_Z^2}\right)(1+\Delta\kappa_f)\,,
\end{equation} 
where the first part takes into account the relation between the Fermi
constant $G_F$ and the $W$ boson mass. The second part takes into account the corrections to the $Z$-fermion vertex form factors and it depends only weakly on the value of $m_W$. The SM numerical result of Eq.~(\ref{eq:sin2theta}) is expressed in terms of a fitting function that depends on the input parameters in (\ref{fitpar}). The fitting function of our theory, which we have to use for the left-right asymmetry parameters $A_f$, will be given by the SM function (with the
appropriate input value $m_{Z,0}$) plus tree-level corrections due to the shift of the $Z$ couplings.

Finally, the forward-backward asymmetries, $A_{FB}^{\ell,}, A_{FB}^{c,0},
A_{FB}^{b,0}$ are given by
\beq
A_{FB}^{f,0}=\frac{3}{4}A_\ell A_f.\eeq

For completeness, we show in Tab.~\ref{exp-values} the collection of experimental
values used in our fit. 

\section{MadGraph implementation of higgsed dark photon model}\label{sec:mgmodel}

We make a fully consistent MadGraph 5 \cite{Alwall:2014hca}
implementation of the higgsed dark photon model, constructed in
FeynRules 2.0 \cite{Alloul:2013bka}, publicly available. It can be
found at the website of the \emph{Exotic Higgs Decay Working Group},
\begin{center}
\url{http://exotichiggs.physics.sunysb.edu/},
\end{center}
or directly at 
\begin{center}
\url{http://insti.physics.sunysb.edu/~curtin/hahm_mg.html}.
\end{center}

\bibliography{zdark}
\bibliographystyle{JHEP}


\end{document}